\documentclass{aa}  
\pdfoutput=1

\usepackage[varg]{txfonts}
\usepackage{natbib}

\usepackage{graphicx}
\usepackage{amsmath}
\usepackage{amssymb}
\usepackage{hyperref}
\hypersetup{
    colorlinks   = true,
    urlcolor     = blue,
    linkcolor    = blue,
    citecolor   = blue
}
\usepackage{multirow}
\usepackage{tabularx}
\usepackage[group-separator={,}]{siunitx}
\usepackage{bm}
\usepackage{xcolor}
\usepackage[export]{adjustbox}
\usepackage{url}

\usepackage[rightcaption]{sidecap}

\newcommand{\rhomBar}{\bar{\rho}_{\rm m}}

\newcommand{\Scr}{\Sigma_{\rm cr}}
\newcommand{\Scrt}{\tilde{\Sigma}_{\rm cr, d}}

\newcommand{\zl}{z_{\rm d}}
\newcommand{\zs}{z_{\rm s}}

\newcommand{\rl}{r_{\rm d}}
\newcommand{\rs}{r_{\rm s}}
\newcommand{\rls}{r_{\rm ds}}

\newcommand{\Mh}{M_{\rm h}}

\newcommand{\nh}{n_{\rm h}}
\newcommand{\bh}{b_{\rm h}}

\newcommand{\Ng}{n_{\rm g}}
\newcommand{\ngBar}{\bar{n}_{\rm g}}

\newcommand{\eone}{\epsilon_{1}}
\newcommand{\etwo}{\epsilon_{2}}
\newcommand{\et}{\epsilon_{\rm t}}
\newcommand{\ex}{\epsilon_{\times}}

\newcommand{\gt}{\gamma_{\rm t}}

\begin{document} 

   \title{Weak lensing constraints on the stellar-to-halo mass relation of galaxy groups with simulation-informed scatter}
   \titlerunning{Stellar-to-halo mass relation of galaxy groups}
   \authorrunning{S.-S. Li et al.}

\author{Shun-Sheng Li\inst{1,2}
        \and
        Henk Hoekstra\inst{1}
        \and
        Konrad Kuijken\inst{1}
        \and
        Matthieu Schaller\inst{1,3}
        \and
        Joop Schaye\inst{1}
}

\institute{Leiden Observatory, Leiden University, Einsteinweg 55, 2333 CC Leiden, the Netherlands\\
            \email{ssli@strw.leidenuniv.nl}
\and
Aix-Marseille Universit\'{e}, CNRS, CNES, LAM, Marseille, France
\and
Lorentz Institute for Theoretical Physics, Leiden University, PO Box 9506, 2300 RA Leiden, the Netherlands
            }

   \date{Received 5 November 2024 / Accepted 7 July 2025}
 
  \abstract
  {Understanding the scaling relation between baryonic observables and dark matter halo properties is crucial not only for studying galaxy formation and evolution, but also for deriving accurate cosmological constraints from galaxy surveys. In this paper, we constrain the stellar-to-halo mass relation of galaxy groups identified by the Galaxy and Mass Assembly survey, using weak lensing signals measured by the Kilo-Degree Survey. We compare our measured scaling relation with predictions from the FLAMINGO hydrodynamical simulations and the \textsc{L-Galaxies} semi-analytical model. We find a general agreement between our measurements and simulation predictions for halos with masses ${\gtrsim}10^{13.5}\ h_{70}^{-1}{\rm M}_{\odot}$, but observe slight discrepancies with the FLAMINGO simulations at lower halo masses. We explore improvements to the current halo model framework by incorporating simulation-informed scatter in the group stellar mass distribution as a function of halo mass. We find that including a simulation-informed scatter model tightens the constraints on scaling relations, despite the current data statistics being insufficient to directly constrain the variable scatter. We also test the robustness of our results against different statistical models of miscentring effects from selected central galaxies. We find that accounting for miscentring is essential, but our current measurements do not distinguish among different miscentring models.}


   \maketitle

\section{Introduction}

According to the current standard model of cosmology, galaxies form within cold dark matter halos, which originate from small initial density perturbations amplified by gravitational instability. This framework predicts a strong correlation between properties of galaxies and their host dark matter halos (see \citealt{Wechsler2018ARAA..56..435W}, for a review). Dark matter halos dominate the local gravitational potential and provide the environment for galaxy formation and evolution (e.g.~\citealt{Blumenthal1984Natur.311..517B,Davis1985ApJ...292..371D}). Conversely, various baryonic processes associated with galaxy formation and evolution, especially the energetic feedback processes from supernovae and active galactic nuclei (AGNs), impact the matter distribution on small scales (e.g.~\citealt{Daalen2011MNRAS.415.3649V,Hellwing2016MNRAS.461L..11H,Chisari2018MNRAS.480.3962C,Daalen2020MNRAS.491.2424V}). Therefore, a comprehensive understanding of the galaxy-halo connection is crucial not only for studying galaxy formation and evolution, but also for ensuring the accuracy of cosmological constraints inferred from observations of large-scale structures (e.g.~\citealt{Semboloni2011MNRAS.417.2020S,Schneider2020JCAP...04..019S,Castro2021MNRAS.500.2316C,Debackere2021MNRAS.505..593D}).

Given that dark matter halos typically host multiple galaxies, galaxy groups and clusters play a central role in studying the galaxy-halo connection. Although massive galaxy clusters serve as a powerful tool for constraining cosmological models, they are rare and represent extreme conditions in the Universe (see \citealt{Allen2011ARAA..49..409A}, for a review). In contrast, galaxy groups, which host the majority of present-day galaxies and a significant portion of baryons, are more representative (e.g.~\citealt{Robotham2011MNRAS.416.2640R}). Besides, the relatively low gravitational binding energy of galaxy groups makes them particularly valuable for studying the impact of baryonic feedback (e.g.~\citealt{McCarthy2010MNRAS.406..822M, Kettula2015MNRAS.451.1460K}). They also contribute significantly to the cosmic shear signal, making it important to understand their properties in weak lensing studies (e.g.~\citealt{Semboloni2011MNRAS.417.2020S,Debackere2020MNRAS.492.2285D}).

However, robustly identifying galaxy groups is challenging. One approach is to use spectroscopic surveys with high spatial and redshift completeness. The Galaxy and Mass Assembly project (GAMA, \citealt{Driver2011MNRAS.413..971D}) represents one such effort. With a spectroscopic completeness of 95 percent for galaxies down to an $r$-band magnitude of $19.65$, and sky coverage of approximately $250$ square degrees, GAMA offers the highest redshift density available for such an extensive area to date~\citep{Driver2022MNRAS.513..439D}. The galaxy group catalogue, as one of its key products, is an invaluable resource for studying group properties~\citep{Robotham2011MNRAS.416.2640R}.

The next challenge lies in measuring the dark matter properties of galaxy groups. This becomes evident even when inferring basic properties such as halo masses. Unlike massive galaxy clusters, where X-ray measurements of the intracluster medium provide good mass estimates (e.g.~\citealt{Cavaliere1976AA....49..137C,Evrard1996ApJ...469..494E}), galaxy groups have much fainter X-ray signals, limiting the effectiveness of this technique (e.g.~\citealt{Eckmiller2011AA...535A.105E,Pop2022arXiv220511528P,Bahar2022AA...661A...7B}). Additionally, baryonic processes such as cooling, star formation, and feedback cause deviations from hydrostatic equilibrium, which is a common assumption in X-ray mass measurements. These deviations introduce biases in mass estimates that rely on this assumption (e.g.~\citealt{Rasia2006MNRAS.369.2013R,Biffi2016ApJ...827..112B,Barnes2021MNRAS.506.2533B,Logan2022AA...665A.124L}).

Weak gravitational lensing provides an alternative approach to estimate halo masses. It measures the subtle yet coherent distortions in the shapes of background galaxies, caused by the gravitational field of a foreground lens (see \citealt{Bartelmann2001PhR...340..291B}, for a review). These distortions directly trace the matter distribution of the foreground lens, enabling mass estimation without assumptions about its dynamical state (e.g.~\citealt{Tyson1990ApJ...349L...1T,Kaiser1993ApJ...404..441K,Linden2014MNRAS.443.1973V,Hoekstra2015MNRAS.449..685H,Robertson2024AA...681A..87R}).

However, the weak lensing signals induced by individual galaxy groups often have low signal-to-noise ratios (S/Ns), limiting the precision of individual mass estimates. Therefore, in practice, we select the lens sample based on certain observable properties, then measure the averaged signals to estimate the average masses. This approach boosts the measurement of the S/N and provides a statistical description of the scaling relation between halo mass and the corresponding observable properties (e.g.~\citealt{Johnston2007arXiv0709.1159J,Leauthaud2010ApJ...709...97L,Han2015MNRAS.446.1356H,Viola2015MNRAS.452.3529V,Rana2022MNRAS.510.5408R}).

To interpret the averaged weak lensing signal, we need a statistical model to describe the galaxy and dark matter properties. The halo model, combined with halo occupation statistics, offers such a theoretical framework (e.g.~\citealt{Seljak2000MNRAS.318..203S,Peacock2000MNRAS.318.1144P,Cooray2002PhR...372....1C,Berlind2002ApJ...575..587B,Yang2003MNRAS.339.1057Y,Vale2004MNRAS.353..189V,Cooray2006MNRAS.365..842C,Bosch2013MNRAS.430..725V}). It assumes that all dark matter exists within virialised halos and that galaxies are populated within these dark matter halos. After parameterising the desired scaling relation between galaxy observables and halo properties within the model, we can directly constrain it by fitting to the averaged weak lensing signals (e.g.~\citealt{Guzik2002MNRAS.335..311G,Mandelbaum2006MNRAS.368..715M,Uitert2011AA...534A..14V,Cacciato2014MNRAS.437..377C}).

In practice, the halo model contains many theoretically motivated and empirically informed ingredients, each governed by a set of free parameters that may not be well constrained by the available data. Furthermore, the potential degeneracy among these parameters can complicate the interpretation of results derived from the halo model (e.g.~\citealt{Viola2015MNRAS.452.3529V}). To address these challenges, we can improve the halo model by refining certain components or providing informed priors for model parameters, using our knowledge of galaxy formation and dark matter properties (e.g.~\citealt{Smith2003MNRAS.341.1311S,Mead2015MNRAS.454.1958M,Fortuna2021MNRAS.501.2983F}).

This knowledge is often encoded in cosmological simulations, which numerically model various physical processes to track the non-linear evolution of matter in an expanding cosmological background, using initial conditions constrained by the well-measured cosmic microwave background. With decades of development and vastly improved computational power, modern cosmological simulations can solve gravitational $N$-body problems with high accuracy for volumes that are sufficient for interpreting observations from the next generation of galaxy surveys~(see \citealt{Angulo2022LRCA....8....1A}, for a recent review). Nevertheless, there are still challenges in accounting for the more complicated baryonic effects in these simulations.

The two main strategies to address baryonic effects currently in active development are semi-analytical modelling and hydrodynamical simulations. The former still relies on gravity-only simulations but populates galaxies based on a semi-analytical model (SAM, e.g.~\citealt{White1978MNRAS.183..341W,White1991ApJ...379...52W,Cole1991ApJ...367...45C}). This SAM consists of a set of simplified equations to account for the key baryonic processes that affect the formation and evolution of galaxies. Because the ingredients in a SAM are theoretically simplified and often phenomenological, not all parameters can be rigidly determined from physical arguments. This results in a number of assumptions and free parameters that require calibration based on observational data.

The latter, hydrodynamical simulations, adopt a more sophisticated approach by self-consistently solving the co-evolution of dark matter and baryons from the outset~(see \citealt{Vogelsberger2020NatRP...2...42V,Crain2023ARAA..61..473C}, for some recent reviews). However, due to the vast range of physical processes involved and the limited resolution of simulations, some effective models are still necessary to capture subgrid baryonic processes that are not resolved by the numerical calculations. These subgrid models also contain free parameters that require calibration based on observations (see \citealt{Schaye2015MNRAS.446..521S}, for a discussion).

We study this intriguing interplay between cosmological simulations and weak lensing analysis in this paper. First, we constrain the stellar-to-halo mass scaling relation of GAMA galaxy groups through a halo model-based analysis of weak lensing signals measured from the Kilo-Degree Survey (KiDS, \citealt{Jong2013ExA....35...25D,Kuijken2015MNRAS.454.3500K}). The constrained scaling relation is then compared with predictions from the latest FLAMINGO cosmological hydrodynamical simulations~\citep{Schaye2023MNRAS.526.4978S,Kugel2023MNRAS.526.6103K} and the \textsc{L-Galaxies} SAM run on the IllustrisTNG gravity-only simulations~\citep{Henriques2015MNRAS.451.2663H,Ayromlou2021MNRAS.502.1051A} to assess the reliability of the simulation predictions. Alternatively, one can directly compare the averaged weak lensing signal between observational data and simulations by including measurement effects into simulations (e.g.~\citealt{Velliscig2017MNRAS.471.2856V,Jakobs2018MNRAS.480.3338J,Gouin2019AA...626A..72G}).

After validating the simulation predictions, we explore how insights from these recent simulations can be used to improve our current halo model. For this, we also use results from the IllustrisTNG hydrodynamical simulations~\citep{Nelson2019ComAC...6....2N}, which cover the low halo mass range not constrained by our data but necessary for our halo modelling. We focus on improving one of the key halo model ingredients: the scatter in the group stellar mass distribution as a function of halo mass. Besides, we assess the impact of different statistical models for the miscentring of identified central galaxies. These investigations guide the future development of more robust, simulation-informed halo models, which will be crucial for interpreting the significantly improved weak lensing measurements from upcoming surveys like the ESA \textit{Euclid} space mission~\citep{Euclid2025AA...697A...1E} and the \textit{Rubin} Observatory Legacy Survey of Space and Time~(LSST, \citealt{Ivezic2019ApJ873111I}).

The rest of the paper is structured as follows. Sections~\ref{Sec:data} and \ref{Sec:sim} describe the data and simulations, respectively. Section~\ref{Sec:measure} details the measurement of weak lensing signals and the covariance matrix. Our baseline model is outlined in Sect.~\ref{Sec:model}, and its results are compared to previous studies and simulation predictions in Sect.~\ref{Sec:res}. We introduce a simulation-informed scatter model and update the scaling relation constraints in Sect.~\ref{Sec:simScatter}. Section~\ref{Sec:sensiMis} tests different miscentring models, and we conclude in Sect.~\ref{Sec:conclusion}. Appendix~\ref{Sec:nonlinear} investigates the higher-order shear biases.

The overdensity threshold of a virialised dark matter halo is defined such that the average density within the virial radius is 200 times the mean matter density of the Universe at the redshift of the halo. When reporting values dependent on Hubble's constant, we use $h_{70}=H_0/70~{\rm km}~{\rm s}^{-1}{\rm Mpc}^{-1}$ to facilitate comparison of results derived from observations and different simulations. All measurements are presented in comoving units, and the logarithm base is 10.

\section{Data}
\label{Sec:data}

Our lens sample is from the GAMA survey\footnote{\url{https://www.gama-survey.org/dr4}}, while the source sample is from the KiDS survey\footnote{\url{https://kids.strw.leidenuniv.nl/DR4}}. In this section, we provide an overview of the catalogues used in our study. For technical details, we direct interested readers to the relevant data release papers.

\subsection{Lenses: GAMA groups}
\label{Sec:dataGAMA}

GAMA is a high-density, high-completeness spectroscopic survey conducted using the AAOmega instrument on the Anglo-Australian Telescope~\citep{Driver2011MNRAS.413..971D}. We used data from three equatorial fields of the GAMA II phase (G09, G12, G15), each covering a sky area of 60 square degrees~\citep{Liske2015MNRAS.452.2087L}. The GAMA data in these fields have a spectroscopic completeness of 98 percent for galaxies within the observed magnitude limit of $r{<}19.58$~\citep{Driver2022MNRAS.513..439D}. In particular, we used three key GAMA products: the ${\rm G}^3{\rm C}$ group catalogue (version 10, \citealt{Robotham2011MNRAS.416.2640R}), the \texttt{StellarMassesLambdar} catalogue (version 24, \citealt{Taylor2011MNRAS.418.1587T}), and the random catalogue (version 2, \citealt{Farrow2015MNRAS.454.2120F}).

The ${\rm G}^3{\rm C}$ group catalogue (version 10) includes \num{26194} groups identified using an implementation of the friends-of-friends (FoF) algorithm based on galaxy separations. It separately treats the projected and radial separations to account for line-of-sight effects caused by peculiar velocities within groups. The two key linking parameters, the linking length (which defines the overdensity) and the radial expansion factor (which accounts for the peculiar motions of galaxies within groups), scale with the observed density contrast and depend on the galaxy positions and the magnitude limit of the survey. These parameters are optimised for quality of group finding through tests on GAMA lightcone mock data derived from semi-analytical simulations (see \citealt{Robotham2011MNRAS.416.2640R}, for further details). To minimise the impact of interlopers, we used groups with at least five identified members, resulting in a final sample of \num{2752} groups.

We used the sky position of the brightest group (or cluster) galaxy (BCG) as the group centre for our lensing measurements. Another commonly used method for selecting the central galaxy is by iteratively removing group members that are furthest from the light centre of the group. However, \citet{Robotham2011MNRAS.416.2640R} found that for groups with at least five members, this iterative procedure converges on the BCG $95\%$ of the time. Given the current measurement uncertainties, the subtle difference between these two methods is even more negligible (see Appendix A of \citealt{Viola2015MNRAS.452.3529V}). Therefore, we opted not to repeat measurements with alternative centres and instead modelled the miscentring statistically within our analysis (see Sect.~\ref{Sec:model}).

The stellar masses of galaxies were obtained from the \texttt{StellarMassesLambdar} catalogue (version 24), which uses stellar population synthesis models from \citet{Bruzual2003MNRAS.344.1000B}, assuming a \citet{Chabrier2003PASP..115..763C} initial mass function. The model fits were applied over a fixed rest-frame wavelength range ($300{-}11000~\AA$) using matched aperture photometry from the Lambda Adaptive Multi-Band Deblending Algorithm in R (LAMBDAR, \citealt{Wright2016MNRAS.460..765W}). The LAMBDAR code is designed to ensure consistent photometry and uncertainty estimation across a wide range of photometric imaging for calculating spectral energy distributions. It uses predefined elliptical apertures, initially estimated using SExtractor~\citep{Bertin1996AAS..117..393B} runs on SDSS $r$-band and VIKING $Z$-band imaging, followed by visual inspection and manual adjustments for objects flagged with poor aperture determinations. We used the \texttt{logmstar} value, which represents the total mass of luminous material and remnants, excluding mass recycled into the interstellar medium. We opted not to adjust the aperture-based stellar mass to total stellar mass because not all galaxies in the GAMA survey have accurate total flux estimates. This also aligns with our use of aperture stellar mass from simulations (see Sect.~\ref{Sec:sim}).

To calculate the total stellar masses of galaxy groups, we applied the correction method from \citet{Robotham2011MNRAS.416.2640R} used for estimating the $r$-band total luminosity. This method, based on the global GAMA galaxy luminosity function, accounts for missing flux due to the flux limit of the survey and includes a global optimisation factor to address biases from the luminosity function-based corrections, environmental effects, and extrapolation. The final corrections are redshift-dependent, ranging from a few percent at low redshift to a few factors at high redshift. We applied the same corrections to the observed group stellar masses, assuming a comparable stellar mass-to-light ratio between observed and intrinsic group properties. This assumption is valid given the depth of the GAMA survey, which recovers nearly all of the luminosity (or stellar mass) density~\citep{Loveday2012MNRAS.420.1239L}.

We used the GAMA random catalogue (version 2) to assess additive shear biases in measured weak lensing signals, as is detailed in Sect.~\ref{Sec:ESD}. This catalogue consists of randomly distributed points, reflecting the same selection function as the main spectroscopic survey. For our analysis, we randomly selected one million points from this catalogue for each of the GAMA fields under consideration. We measured weak lensing signals around these random points to quantify and correct potential additive shear biases in our measurements, following \citet{Dvornik2017MNRAS.468.3251D}.

\subsection{Sources: KiDS galaxies}

KiDS is a wide-field imaging survey, with weak gravitational lensing analysis as its primary scientific objective~\citep{Jong2013ExA....35...25D,Kuijken2015MNRAS.454.3500K}. The complete survey covers \num{1350}~{${\rm deg}^2$} of the sky, with optical images in the $ugri$ bands taken from the ESO VLT Survey Telescope. Among these, the $r$-band images, offering the highest imaging quality, are used for measuring galaxy shapes. In collaboration with the VISTA Kilo-degree INfrared Galaxy survey (VIKING, \citealt{Edge2013Msngr.154...32E}) conducted with the nearby ESO VISTA telescope, the KiDS shear catalogue also contains photometry from five $ZYJHK_s$ near-infrared bands. This additional near-infrared dataset significantly enhances the accuracy of photometric redshift estimates.

\begin{figure*}
   \centering
   \includegraphics[width=\textwidth]{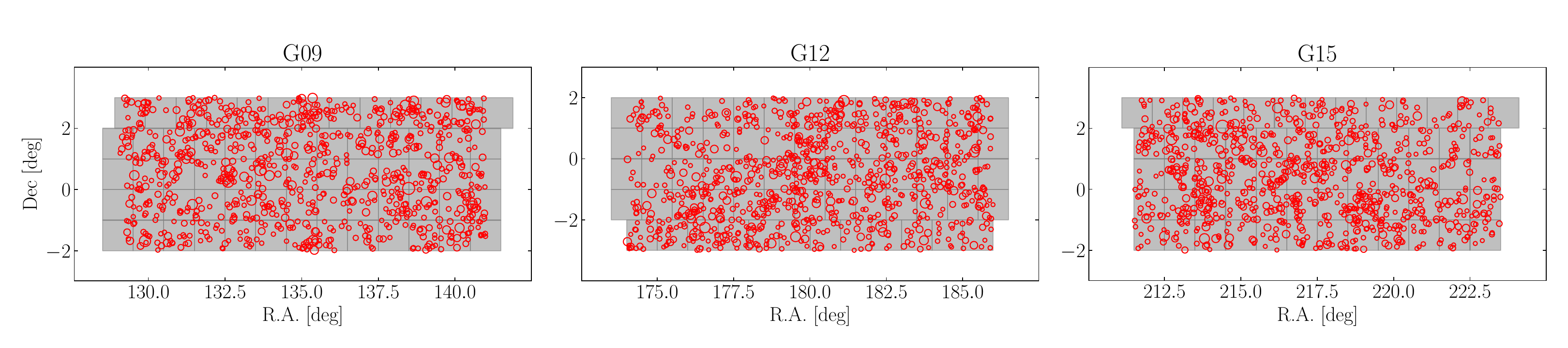}
   \caption{Sky coverage of the KiDS-1000-v2 shear catalogue for the three equatorial GAMA fields (G09, G12, G15). The grey boxes represent KiDS tile images, each covering $1~{\rm deg}^2$. The red circles indicate the selected GAMA groups, each consisting of at least five members. The size of these circles corresponds to the logarithm of the group richness.}
    \label{fig:footprint}
\end{figure*}

For our analysis, we used the latest KiDS-1000 shear catalogue (v2) from \citet{Li2023AA...679A.133L}. This catalogue is based on the fourth data release of KiDS~\citep{Kuijken2019AA...625A...2K,Giblin2021AA...645A.105G} with enhanced redshift calibration from \citet{Busch2022AA...664A.170V} and updated shear measurement and calibration from \citet{Li2023AA...670A.100L}. It fully covers the three equatorial fields of GAMA, as illustrated in Fig.~\ref{fig:footprint}. Thanks to this complete coverage, we are now able to measure weak lensing signals around all \num{2752} selected GAMA groups, approximately doubling the number used in previous similar analyses by \citet{Viola2015MNRAS.452.3529V} and \citet{Rana2022MNRAS.510.5408R}.

\section{Simulations}
\label{Sec:sim}

In this study, we used two state-of-the-art cosmological (magneto)hydrodynamical simulations, namely FLAMINGO\footnote{\url{https://flamingo.strw.leidenuniv.nl/}} and IllustrisTNG\footnote{\url{https://www.tng-project.org/}}, which offer complementary combinations of volume and resolution. Additionally, we used the \textsc{L-Galaxies} SAM\footnote{\url{https://lgalaxiespublicrelease.github.io/}} run on the IllustrisTNG gravity-only simulations as a representation of the other simulation technique, so we have a broad coverage of simulation uncertainties that we can account for in our halo model development. This section provides an overview of these simulations and describes how we construct mock group catalogues with the desired properties from them. Units in this section are relative to the cosmology adopted by each specific simulation, except when comparing properties across different simulations or with observational data. In such cases, we scaled the quantities to account for differences in the Hubble constant, while ignoring the impact of differences in other cosmological parameters.

\subsection{FLAMINGO simulations}
\label{Sec:FLAMINGO}

FLAMINGO is a suite of hydrodynamical cosmological simulations generated using the smoothed particle hydrodynamics-based code \textsc{swift}~\citep{Schaller2024MNRAS.530.2378S}. The FLAMINGO fiducial cosmology is based on the Dark Energy Survey year three results~($3{\times}2$pt plus external constraints, \citealt{Abbott2022PhRvD.105b3520A}; $\Omega_{\Lambda}{=}0.694$, $\Omega_{\rm m}{=}0.306$, $\Omega_{\rm b}{=}0.0486$, $\sigma_8{=}0.807$, $n_{\rm s}{=}0.967$, and $h{=}0.681$) and includes a single massive neutrino species with a mass of $0.06~{\rm eV}$ and two massless species. Its galaxy formation model builds upon those developed for the OWLS~\citep{Schaye2010MNRAS.402.1536S}, cosmo-OWLS~\citep{Brun2014MNRAS.441.1270L}, EAGLE~\citep{Schaye2015MNRAS.446..521S}, and BAHAMAS~\citep{McCarthy2017MNRAS.465.2936M} projects, including radiative cooling and heating, star formation and evolution, stellar energy feedback, growth of supermassive black holes, and AGN feedback. A notable advancement of the FLAMINGO galaxy formation model is the calibration of its free parameters to the observed present-day galaxy stellar mass function (GSMF) and low-redshift cluster gas fraction using Gaussian process emulators, which explicitly accounts for observational uncertainties and biases~(see \citealt{Kugel2023MNRAS.526.6103K} and \citealt{Schaye2023MNRAS.526.4978S}, for a detailed description of the model).

The simulation snapshots are saved at various redshifts from $10$ to $0$. In each snapshot, halos and substructures are identified using the \textsc{velociraptor} subhalo finder~\citep{Elahi2019PASA...36...21E}. In short, it first identifies halos using a standard 3D FoF algorithm with a linking length of 0.2~\citep{Davis1985ApJ...292..371D}. Then, within each FoF halo, it iteratively searches for subhalos that are dynamically distinct from the mean background halo using a 6D FoF algorithm in position-velocity phase space. Finally, (sub)halo properties are computed for a range of apertures using SOAP (Spherical Overdensity Aperture Processor)\footnote{\url{https://github.com/SWIFTSIM/SOAP.git}}. These aperture property estimates are essential for building mock galaxy group catalogues, which we compare to observations.

In this work, we used two FLAMINGO simulations with volumes of $(1~{\rm Gpc})^3$, and initial mean baryonic particle masses of $1.34\times 10^8~{\rm M}_{\odot}$ (L1\_m8) and $1.07\times 10^9~{\rm M}_{\odot}$ (L1\_m9). Both simulations employ the fiducial galaxy formation model and assume the fiducial cosmology. However, despite the resolution-dependent calibration of the FLAMINGO model, which uses different mass ranges for calibration, the resulting predictions still exhibit some variation across different resolutions~\citep{Kugel2023MNRAS.526.6103K}. By using simulations from both resolutions, we can account for the impact of these resolution-related differences in our halo model development.

\subsection{IllustrisTNG simulations}

IllustrisTNG is a suite of magnetohydrodynamical cosmological simulations generated with the moving-mesh code \textsc{arepo}~\citep{Springel2010MNRAS.401..791S}. It assumes a cosmology consistent with the \citet{Planck2016AA...594A..13P} results ($\Omega_{\Lambda}{=}0.6911$, $\Omega_{\rm m}{=}0.3089$, $\Omega_{\rm b}{=}0.0486$, $\sigma_8{=}0.8159$, $n_{\rm s}{=}0.9667$, and $h{=}0.6774$). Its galaxy formation model builds upon the framework used in the Illustris project~\citep{Vogelsberger2014MNRAS.444.1518V,Genel2014MNRAS.445..175G}, including all the main ingredients mentioned in Sect.~\ref{Sec:FLAMINGO} but with different implementations (see \citealt{Vogelsberger2013MNRAS.436.3031V} for details). The free parameters in the IllustrisTNG model are manually tuned to alleviate the most striking observational tensions identified in the Illustris simulations~\citep{Nelson2015AC....13...12N}. Key advancements in IllustrisTNG include the inclusion of a seed magnetic field, a revised galactic wind implementation, and a kinetic AGN feedback model for low-accretion states (see \citealt{Pillepich2018MNRAS.473.4077P} for details).

The simulation snapshots are saved at redshifts from $20$ to $0$. For each snapshot, halos are first identified using a standard 3D FoF group finder with a linking length of 0.2. Then, within each FoF halo, the gravitationally bound substructures are located and characterised hierarchically using the \textsc{subfind} algorithm~\citep{Springel2001MNRAS.328..726S}.

In this work, we used their flagship run with a box side length of $110.7$ Mpc and an effective mass resolution of $1.39\times 10^6~{\rm M}_{\odot}$ for baryons and $7.46\times 10^6~{\rm M}_{\odot}$ for dark matter (TNG100-1, \citealt{Nelson2019ComAC...6....2N}). The TNG100-1 simulation provides a different balance between volume and resolution, covering the low halo mass range missed by the FLAMINGO simulations. Most importantly, it includes aperture stellar mass estimates produced by \citet{Engler2021MNRAS.500.3957E}, which are essential for our study (see Sect.~\ref{Sec:mock}).

\subsection{L-Galaxies semi-analytical model}

\textsc{L-Galaxies} is a SAM of galaxy formation designed to run on subhalo merger trees produced by $N$-body simulations. The model is based on seminal works of \citet{White1991ApJ...379...52W}, \citet{Kauffmann1993MNRAS.264..201K,Kauffmann1999MNRAS.303..188K} and \citet{Springel2001MNRAS.328..726S}, with its first relatively mature implementation in the Millennium Simulations by \citet{Springel2005Natur.435..629S}. Since then, the model has undergone a series of updates driven by discrepancies identified between model predictions and observations, resulting in a series of public releases of mock catalogues.

In our study, we used the catalogues produced by \citet{Ayromlou2021MNRAS.502.1051A}, who ran the \citet{Henriques2015MNRAS.451.2663H} version of the \textsc{L-Galaxies} model on the companion gravity-only simulations of IllustrisTNG. The \citet{Henriques2015MNRAS.451.2663H} version of \textsc{L-Galaxies} builds on \citet{Guo2011MNRAS.413..101G}, aiming for a better representation of the observed evolution of low-mass galaxies. The model contains a set of coupled differential equations to follow the evolution of baryonic components in each hierarchical merger tree. A complete description of the model treatment can be found in the supplementary material associated with the arXiv version of \citet{Henriques2015MNRAS.451.2663H}. The free parameters in the model are calibrated to match the observed GSMF at $z = 0$, 1, 2, and 3, as well as the fraction of red galaxies as a function of stellar mass at $z = 0$, 0.4, 1, 2, and 3.

We used two \textsc{L-Galaxies} runs with box sizes of $110.7$ Mpc (LGal100-1) and $302.6$ Mpc (LGal300-1). The former shares the same mass coverage as TNG100-1, enabling a direct comparison between hydrodynamical simulations and SAM results, while the latter overlaps in mass range with FLAMINGO simulations and is better aligned with the mass range of our weak lensing measurements.

\subsection{Construction of mock group catalogues}
\label{Sec:mock}

To perform a consistent comparison between simulations and observations, it is crucial to translate simulated properties into mock observables that are comparable to those from observations. In our study, these observables mainly concern the virial mass of halos and the total stellar mass of galaxy groups. The primary factor influencing halo properties is the particle resolution of the simulations. Therefore, we implemented a halo mass cut in the simulations based on their mean dark matter particle masses. The detailed mass cut selections are summarised in Table~\ref{table:simCut}. These thresholds correspond to a minimum of ${\sim}5000$ particles for IllustrisTNG and ${\sim}1000$ particles for FLAMINGO, ensuring that only well-resolved halos are included in our mock catalogues.

To estimate the total stellar masses of galaxy groups, we considered both observational effects and the particle resolution of simulations. As is described in Sect.~\ref{Sec:dataGAMA}, the total stellar masses of GAMA galaxy groups are estimated using the stellar mass measurements of individual galaxies and the global galaxy luminosity function. Ideally, we would mimic these observational estimates by deriving stellar mass from measured aperture flux and galaxy profile fitting, but this is challenging, particularly given the limited resolution of simulations used in this study~\citep{Graaff2022MNRAS.511.2544D,Kugel2023MNRAS.526.6103K}.

Therefore, we adopted a simplified approach while maintaining the principle of using individual galaxy stellar masses to estimate the total stellar masses of galaxy groups. In the case of SAM, the galaxy stellar mass is relatively well defined as the sum of disc and bulge stellar masses. For hydrodynamical simulations, there are several different methods for estimating galaxy stellar masses. Following previous studies comparing simulation and observational results~\citep{Schaye2015MNRAS.446..521S,Graaff2022MNRAS.511.2544D}, we opted to use the 3D physical aperture stellar mass estimation as the galaxy stellar mass, which is defined as the sum of gravitationally bound stellar particles within a given radius.

For TNG100-1, we used the $30~{\rm kpc}$ aperture estimates provided in their supplementary data release~\citep{Engler2021MNRAS.500.3957E}. For FLAMINGO, we used the $50~{\rm kpc}$ aperture estimates, following their model calibration choice. As shown by \citet{Kugel2023MNRAS.526.6103K}, the impact of this difference in aperture size is negligible for galaxies with stellar masses below $10^{11}\ {\rm M}_{\odot}$. We confirmed that our results remain unchanged when using the $30~{\rm kpc}$ aperture estimates for all simulations.
 
\begin{table}
\renewcommand{\arraystretch}{1.5}
\caption{Sample selection for construction of mock group catalogues.}
\label{table:simCut}      
\centering          
\begin{tabular}{lcc} 
\hline\hline      
Simulations & Halo mass cut & Stellar mass cut\\ 
\hline                  
FLAMINGO (L1\_m9) & $\geq 10^{12.7}\ {\rm M}_{\odot}$ & $\geq 10^{10}\ {\rm M}_{\odot}$\\
\hline            
FLAMINGO (L1\_m8) & $\geq 10^{11.8}\ {\rm M}_{\odot}$ & $\geq 10^{8.5}\ {\rm M}_{\odot}$\\
\hline              
LGal300-1 & $\geq 10^{11.5}\ {\rm M}_{\odot}$ & $-$\\
\hline              
LGal100-1 & $\geq 10^{10.6}\ {\rm M}_{\odot}$ & $-$\\
\hline              
TNG100-1 & $\geq 10^{10.6}\ {\rm M}_{\odot}$ & $-$\\
\hline              
\end{tabular}
\end{table}

\begin{figure}
  \centering
  \includegraphics[width=\hsize]{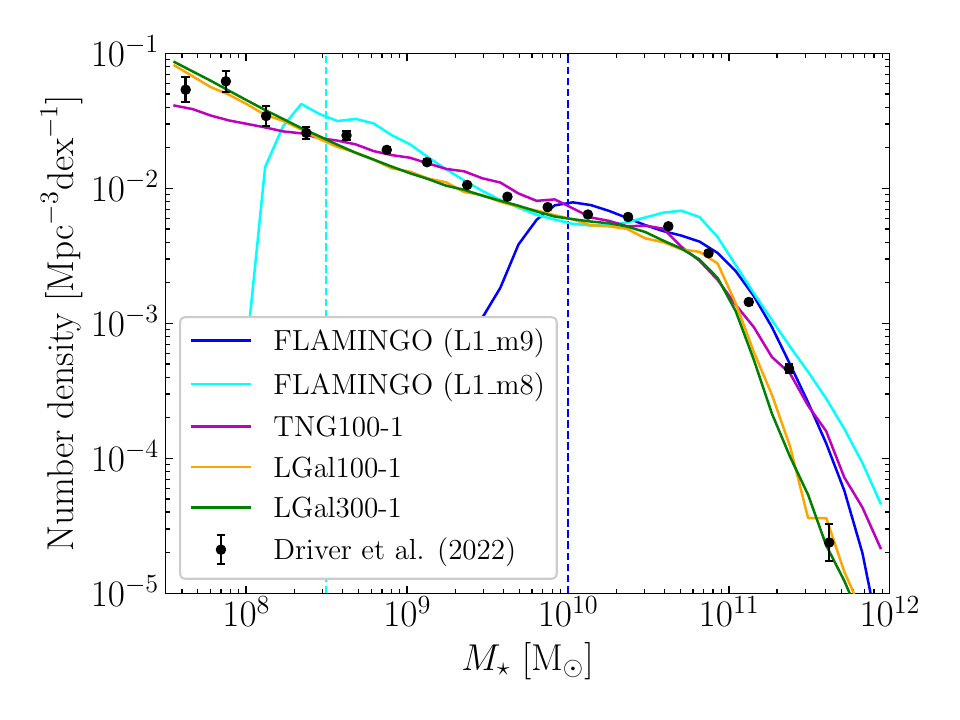}
      \caption{Galaxy stellar mass function at redshift $z{=}0$ from simulations and GAMA observations~\citep{Driver2022MNRAS.513..439D}. For comparison, all properties have been converted to a $h_{70}$ cosmology, with masses from simulations scaled as $h_{70}^{-1}$ and masses from observational data scaled as $h_{70}^{-2}$. The vertical dashed lines indicate the stellar mass cuts applied to the two FLAMINGO simulations, based on their respective resolutions. The overall agreement between simulations and observations supports our approach for estimating the total stellar masses of mock galaxy groups, while the resolution limits of the FLAMINGO simulations highlight the need for careful treatment, as is detailed in Sect.~\ref{Sec:mock}.}

         \label{fig:SMF}
\end{figure}

Figure~\ref{fig:SMF} compares the present-day GSMF between the simulations and the latest GAMA observations from \citet{Driver2022MNRAS.513..439D}. Overall, the simulations show good agreement with the observations, which is expected given that the present-day GSMF is one of the key observational constraints used to tune the free parameters in the simulations~\citep{Henriques2015MNRAS.451.2663H,Pillepich2018MNRAS.473.4077P,Kugel2023MNRAS.526.6103K}. However, the FLAMINGO simulations exhibit a drop at the low-mass end, attributed to their resolution limit. To account for this low-mass drop, we took extra care when calculating the total stellar mass of FLAMINGO galaxy groups. Specifically, we first implemented a stellar mass cut in the FLAMINGO galaxies, setting a minimum of $10^{8.5}~\mathrm{M}_{\odot}$ and $10^{10}~\mathrm{M}_{\odot}$ for L1\_m8 and L1\_m9, respectively, based on the visual inspection of the dropping points (vertical dashed lines in Fig.~\ref{fig:SMF}). These cuts correspond to at least three bounded particles in L1\_m8 and ten bounded particles in L1\_m9. We tested a more conservative stellar mass cut, with a minimum of $10^{9.3}~\mathrm{M}_{\odot}$ for the L1\_m8 simulations and found consistent results. After this selection, we estimated the total stellar mass of each galaxy group using
\begin{equation}
M^{\rm grp}_{\star} = \left(\sum_i M_{\star,~i}\right) ~ \frac{\int_{0}^{\infty}~\mathrm{d}M_{\star}~\phi(M_{\star})~M_{\star}}{\int_{M_{\star, \rm min}}^{\infty}~\mathrm{d}M_{\star}~\phi(M_{\star})~M_{\star}}~.
\end{equation}
In practice, we approximated zero and infinity by using $1~{\rm M}_{\odot}$ and $10^{13}~{\rm M}_{\odot}$, respectively, as the contribution to the stellar mass density from galaxies outside this mass range is negligible. For the GSMF, $\phi(M_{\star})$, we used the double Schechter function fit from the latest GAMA observations (\citealt{Driver2022MNRAS.513..439D}, Table 7).

Since TNG100-1 resolves galaxies down to $10^7~{\rm M}_{\odot}$ (corresponding to ${\sim}10$ stellar particles), these additional steps of stellar mass cut and boost factor are unnecessary. This also holds true for the \textsc{L-Galaxies} SAM, where completeness is maintained down to $10^6~{\rm M}_{\odot}$. Therefore, for these simulations, we simply sum the stellar masses of all member galaxies to obtain the total stellar mass of galaxy groups.

For all simulations, we constructed mock group catalogues only from their present-day snapshot ($z{=}0$), as the GAMA galaxy groups are local, with a mean redshift of ${\sim}0.2$. We examined the evolution of the desired galaxy group and halo properties from redshift 0 to 0.2 in all simulations and found negligible differences.

\section{The weak lensing signals}
\label{Sec:measure}

The weak lensing effect introduces coherent tangential distortions in the observed shapes of background galaxies. These distortions, known as the tangential shear, $\gt$, correlate with the projected mass density contrast of the foreground lens\footnote{Throughout this work, we do not distinguish between the original shear $\gamma$ and the reduced shear $g\equiv\gamma/(1-\kappa)$, given that the convergence $\kappa$ is much less than one in the weak lensing regime.} (e.g.~\citealt{Bartelmann2001PhR...340..291B}):
\begin{equation}
\label{eq:ESD}
\Delta\Sigma(R) \equiv \Bar{\Sigma}(\le R) - \Sigma(R) = \Scr\gt(R)\ ,
\end{equation}
where the mass density contrast, $\Delta\Sigma(R)$, is also commonly referred to as the excess surface density (ESD). The $\Sigma(R)$ represents the local surface mass density at a projected comoving separation, $R$, between the lens and source, while $\Bar{\Sigma}(\le R)$ denotes the mean surface density within this radius. The critical surface density, $\Scr$, serves as a measure of lensing efficiency and is defined as
\begin{equation}
\label{eq:Scr}
\Scr \equiv \frac{c^2}{4\pi G}\frac{\rs}{(1+\zl)\ \rl\rls}\ ,
\end{equation}
where $G$ and $c$ denote the gravitational constant and the speed of light, respectively. The $\rl$, $\rs$, and $\rls$ are the comoving distances to the lens, source and between these two, respectively, and $\zl$ is the redshift of the lens. The factor $(1+\zl)$ arises due to our use of comoving distances and co-ordinates (see Appendix C of \citealt{Dvornik2018MNRAS.479.1240D} for a detailed derivation).

Therefore, by assuming a certain density profile for an object, we can infer its mass by measuring the ESD signals around it. In this section, we detail how we estimate ESD for the selected GAMA galaxy groups using the KiDS shear measurements and the corresponding covariance matrix necessary for modelling.

\subsection{ESD measurements}
\label{Sec:ESD}

We estimated the tangential shear using the azimuthal average of the tangential projection, $\et$, of the \textit{lens}fit measured ellipticities of the KiDS source galaxies. It is defined as 
\begin{equation}
\begin{bmatrix}
   \et \\
    \ex
\end{bmatrix} \equiv \begin{bmatrix}
    -\cos(2\varphi) & -\sin(2\varphi)\\
    \sin(2\varphi) & -\cos(2\varphi)
\end{bmatrix} \cdot \begin{bmatrix}
    \eone \\
    \etwo
\end{bmatrix}~,
\end{equation}
where $\varphi$ denotes the relative position angle of the source with respect to the lens. The azimuthal average of the cross projection, $\ex$, can serve as an indicator of potential systematic contamination, given that the lensing effect only introduces tangential shear to the leading order.

To account for both measurement uncertainties and lensing efficiency, a weight was assigned to each lens-source pair when calculating the azimuthal average. This weight is given by
\begin{equation}
    \label{eq:wei}
    w_{\rm ds} \equiv w_{\rm s}\ \Scrt^{-2}~,
\end{equation}
where $w_{\rm s}$ is the \textit{lens}fit weight, reflecting the individual galaxy shape measurement uncertainties, and $\Scrt$ is the `effective critical surface density', which down-weights lens-source pairs that are close in redshift and thus carry lower lensing signals. 

Following \citet{Dvornik2017MNRAS.468.3251D}, we calculated $\Scrt$ for each lens by integrating the redshift distribution of the whole source sample behind the given lens. This averaging approach aligns with the KiDS-1000 redshift calibration~\citep{Hildebrandt2021AA...647A.124H}. An alternative way to determine source distances is by using the individual posterior redshift distributions of each source galaxy, as in \citet{Viola2015MNRAS.452.3529V}. \citet{Dvornik2017MNRAS.468.3251D} verified that these two approaches yield consistent signals within the error budget for the lensing signal of the GAMA sample. Following Eq.~(\ref{eq:Scr}), the calculation is formulated as
\begin{equation}
    \label{eq:SigmaCR}
    \Scrt^{-1} = \frac{4\pi G}{c^2}(1+\zl)\rl\int_{\ \zl + \delta_z}^{\ \infty}{\rm d}\zs\ \frac{\rls}{\rs}\ n(\zs)\ ,
\end{equation}
where the source redshift distribution, $n(\zs)$, was determined from the redshift calibration reference sample of \citet{Li2023AA...679A.133L}, which is based on the fiducial spectroscopic sample of \citet{Busch2022AA...664A.170V}. A redshift difference threshold, $\delta_z{=}0.2$, is introduced to mitigate contamination from group members to the source sample. This redshift cut-off, $\zs > \zl + \delta_z$, is applied to the source galaxies involved in the calculation as well as to the reference spectroscopic sample.

The median velocity dispersion of the GAMA galaxy groups used in our study is ${\sim}300~{\rm km~s^{-1}}$, which is not massive enough to enable the lensing measurement for individual groups. Therefore, we use a stacking process to enhance the S/N, following the methodology of previous KiDS analyses (e.g.~\citealt{Viola2015MNRAS.452.3529V, Dvornik2017MNRAS.468.3251D}). It estimates the stacked ESD profile for an ensemble of galaxy groups as 
\begin{equation}
    \label{eq:ESDmeasure}
    \Delta\Sigma(R) = \left[\frac{\sum_{\rm ds}w_{\rm ds}\ \et\ \Scrt}{\sum_{\rm ds}w_{\rm ds}}\right]\ \frac{1}{1+K}~,
\end{equation}
where the correction 
\begin{equation}
    \label{eq:K}
    K = \frac{\sum_{\rm ds}w_{\rm ds}\ m_{\rm s}}{\sum_{\rm ds}w_{\rm ds}}~,
\end{equation}
accounts for the multiplicative bias, $m_{\rm s}$, of our \textit{lens}fit shear measurements. 

We estimated the multiplicative bias, $m_{\rm s}$, using the latest SKiLLS image simulations developed by \citet{Li2023AA...670A.100L}. To capture the variation in $m_{\rm s}$, we first divided SKiLLS galaxies into small bins based on redshift, resolution (the ratio of galaxy size to PSF size), and S/N. Specifically, we used uniform redshift bins with a width of 0.1, and for each redshift bin, $20 \times 20$ weighted quantile bins in resolution and S/N. We then calculated the average $m_{\rm s}$ within each bin and assigned these values to KiDS galaxies based on their corresponding properties. This approach accounts for the strong dependence of $m_{\rm s}$ on redshift, resolution, and S/N. However, it implicitly assumes that the multiplicative bias does not depend on the shear, which may not always be correct, depending on the magnitude of the shear signal and the specific shear measurement algorithm (e.g.~\citealt{Kitching2022OJAp....5E...6K,Jansen2024AA...683A.240J}). In Appendix~\ref{Sec:nonlinear}, we investigate the potential violation of this linear shear bias assumption using dedicated simulations. We find that the multiplicative bias from \textit{lens}fit remains fairly linear within the typical shear amplitude range ($|\gamma|{\lesssim}0.1$). However, in higher-shear regimes, higher-order terms up to the third order are needed to capture the non-linear behaviour if sub-percent level calibration accuracy is required. We also detect a small shear bias (${\sim}10^{-4}$) responding to the shear in the other component. The overall correction factor, $K$, is at the sub-percent level, with small variations across angular separation and lens observable bins.

Besides the multiplicative shear bias, we accounted for the additive shear bias by measuring lensing signals around one million random points selected from the GAMA random catalogue (version 2, \citealt{Farrow2015MNRAS.454.2120F}). The additive bias is scale-dependent, with larger biases at larger lens-source separations, and depend on the GAMA patch (see Appendix A of \citealt{Dvornik2017MNRAS.468.3251D}). Thus, we corrected the three GAMA patches (G9, G12, and G15), separately. The overall correction is small, with values at the sub-percent level, attributable to the relatively small scales our study focused on and the complete coverage of the GAMA fields by the current KiDS observations.

\begin{figure}
  \centering
  \includegraphics[width=\hsize]{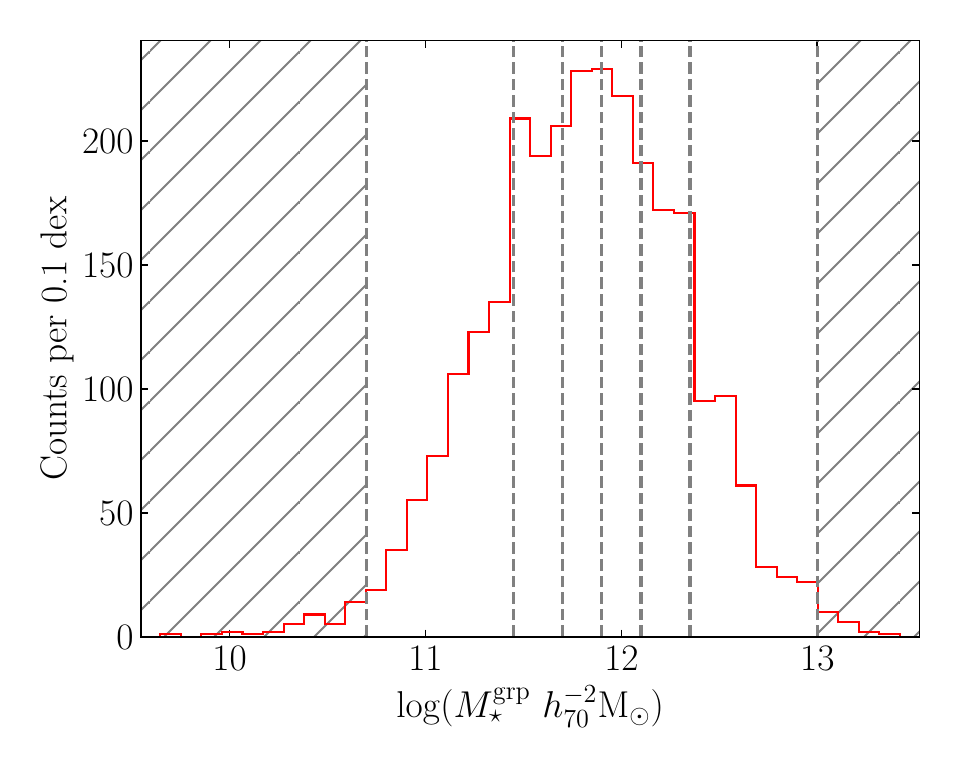}
  \includegraphics[width=\hsize]{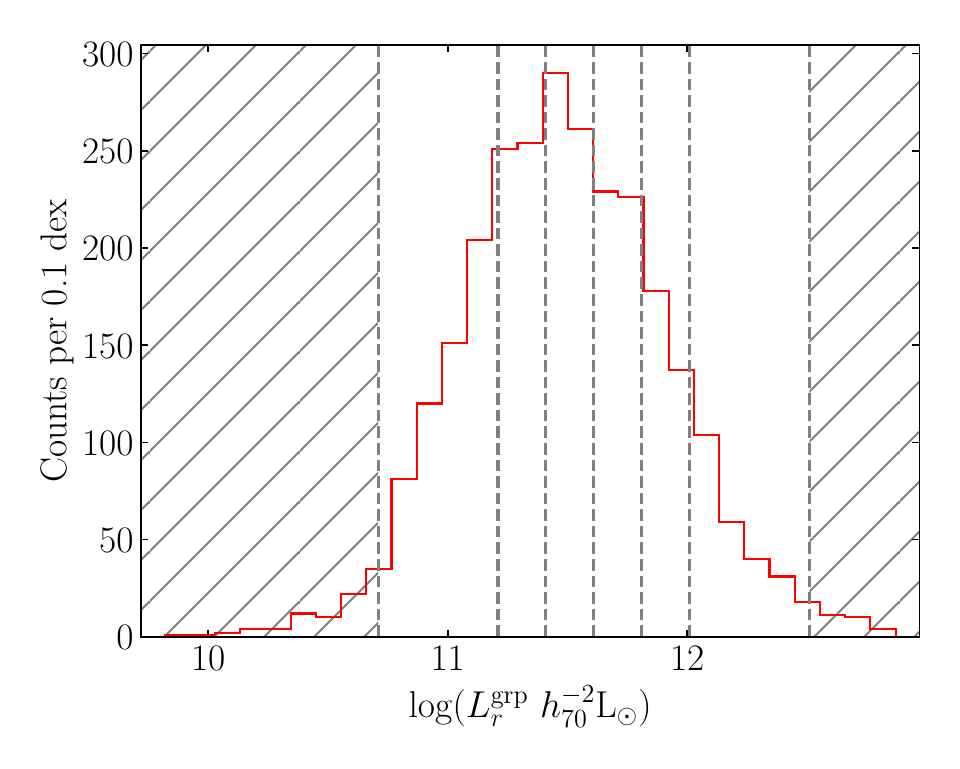}
      \caption{Distributions of the group total stellar mass (upper panel) and the group $r$-band total luminosity (bottom panel). The vertical lines represent the boundaries of the bins for measuring stacked ESD profiles. The corresponding values are listed in Table~\ref{table:bins}. Objects falling within the hatched regions are excluded from our analysis.} 
         \label{fig:distri}
\end{figure}

In this study, we focus on the halo mass relation with the total stellar mass of galaxy groups, which can be directly compared to the properties extracted from the mock group catalogues as described in Sect.~\ref{Sec:mock}. Besides, we verify our results, which feature updated shear measurements and new halo model ingredients, against previous similar studies that measured the halo mass relation with the $r$-band total luminosity of galaxy groups~\citep{Viola2015MNRAS.452.3529V,Rana2022MNRAS.510.5408R}. Therefore, we performed two sets of ESD measurements by dividing the GAMA groups into six bins based on either their group stellar masses or their $r$-band total luminosity. We set lower and upper limits to exclude the tails of the distribution, as shown in Fig.~\ref{fig:distri}, to mitigate selection effects at both ends and facilitate the modelling. Table~\ref{table:bins} provides detailed information about the defined bins.

\begin{table}
\renewcommand{\arraystretch}{1.5}
\caption{Summary of the binning boundaries, number of groups, mean redshift of the groups, and mean stellar mass of the BCGs for each bin used in the stacked ESD measurements.}
\label{table:bins}      
\centering          
\begin{tabular}{llcll}     
\hline\hline      
Observable & Range & $N_{\rm groups}$ &  $z_{\rm mean}$ & $\log(\bar{M}_{\star}^{\rm BCG})$
\\ 
\hline            
$\log(M^{\rm grp}_{\star})$ & (10.70, 11.45] & 589 & 0.12 & 10.83 \\
& (11.45, 11.70] & 470 & 0.16 & 11.07 \\
& (11.70, 11.90] & 429 & 0.20 & 11.19 \\
& (11.90, 12.10] & 416 & 0.24 & 11.28 \\
& (12.10, 12.35] & 419 & 0.28 & 11.36 \\
& (12.35, 13.00] & 368 & 0.32 & 11.49 \\
\hline   
$\log(L^{\rm grp}_{r})$ & (10.71, 11.21] & 628 & 0.12 & 10.90 \\
& (11.21, 11.41] & 477 & 0.17 & 11.10 \\
& (11.41, 11.61] & 528 & 0.21 & 11.22 \\
& (11.61, 11.81] & 432 & 0.26 & 11.31 \\
& (11.81, 12.01] & 312 & 0.29 & 11.39 \\
& (12.01, 12.50] & 267 & 0.32 & 11.49 \\
\hline                  
\end{tabular}
\tablefoot{The unit of stellar mass is $h_{70}^{-2}{\rm M}_{\odot}$, and the unit of luminosity is $h_{70}^{-2}{\rm L}_{\odot}$.}
\end{table}

\begin{figure*}
  \centering
  \includegraphics[width=\hsize]{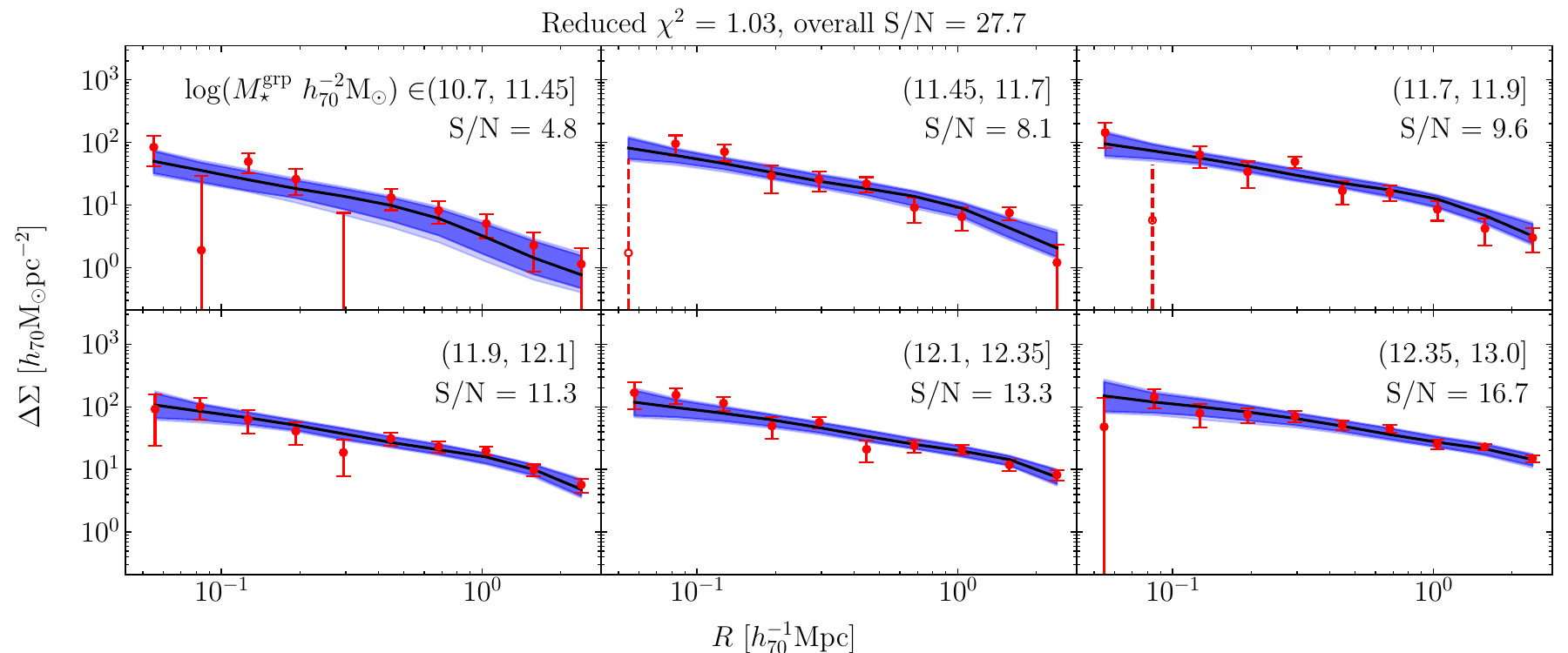}
      \caption{Stacked ESD profiles in the six bins of group total stellar mass ($M^{\rm grp}_{\star}$). The error bars correspond to the square root of the diagonal elements of the covariance matrix. We use open circles with dashed bars for negative values of the ESD. The black lines show the best-fit results from our baseline model (Sect.~\ref{Sec:model}), with the shaded dark and light blue regions indicating the 68\% and 95\% credible intervals, respectively. The S/N for each $M^{\rm grp}_{\star}$ bin only accounts for correlations within that bin across different radial bins, while the overall S/N also accounts for correlations between different $M^{\rm grp}_{\star}$ bins. The reduced $\chi^2$ value of $1.03$ (considering 7 independent fitting parameters and 60 data points) for the best-fit results suggests an overall good fit to the data.}
         \label{fig:ESDMgrp}
\end{figure*}

\begin{figure*}
  \centering
  \includegraphics[width=\hsize]{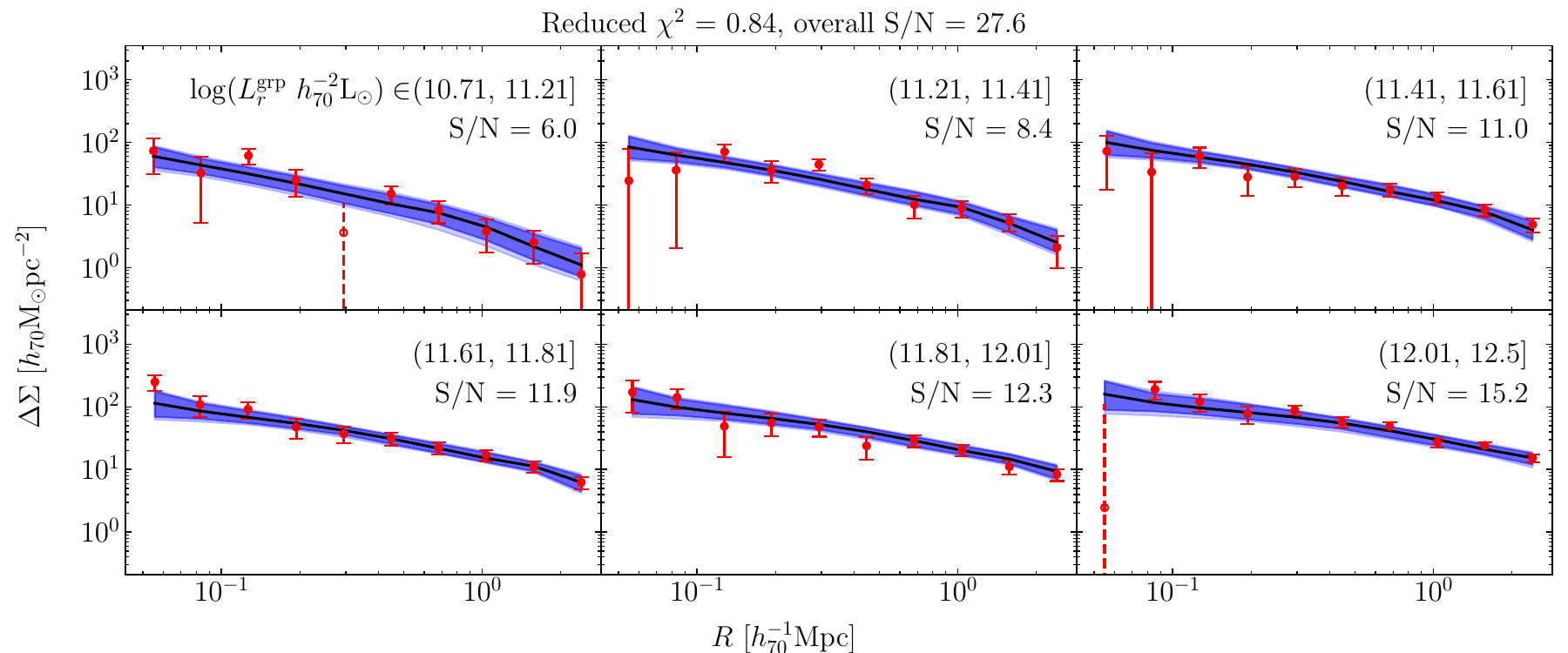}
      \caption{Same as Fig.~\ref{fig:ESDMgrp}, but for the six bins of group $r$-band total luminosity ($L^{\rm grp}_{r}$).}
         \label{fig:ESDLgrp}
\end{figure*}

We measured the ESD profiles in ten logarithmically spaced comoving radial bins, ranging from $0.04$ to $2.86\ h_{70}^{-1}\ {\rm Mpc}$. The centre of the measured ESD profile was defined as the location of the identified BCG for each group. The lower limit was chosen to balance the S/Ns and the impact of blending effects, while the upper limit is set to mitigate large-scale systematics~\citep{Viola2015MNRAS.452.3529V}. Figures~\ref{fig:ESDMgrp} and \ref{fig:ESDLgrp} show the resulting ESD profiles for $M^{\rm grp}_{\star}$ and $L^{\rm grp}_{r}$ binning, respectively. The signals have been corrected for additive and multiplicative shear biases as previously described. The overall S/N, accounting for the full correlation among data points (see Sect.~\ref{Sec:cov}), is 27.7 for $M^{\rm grp}_{\star}$ binning and 27.6 for $L^{\rm grp}_{r}$ binning. The S/N for each bin is shown in the plots.

\subsection{Covariance matrix estimation}
\label{Sec:cov}

The source galaxies can be used multiple times to estimate $\Delta\Sigma(R)$ across different radial and observable bins, leading to correlations between the stacked ESD measurements. To account for these correlations in our modelling, we adopted the covariance matrix estimation method developed by \citet{Viola2015MNRAS.452.3529V}. It analytically calculates the covariance directly from the data, accounting for correlations introduced by repeated entries of source galaxies and contributions from shape noise, while ignoring cosmic variance (see Section 3.4 of \citealt{Viola2015MNRAS.452.3529V} for details). This method has been used in other KiDS+GAMA analyses and shown to be sufficiently accurate for measurements within our adopted scale range of $R \leq 2.86\ h_{70}^{-1}\ {\rm Mpc}$ (e.g.~\citealt{Sifon2015MNRAS.454.3938S, Brouwer2016MNRAS.462.4451B}).

\section{Halo model and occupation statistics}
\label{Sec:model}

From a statistical perspective, the ESD profile, $\Delta\Sigma(R)$, of an ensemble of lenses is related to the galaxy-matter power spectrum, $P_{\rm gm}(k)$, through~(e.g.~\citealt{Murata2018ApJ...854..120M})
\begin{equation}
\label{eq:DeltaSigma}
\Delta\Sigma(R) = \frac{\bar{\rho}_{\rm m}}{2\pi}\int_{0}^{\infty}{\rm d}k\ k \ P_{\rm gm}(k)\ J_2(kR)~,
\end{equation}
where $k$ represents the comoving wavenumber, $\bar{\rho}_{\rm m}$ is the current mean matter density of the Universe, and $J_2(kR)$ is the second-order Bessel function of the first kind. The specific form of $P_{\rm gm}(k)$ depends on the redshift of the lens and the types of galaxies contributing to the lens sample, such as central or satellite galaxies. Therefore, we can interpret the measured $\Delta\Sigma(R)$ signal if we have a model to describe $P_{\rm gm}(k)$, tailored to the characteristics of the lens sample. The halo model, complemented by the halo occupation statistics, offers such a theoretical framework~(e.g.~\citealt{Seljak2000MNRAS.318..203S,Cooray2002PhR...372....1C,Peacock2000MNRAS.318.1144P,Berlind2002ApJ...575..587B,Yang2003MNRAS.339.1057Y,Bosch2013MNRAS.430..725V}).

In this section, we detail our approach to interpreting the stacked ESD measurements using the halo model framework. First, we provide a concise overview of the halo model formalism, primarily following the notation of \citet{Bosch2013MNRAS.430..725V} and \citet{Uitert2016MNRAS.459.3251V}. We then outline our baseline model ingredients, drawing primarily on previous KiDS studies (e.g.~\citealt{Viola2015MNRAS.452.3529V, Uitert2016MNRAS.459.3251V, Dvornik2017MNRAS.468.3251D}), but incorporating some improvements motivated by recent progress. 

\subsection{Halo model with galaxy population statistics}
\label{Sec:HaloModel}

The halo model framework is built on the assumption that all dark matter resides within virialised halos, whose sizes and masses are determined by a chosen overdensity threshold. By adopting models for the internal density profiles of these halos, we can use them to describe the matter-matter power spectrum of the Universe. Furthermore, by incorporating statistical models of how galaxies populate these dark matter halos, often referred to as halo occupation statistics or the halo occupation distribution (HOD), the halo model framework can also be extended to describe the galaxy-matter power spectrum (e.g.~\citealt{Bosch2013MNRAS.430..725V}).

Following the notation of \citet{Bosch2013MNRAS.430..725V} and \citet{Uitert2016MNRAS.459.3251V}, we formulate the galaxy-matter power spectrum as
\begin{equation}
    P_{\rm gm}(k) = P^{1\rm h}_{\rm gm}(k) + P^{2\rm h}_{\rm gm}(k)\ ,
\end{equation}
where the one-halo term, describing correlations within a single halo, is defined as 
\begin{equation}
    P^{1\rm h}_{\rm gm}(k) \equiv \int{\rm d}\Mh\ \mathcal{H}_{\rm m}(k, \ \Mh)\ \mathcal{H}_{\rm g}(k,\ \Mh)\ \nh(\Mh)\ ,
\end{equation}
and the two-halo term, representing correlations between different halos, is defined as
\begin{equation}
\label{eq:Ph2}
\begin{split}
    P^{2\rm h}_{\rm gm}(k) \equiv P^{\rm lin}_{\rm m}(k) & \int{\rm d}M_{\rm h, 1}\ \mathcal{H}_{\rm m}(k,\ M_{\rm h, 1})\ \nh(M_{\rm h, 1})\ \bh(M_{\rm h, 1})\\
    & \int{\rm d}M_{\rm h, 2}\ \mathcal{H}_{\rm g}(k,\ M_{\rm h, 2})\ \nh(M_{\rm h, 2})\ \bh(M_{\rm h, 2})\ .
\end{split}
\end{equation}
Here, $P^{\rm lin}_{\rm m}(k)$ is the linear matter power spectrum, $\nh(\Mh)$ is the halo mass function, and $\bh(\Mh)$ is the associated large-scale halo bias, both with respect to the halo mass, $\Mh$. These properties are redshift-dependent, and in our analysis, we used the mean redshift of the lens samples for each stacked bin, as is shown in Table~\ref{table:bins}. The subscript `g' denotes galaxies, and `m' denotes matter, corresponding to different forms of $\mathcal{H}_x(k,\ \Mh)$:
\begin{equation}
    \mathcal{H}_{\rm m}(k,\ \Mh) \equiv \frac{\Mh}{\rhomBar}\ \tilde{u}_{\rm m}(k|\Mh)~,
\end{equation}
or
\begin{equation}
\label{eq:Hg}
    \mathcal{H}_{\rm g}(k,\ \Mh) \equiv \frac{\langle \Ng | \Mh \rangle}{\ngBar}\ \tilde{u}_{\rm g}(k|\Mh)~.
\end{equation}
The terms $\tilde{u}_{\rm m}(k|\Mh)$ and $\tilde{u}_{\rm g}(k|\Mh)$ describe the normalised density profiles of dark matter halos and galaxy distributions within a halo, respectively, in Fourier space. The term $\langle \Ng | \Mh \rangle$ is the average number of galaxies in a halo of mass $\Mh$, and $\ngBar$ is the average galaxy number density, given by
\begin{equation}
\label{eq:ngBar}
    \ngBar = \int {\rm d}\Mh\ \langle \Ng | \Mh \rangle\ \nh(\Mh)\ .
\end{equation}

\subsection{Baseline model ingredients}
\label{Sec:ModelIng}

We define the overdensity threshold of a virialised dark matter halo such that the average density within the virial radius is 200 times the mean matter density of the Universe. Consequently, the mass of a specific halo is formulated as
\begin{equation}
\label{eq:MassHalo}
\Mh=\frac{4\pi}{3}\ 200\ \rhomBar\ r^3_{\rm 200m}\ .
\end{equation}
For the internal density profile of these halos, we adopted the Navarro-Frenk-White (NFW, \citealt{Navarro1997ApJ...490..493N}) profile truncated at the virial radius, following the Fourier transform form from \citet{Takada2003MNRAS.340..580T}. Alternatively, the profile can also be smoothly truncated in the manner proposed by \citet{Baltz2009JCAP...01..015B}, which has been shown to perform better in the transition region between one-halo and two-halo terms~\citep{Oguri2011MNRAS.414.1851O}. However, since our measurements are dominated by the one-halo term, we are not sensitive to the subtle differences between these truncation strategies. The mass-concentration relation of the NFW profile is based on \citet{Duffy2008MNRAS.390L..64D}:
\begin{equation}
\label{eq:concentration}
c_{\rm m} = f_{\rm c} \times 10.14 \left(\frac{M_{\rm h}}{2.86\times 10^{12}\ h_{70}^{-1}{\rm M}_{\odot}}\right)^{-0.081} (1+z_{\rm d})^{-1.01}\ ,
\end{equation}
where $f_{\rm c}$ is a free scaling parameter introduced to account for potential deviations of the weak lensing measurements from the original simulation-based fitting results. We do not vary the redshift or mass dependence in this equation, as more complex dependencies are predominantly found at redshifts greater than one (e.g.~\citealt{Munoz2011MNRAS.411..584M,Wang2024MNRAS.527.1580W}), which exceed the highest lens redshift in our study.

To account for the mass contribution from central galaxies residing in the innermost region of the dark matter halo, we incorporated a point mass into the NFW density profile. This mass was set to be linearly scaled with the mean stellar mass of the BCGs for each stacked bin (see Table~\ref{table:bins}):
\begin{equation}
   M_{\rm p} \equiv A_{\rm p}\bar{M}_{\star}^{\rm BCG}\ ,
\end{equation}
where the scaling factor, $A_{\rm p}$, is one of the free parameters we vary during the model fitting. Considering the scales of our ESD measurements, our analysis is insensitive to the detailed stellar mass distributions within the innermost part of the dark matter halo.

For the halo mass function and halo bias, we used the calibrated fitting functions from \citet{Tinker2010ApJ...724..878T}, which are derived from a series of cosmological $N$-body simulations within the $\Lambda$CDM framework. Other halo mass function estimates based on different cosmological simulations, including both $N$-body and hydrodynamical simulations, exist in the literature, and their predictions can vary significantly, often exceeding Poisson errors~\citep{Schaye2023MNRAS.526.4978S}. However, these discrepancies are likely dominated by the different halo definitions used by various halo finders~\citep{Euclid2023AA...671A.100E}. Therefore, to ensure consistency in halo definitions with our halo model, we chose the \citet{Tinker2010ApJ...724..878T} halo mass function, as it employs the same mass definitions. Our analysis is also insensitive to the exact form of the halo bias function given the current measurement uncertainties. This is particularly true as we fit the ESD profiles only up to $2.86\ h_{70}^{-1}\ {\rm Mpc}$, and the halo bias only affects our calculations through the two-halo term (Eq.~\ref{eq:Ph2}).

When calculating the halo mass function, we adopted the \citet{Planck2020AA...641A...6P} cosmological parameters ($\Omega_{\Lambda}{=}0.6842$, $\Omega_{\rm m}{=}0.3158$, $\Omega_{\rm b}{=}0.04939$, $\sigma_8{=}0.8120$, $n_{\rm s}{=}0.96605$, and $h{=}0.6732$). To test the robustness of our results against uncertainties in these cosmological parameters, we also used the latest KiDS cosmic shear results~\citep{Li2023AA...679A.133L}, which feature a lower $\sigma_8(\Omega_{\rm m}/0.3)^{0.5}$ value. We find consistent outcomes, confirming that our analysis is insensitive to the current level of cosmological uncertainties. This test also addresses some of the concerns about the uncertainty in the halo mass function used in our model, as the variation in cosmological parameters tested here is more extreme than the current differences in halo mass function estimates in the literature~\citep{Bocquet2016MNRAS.456.2361B}.

If the selected central galaxy (in our case, the BCG) resides exactly at the centre of its host halo, the $\tilde{u}_{\rm g}(k|\Mh)$ term shown in Eq.~(\ref{eq:Hg}) would be unity. However, in reality, the BCGs are not always located at the true gravitational centre of the host halo due to physical factors such as galaxy evolution and interaction (see, e.g.~\citealt{Cui2016MNRAS.456.2566C,Zhang2019MNRAS.487.2578Z} and references therein), as well as observational effects such as the misidentification of central galaxies or fragmentation and aggregation from group-finding algorithms (e.g.~\citealt{Jakobs2018MNRAS.480.3338J,Ahad2023MNRAS.518.3685A,Kelly2024MNRAS.533..572K}). Thus, we need a proper model to account for the miscentring of the selected central galaxies.

In our baseline model, we adopt a two-component miscentring model defined as
\begin{equation}
\label{eq:miscen}
\tilde{u}_{\rm g}(k|\Mh) = (1 - p_{\rm off}) + p_{\rm off} \times \tilde{P}_{\rm off}(k|\mathcal{R}_{\rm off})\ ,
\end{equation}
where 
\begin{equation}
\label{eq:RayF}
\tilde{P}_{\rm off}(k|\mathcal{R}_{\rm off}) = \left(\frac{1}{2x} - x\right) D_{+}(x) + \frac{1}{2}\ ,
\end{equation}
with $x \equiv (k\ \mathcal{R}_{\rm off} r_{\rm s})/\sqrt{2}$, and $D_{+}(x)$ being the Dawson integral. This model assumes that a fraction $p_{\rm off}$ of BCGs is miscentred, with the normalised radial distribution of these miscentred galaxies relative to the true halo centre following a Rayleigh distribution with a scatter of $\mathcal{R}_{\rm off}$ times the scale radius of the halo, $r_{\rm s}$. The choice of the Rayleigh distribution follows \citet{Johnston2007arXiv0709.1159J}. We explore other statistical distributions in Sect.~\ref{Sec:sensiMis}. Both $p_{\rm off}$ and $\mathcal{R}_{\rm off}$ are free parameters.

The HOD term $\langle \Ng | \Mh \rangle$ in Eq.~(\ref{eq:Hg}) is modelled using the conditional stellar mass function (CSMF, \citealt{Yang2003MNRAS.339.1057Y}). This choice is motivated by the direct connection of the CSMF to the relation between baryonic properties and halo mass, which is the focus of this study. Additionally, it facilitates the implementation of simulation predictions, which is the other key aspect of our study. Specifically, we assume a log-normal distribution for the group CSMF based on previous studies (\citealt{Yang2008ApJ...676..248Y,Cacciato2009MNRAS.394..929C,Bosch2013MNRAS.430..725V,Uitert2016MNRAS.459.3251V}):
\begin{equation}
\label{eq:CLF}
\begin{aligned}
\Phi(M_{\star}^{\rm grp}|\Mh)\  =\ & \frac{1}{\ln(10)\sqrt{2\pi}\ M_{\star}^{\rm grp}\ \sigma_{\log M_{\star}^{\rm grp}}}\\
& \times\exp\left[-\frac{(\log M_{\star}^{\rm grp} - \mu_{\log M_{\star}^{\rm grp}})^2}{2\sigma_{\log M_{\star}^{\rm grp}}^2}\right]\ .
\end{aligned}
\end{equation}
We further validated the log-normal form using FLAMINGO simulations. Although these simulation-based tests do not account for stellar mass measurement uncertainties, previous studies have shown that the distribution of these measurement uncertainties for a stacked ensemble is also well approximated by a log-normal distribution~\citep{Yang2009ApJ...695..900Y,Behroozi2010ApJ...717..379B}. Therefore, the log-normal form remains appropriate even in the presence of measurement uncertainties, although the inferred scatter will reflect a combination of intrinsic scatter and measurement uncertainty~\citep{Moster2010ApJ...710..903M,Leauthaud2012ApJ...744..159L}.

Equation~(\ref{eq:CLF}) has two free parameters: the logarithmic scatter $\sigma_{\log M_{\star}^{\rm grp}}$ and the logarithmic mean $\mu_{\log M_{\star}^{\rm grp}}$ of the group stellar mass distribution for a given halo mass $\Mh$. We model the logarithmic mean using a power-law scaling relation:
\begin{equation}
\label{eq:scaling}
\mu_{\log M_{\star}^{\rm grp}} = 11.8 + \log(A_{\rm s}) + \alpha_{\rm s}\log\left(\frac{\Mh}{10^{14.15}\ h_{70}^{-1}{\rm M}_{\odot}}\right)\ ,
\end{equation}
where the amplitude $A_{\rm s}$ and index $\alpha_{\rm s}$ are free parameters. The normalisation $11.8$ corresponds to the logarithmic mean group stellar mass of the full GAMA sample, and the choice of $10^{14.15}\ h_{70}^{-1}{\rm M}_{\odot}$ follows \citet{Viola2015MNRAS.452.3529V}. In our baseline model, we assume the logarithmic scatter to be a halo mass-independent free parameter. However, we explore a more realistic, simulation-informed scatter model in Sect.~\ref{Sec:simScatter}.

Under the assumption of sample completeness, which is valid given the high completeness of the GAMA survey and our exclusion of distribution tails (see Fig.~\ref{fig:distri}), we can calculate the mean number of groups per specific observable bin as follows:
\begin{equation}
\label{eq:ngMh}
\langle \Ng | \Mh \rangle = \int_{M_{\star, \rm min}^{\rm grp}}^{M_{\star, \rm max}^{\rm grp}} \mathrm{d}M_{\star}^{\rm grp}~\Phi(M_{\star}^{\rm grp}|\Mh)\ ,
\end{equation}
where the integral limits, $M_{\star, \rm min}^{\rm grp}$ and $M_{\star, \rm max}^{\rm grp}$, correspond to the stacked bin boundaries, as detailed in Table~\ref{table:bins}.

We tested the impact of potential sample incompleteness by introducing an additional incompleteness function into Eq.~(\ref{eq:ngMh}), following \citet{Uitert2016MNRAS.459.3251V}. This function, based on an error function, includes two additional free parameters. We find that these incompleteness parameters are not constrained by the current data, while constraints on the other parameters remain fully consistent with our baseline model. This confirms the assumption of high completeness in our sample and indicates that introducing additional incompleteness parameters into the model is unnecessary.

We used the same HOD form for the relation between halo mass and group luminosity, namely a log-normal distribution for the group conditional luminosity function (Eq.~\ref{eq:CLF}), with the logarithmic mean of the group luminosity scaling with halo mass through a power-law relation (Eq.~\ref{eq:scaling}). It is worth noting that this conditional luminosity function-based HOD model differs from those used by \citet{Viola2015MNRAS.452.3529V} and \citet{Rana2022MNRAS.510.5408R}, who defined the HOD directly based on the average number of galaxies as a function of halo mass. Moreover, they constrained the scaling relation as the mean halo mass for a given luminosity, which differs from the logarithmic mean luminosity for a given halo mass due to the scatter. In order to compare our results with these previous results, we derived the mean halo mass for a given luminosity using Bayes theorem (e.g.~\citealt{Coupon2015MNRAS.449.1352C}):
\begin{equation}
\label{eq:meanHalo}
    \langle \Mh | L_{r}^{\rm grp} \rangle = \frac{\int\mathrm{d}\Mh\ \nh(\Mh)\ \Phi(L_{r}^{\rm grp}|\Mh)\ \Mh}{\int\mathrm{d}\Mh\ \nh(\Mh)\ \Phi(L_{r}^{\rm grp}|\Mh)}~,
\end{equation}
where $\Phi(L_{r}^{\rm grp}|\Mh)$ is the group conditional luminosity function, following the same form as Eq.~(\ref{eq:CLF}).

\subsection{Model fitting}
\label{Sec:fit}

The baseline model outlined above contains seven free parameters, each with broad, uninformative priors, as detailed in Tables~\ref{table:para} and \ref{table:paraLgrp}. We performed a joint fit to the stacked ESD measurements of the six observable bins using the specified halo model, accounting for the full correlations between the 60 data points (see Sect.~\ref{Sec:cov}). The posterior parameter space was sampled using the \texttt{emcee} code~\citep{Foreman2013PASP..125..306F}, an implementation of the affine-invariant Markov chain Monte Carlo (MCMC) ensemble sampler~\citep{Goodman2010CAMCS...5...65G}. The convergence of the MCMC chains was assessed using the integrated autocorrelation time (e.g.~\citealt{Goodman2010CAMCS...5...65G}). 

\begin{table}
\renewcommand{\arraystretch}{1.5}
\caption{Free parameters in our baseline model for fitting stacked ESD profiles binned by group total stellar mass.}
\label{table:para}      
\centering          
\begin{tabular}{llcc} 
\hline\hline      
Parameter & Prior & Constraints & Best-fit values \\ 
\hline            
$f_{\rm c}$ & [0.2, 3] & $0.88_{-0.24}^{+0.57}$ & $1.38$\\
$A_{\rm p}$ & [0.1, 5] & $2.19_{-1.32}^{+1.44}$ & $1.16$\\
$p_{\rm off}$ & [0, 1] & $0.34_{-0.19}^{+0.18}$ & $0.50$\\
$\mathcal{R}_{\rm off}$ & [0, 5] & $2.47_{-0.92}^{+1.31}$ & $2.26$\\
\hline
$\sigma_{\log M_{\star}^{\rm grp}}$ & [0.01, 1] & $0.14_{-0.09}^{+0.11}$ & $0.09$\\
$A_{\rm s}$ &       [0, 5]  & $2.36_{-0.63}^{+0.60}$ & $2.72$\\
$\alpha_{\rm s}$ &  [0, 5]   & $1.02_{-0.11}^{+0.11}$ & $1.04$\\
\hline                  
\end{tabular}
\tablefoot{Constraints are presented as the median of the marginalised posterior distributions, with 68\% credible intervals. Best-fit values correspond to the parameter set that minimises the $\chi^2$ value of the model fit.}
\end{table}

\begin{table}
\renewcommand{\arraystretch}{1.5}
\caption{Same as Table~\ref{table:para}, but for fitting measurements binned by group $r$-band total luminosity.}
\label{table:paraLgrp}      
\centering          
\begin{tabular}{llcc} 
\hline\hline      
Parameter & Prior & Constraints & Best-fit values \\ 
\hline            
$f_{\rm c}$ & [0.2, 3] & $0.92_{-0.24}^{+0.41}$ & $0.93$\\
$A_{\rm p}$ & [0.1, 5] & $2.15_{-1.29}^{+1.42}$ & $2.00$\\
$p_{\rm off}$ & [0, 1] & $0.36_{-0.17}^{+0.14}$ & $0.38$\\
$\mathcal{R}_{\rm off}$ & [0, 5] & $2.82_{-0.95}^{+1.11}$ & $2.66$\\
\hline
$\sigma_{\log L_{r}^{\rm grp}}$ & [0.01, 1] & $0.13_{-0.08}^{+0.10}$ & $0.06$\\
$A_{\rm s}$ &       [0, 3]  & $0.87_{-0.23}^{+0.22}$ & $0.94$\\
$\alpha_{\rm s}$ &  [0, 5]   & $0.88_{-0.10}^{+0.10}$ & $0.89$\\
\hline                  
\end{tabular}
\end{table}

\section{Results from the baseline model}
\label{Sec:res}

Figures~\ref{fig:ESDMgrp} and \ref{fig:ESDLgrp} show the best-fit ESD profiles along with their $68\%$ and $95\%$ credible intervals for $M^{\rm grp}_{\star}$ and $L^{\rm grp}_{r}$ binning, respectively. Assuming independence among the seven free parameters, our baseline model fitting achieves a reduced $\chi^2$ of $1.03$ for $M^{\rm grp}_{\star}$ binning and $0.84$ for $L^{\rm grp}_{r}$ binning, indicating an overall good fit to the data. Tables~\ref{table:para} and \ref{table:paraLgrp} present the marginalised median constraints and best-fit values for these parameters, with uncertainties corresponding to $68\%$ credible intervals. 

\begin{figure}
  \centering
  \includegraphics[width=\hsize]{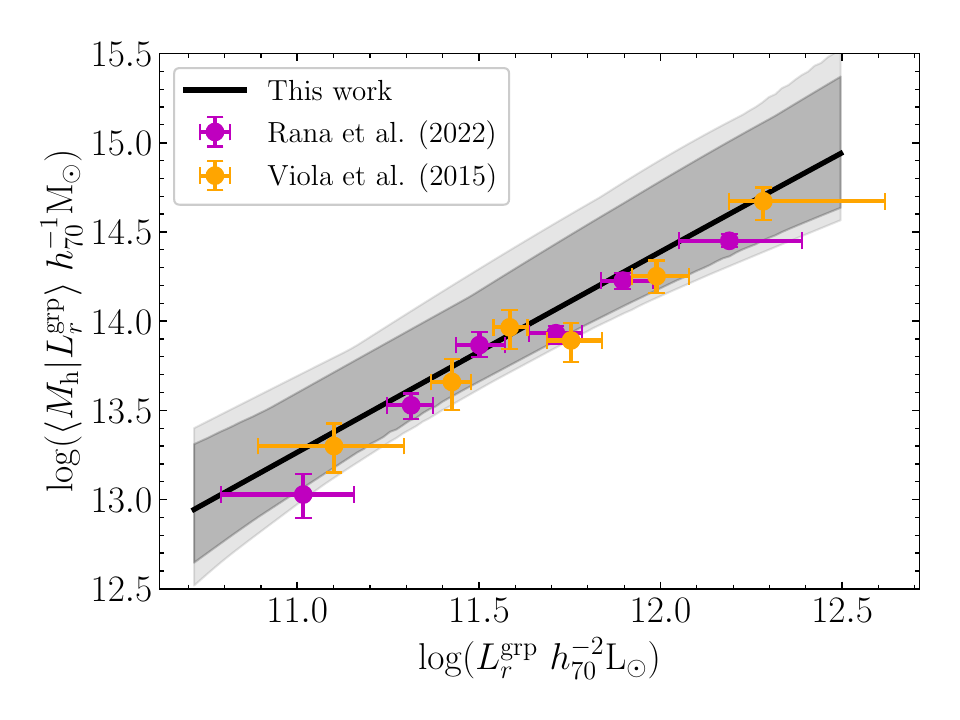}
      \caption{Scaling relation between the halo mass and $r$-band total luminosity of galaxy groups from our baseline model. The black line shows the best-fit results, with the shaded regions illustrating the corresponding $68\%$ and $95\%$ credible intervals. The parameter values are provided in Table~\ref{table:paraLgrp}. The results are compared to previous measurements from \citet{Viola2015MNRAS.452.3529V} (orange points) and \citet{Rana2022MNRAS.510.5408R} (magenta points). All three measurements are based on the GAMA group catalogue, but with different shear measurements and modelling approaches. We note that the scaling relation is demonstrated as the mean halo mass for a given luminosity.} 
         \label{fig:scalingL}
\end{figure}

Figure~\ref{fig:scalingL} shows the scaling relation between the mean halo mass and group $r$-band luminosity, compared to previous studies by \citet{Viola2015MNRAS.452.3529V} and \citet{Rana2022MNRAS.510.5408R}. We re-emphasise that our halo model directly constrains the logarithmic mean of the group luminosity as a function of halo mass (Eq.~\ref{eq:scaling}), while \citet{Viola2015MNRAS.452.3529V} and \citet{Rana2022MNRAS.510.5408R} constrained the scaling relation as the mean halo mass for a given luminosity. To enable a comparison with these previous results, we converted our constraint using Eq.~(\ref{eq:meanHalo}). Our new analysis features updated shear measurements and calibration relative to \citet{Viola2015MNRAS.452.3529V} and employs a different modelling approach compared to these two studies (see Sect.~\ref{Sec:ModelIng}). Despite these changes, we observe good agreement between our measurements and previous studies, verifying the reliability of our new modelling approach.

To be more quantitative, we fitted a power-law relation between the mean halo mass and the group $r$-band luminosity and found
\begin{equation}
\label{eq:scalingRes}
\frac{\langle \Mh | L_{r}^{\rm grp} \rangle}{10^{14.15}\ h_{70}^{-1}{\rm M}_{\odot}} = (0.96_{-0.15}^{+0.24})\left(\frac{L_{r}^{\rm grp}}{10^{11.8}\ h_{70}^{-1}{\rm L}_{\odot}}\right)^{1.08_{-0.09}^{+0.10}},
\end{equation}
where the pivot values followed the choice of \citet{Viola2015MNRAS.452.3529V}. The linear regression is performed on all MCMC chains with the mean halo mass $\langle \Mh | L_{r}^{\rm grp} \rangle$ estimated using Eq.~(\ref{eq:meanHalo}). The reported values correspond to the median of the marginalised distributions, with $68\%$ credible intervals for the uncertainties. These results are fully consistent with those reported by \citet{Viola2015MNRAS.452.3529V} and \citet{Rana2022MNRAS.510.5408R}, who yielded a normalisation and power-law index combination of $0.95\pm 0.14$ and $1.16\pm 0.13$, and $0.81\pm 0.04$ and $1.01\pm 0.07$, respectively.

While our lens sample size is larger than that used by \citet{Viola2015MNRAS.452.3529V}, the uncertainties of our final constraints are comparable to theirs. This is largely because we applied a more stringent scale cut to alleviate blending effects on small scales---we used a scale cut of $0.04\ h_{70}^{-1}{\rm Mpc}$ compared to their $0.03\ h_{70}^{-1}{\rm Mpc}$. Additionally, we excluded the tails of the $L_{r}^{\rm grp}$ distributions to mitigate potential group detection effects, as shown in Fig~\ref{fig:distri}. In this sense, we consciously traded some statistical power for increased robustness.

\begin{figure}
  \centering
  \includegraphics[width=\hsize]{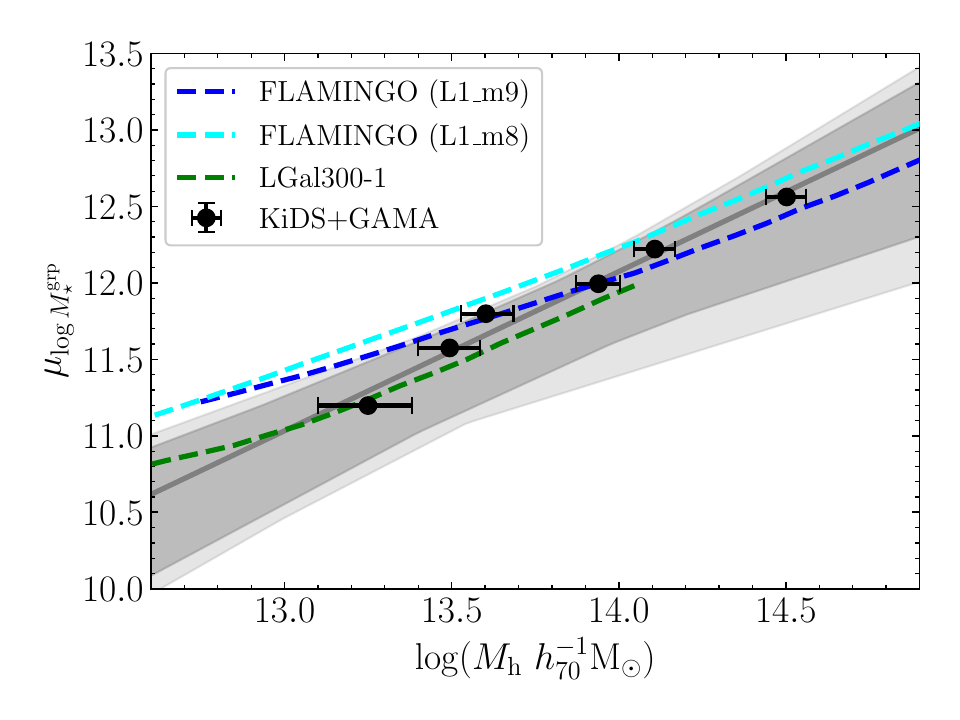}
      \caption{Scaling relation between the total stellar mass of galaxy groups and their halo masses from our baseline model. The grey line shows the best-fit results, with the shaded regions illustrating the corresponding $68\%$ and $95\%$ credible intervals. The parameter values are provided in Table~\ref{table:para}. The black points represent the halo masses calculated by allowing $A_{\rm s}$ to vary for each stacked bin while fixing other parameters to their best-fit values from the joint fit. The error bars correspond to the $68\%$ credible intervals of the new constrained $A_{\rm s}$ distributions. The corresponding $\mu_{\log M_{\star}^{\rm grp}}$ values for these points are the mean log-stellar mass of all groups in the given stacked bin. Predictions from simulations, represented by dashed lines, are estimated from the mock catalogue built in Sect.~\ref{Sec:sim}. All values of $\mu_{\log M_{\star}^{\rm grp}}$ are converted to a $h_{70}$ cosmology for comparison, with $M_{\star}^{\rm grp}$ from simulations scaled as $h_{70}^{-1}$ and those from observations scaled as $h_{70}^{-2}$. We note that the scaling relation is demonstrated as the mean log-stellar mass at a fixed halo mass.} 
         \label{fig:scaling}
\end{figure}

Figure~\ref{fig:scaling} compares the constrained scaling relation to the simulation predictions. Unlike in Fig.~\ref{fig:scalingL}, here we present the direct constraints on the logarithmic mean of the group stellar mass as a function of halo mass without further transformation, as this scaling relation can be easily measured from the mock group catalogues (Sect.~\ref{Sec:mock}). To delineate the mass regions covered by our data measurements from those inferred through model extrapolation, we also include data points for the six observable bins where we measured the stacked ESD signals (Sect.~\ref{Sec:ESD}). These data points were estimated by running separate MCMC chains for each observable bin, fixing the baseline model parameters to their best-fit values from the joint fit, except for the $A_{\rm s}$ parameter. The new $A_{\rm s}$ constraints can provide halo mass estimates for each stacked bin using Eq.~(\ref{eq:scaling}), given that the mean group stellar mass for each stacked bin is well measured. The uncertainties in the halo mass estimates were calculated using the 68\% credible intervals of the new constrained $A_{\rm s}$ distributions. The results indicate that our current sample is sensitive to halos in the mass range of ${\sim}10^{13.1}$ to ${\sim}10^{14.6}~h_{70}^{-1}{\rm M}_{\odot}$. The relation outside this mass range is an extrapolation from our assumed power-law model of Eq.~(\ref{eq:scaling}), thus requiring caution when interpreting these extrapolated regions, especially if the scaling relation deviates from the assumed power-law behaviour.

In general, we find a good agreement between the results inferred from our measurements using our baseline model and all simulation predictions at the high-mass end. At the low-mass end, the FLAMINGO simulations slightly over-predict the mean group stellar mass for a given halo mass, while the \textsc{L-Galaxies} SAM predictions remain closer to our data constraints. The TNG100-1 simulations do not have sufficient volume to cover the halo mass regions measured by our sample, thus cannot be directly tested by our measurements, but they sit between the FLAMINGO and \textsc{L-Galaxies} SAM predictions at the low-mass end.

\section{Simulation-informed scatter model}
\label{Sec:simScatter}

After validating the simulation predictions against the scaling relation constrained by our baseline model, we refine the halo model by incorporating insights from simulations. In this study, we focus on one of the key simplifications in the current halo model, where the scatter in the group stellar mass distribution is assumed to be mass-independent. This assumption contrasts recent findings from semi-empirical models (e.g.~\citealt{Bradshaw2020MNRAS.493..337B}), as well as those from semi-analytical models and hydrodynamical simulations (e.g.~\citealt{Pei2024MNRAS.531.2262P}). With improved measurement statistics and wider coverage of the halo mass range, revisiting this simplification becomes important.

\begin{figure}
  \centering
  \includegraphics[width=\hsize]{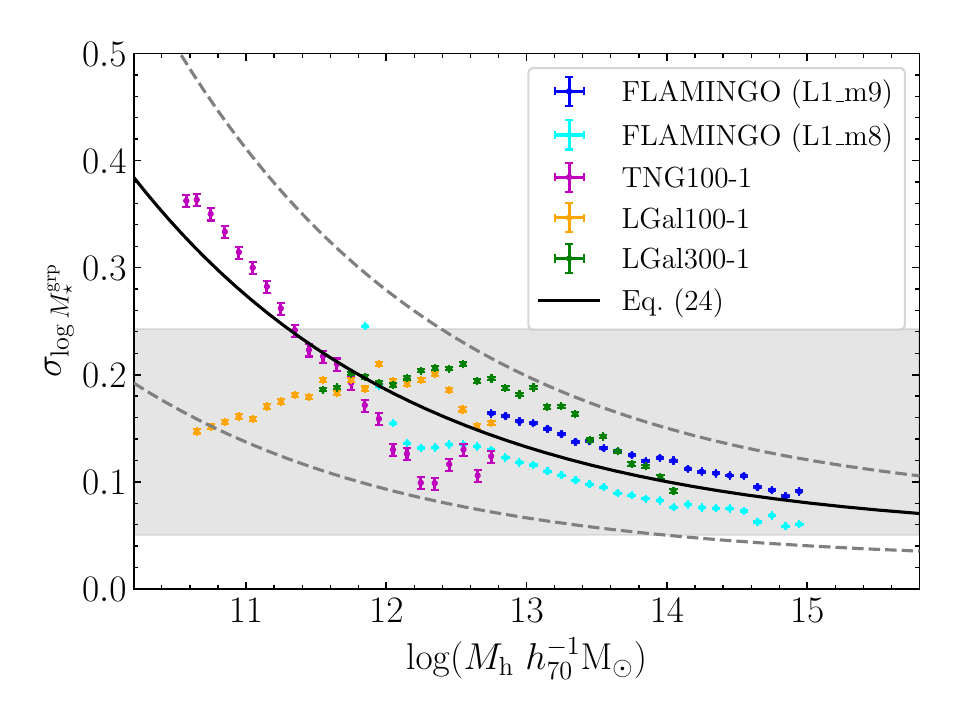}
      \caption{Scatter in the group stellar mass distribution as a function of halo mass, measured from cosmological simulations. The shaded region indicates the 68\% credible interval of the constant scatter constrained by our baseline model (Table~\ref{table:para}). The solid and dashed lines represent the scatter model from Eq.~(\ref{eq:scatter}) with $A_{\sigma}$ set to $0.1$ and $0.1 \pm 0.05$, respectively.} 
         \label{fig:scatter}
\end{figure}

Figure~\ref{fig:scatter} shows the scatter in the group stellar mass distribution as a function of halo mass, measured from the mock group catalogues built from simulations (Sect.~\ref{Sec:mock}). We observe a general decreasing trend in scatter with increasing halo mass, except for the \textsc{L-Galaxies} SAM results, which show an increasing trend at the low halo mass end. However, when considering the uncertainties in current measurements, as indicated by the shaded region representing the 68\% credible interval of the constant scatter constrained by our baseline model, this scatter trend is relatively small within the halo mass range covered by our measurements. Therefore, instead of attempting to directly constrain this scatter trend from our measurements, we opt to incorporate this simulation-informed scatter trend into our halo model and assess how it affects the constrained scaling relation.

Specifically, we model this scatter-halo mass relation with an exponential equation of the form:
\begin{equation}
\label{eq:scatter}
\sigma_{\log M_{\star}^{\rm grp}} \equiv \frac{A_{\sigma}}{2}\left[\exp\left(-0.5\log\left(\frac{\Mh}{10^{14}\ h_{70}^{-1}{\rm M}_{\odot}}\right)\right) + 1\right]~,
\end{equation}
where $A_{\sigma}$ is the amplitude, allowed to vary to account for uncertainties among different simulation predictions. This equation captures the general behaviour of the scatter, as shown in Fig.~\ref{fig:scatter}: a decreasing trend with increasing halo mass at lower mass scales and flattens out at higher mass scales. We find that a Gaussian prior with a mean of 0.1 (solid line in the plot) and a standard deviation of 0.05 (dashed lines in the plot) for $A_{\sigma}$ sufficiently covers the uncertainties among different simulation predictions. We tested using a flat prior and found consistent results, confirming that our current data statistics are insufficient to distinguish subtle differences in the scatter-halo mass relation.

\begin{figure}
  \centering
  \includegraphics[width=\hsize]{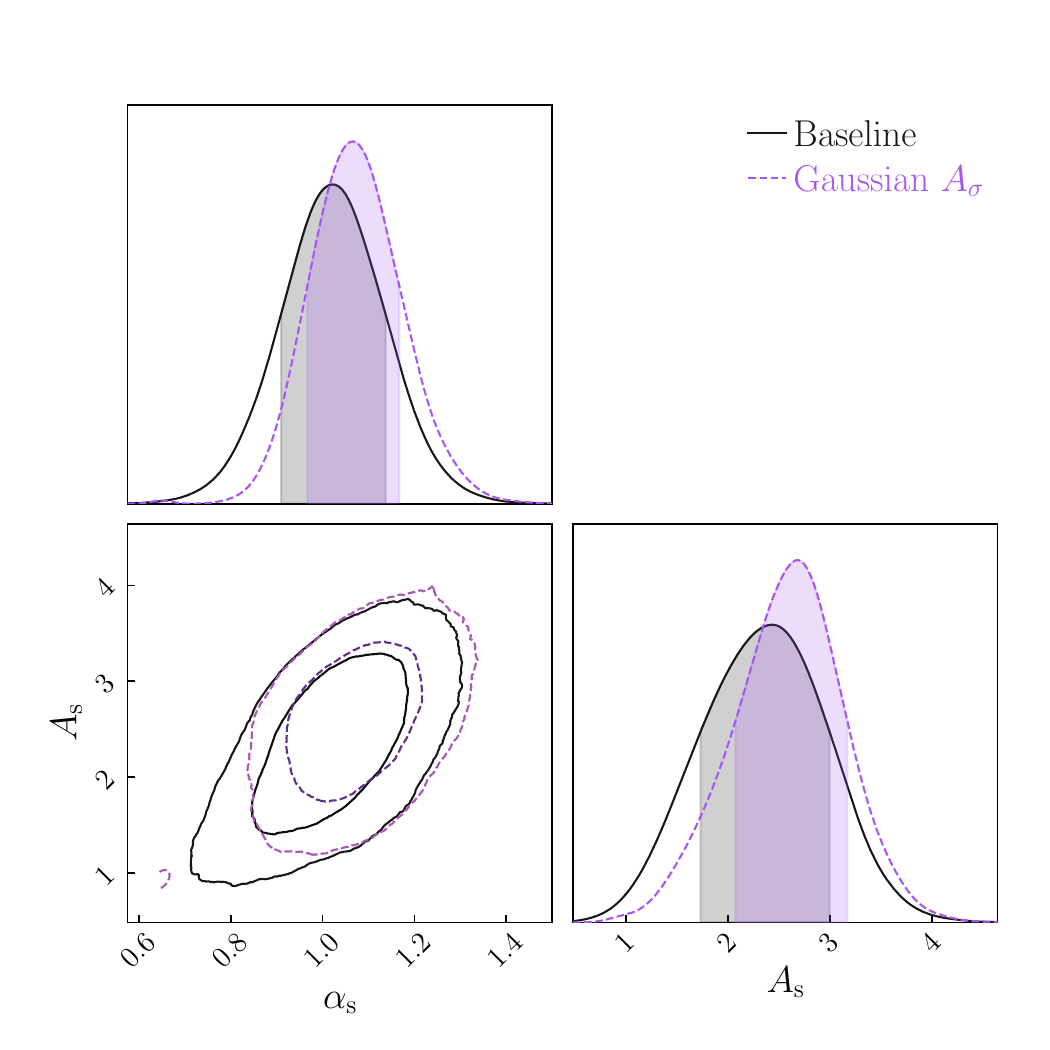}
      \caption{Comparison of projected posterior distributions between the baseline constant scatter model and the simulation-derived scatter model for the two parameters of the scaling relation in Eq.~(\ref{eq:scaling}). The `Gaussian $A_{\sigma}$' refers to the use of scatter-halo mass relation of Eq.~(\ref{eq:scatter}) with a Gaussian prior for $A_{\sigma}$. The contours represent the 68\% and 95\% credible intervals, smoothed with a matched elliptical Gaussian kernel density estimator.} 
         \label{fig:scatterContours}
\end{figure}

Figure~\ref{fig:scatterContours} compares the constraints on the scaling relation parameters between the new scatter model and the baseline constant scatter model. The new results remain consistent with the baseline model. However, we observe tighter and less degenerate constraints, demonstrating the benefits of including the simulation-informed scatter model even with the current data statistics. The consistency between the constant scatter model and the variable scatter model is largely due to the minor scatter variation within the halo mass range covered by our current measurements (${\sim}10^{13.1}$ to ${\sim}10^{14.6}~h_{70}^{-1}{\rm M}_{\odot}$) relative to the measurement uncertainties. With future analyses extending to lower mass ranges and improved statistics, we anticipate a greater impact from the scatter model, warranting continued investigation of the scatter-halo mass relation.

\section{Sensitivity to the miscentring models}
\label{Sec:sensiMis}

To test the importance of miscentring modelling on our results, we focus on two aspects: first, the necessity of accounting for miscentring effects, and second, the sensitivity of the current analysis to different statistical miscentring models. Besides the Rayleigh distribution, the two other commonly used statistical distributions are the Gaussian distribution, formulated as
\begin{equation}
\label{eq:GaussianF}
\tilde{P}_{\rm G}(k|\mathcal{R}_{\rm off}) = \exp\left[-\frac{1}{2}\ k^2\ (r_{\rm s}\mathcal{R}_{\rm off})^2\right]~,
\end{equation}
and the Gamma distribution, formulated as
\begin{equation}
\label{eq:GammaF}
\tilde{P}_{\Gamma}(k|\mathcal{R}_{\rm off}) = \frac{1}{3} \frac{3-k^2(r_{\rm s}\mathcal{R}_{\rm off})^2}{(k^2(r_{\rm s}\mathcal{R}_{\rm off})^2+1)^3}~.
\end{equation}

\begin{figure}
  \centering
  \includegraphics[width=\hsize]{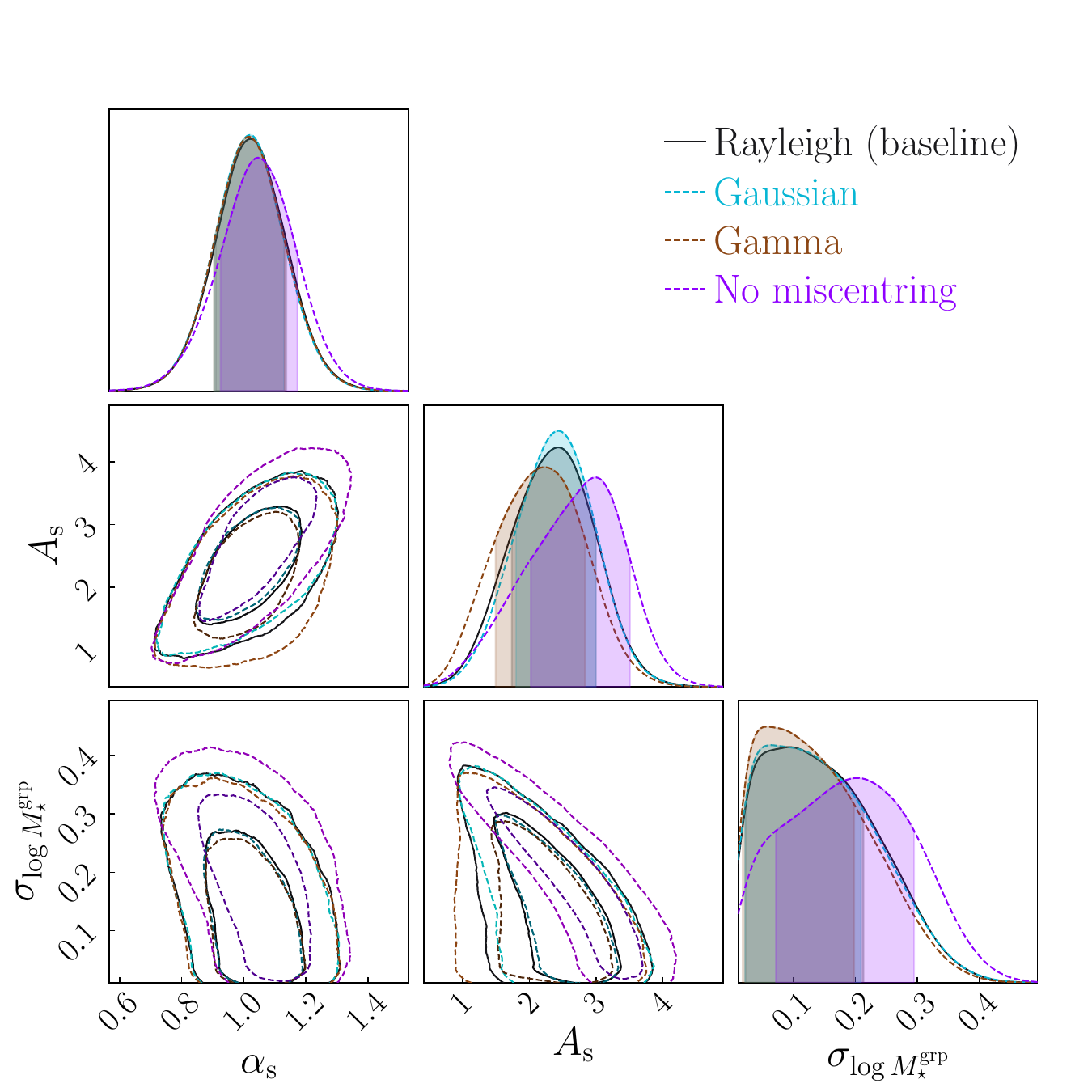}
      \caption{Comparison of projected posterior distributions across different treatments of miscentring effects for the scaling relation parameters and scatter parameter. The contours represent the 68\% and 95\% credible intervals, smoothed with a matched elliptical Gaussian kernel density estimator.} 
         \label{fig:miscenContours}
\end{figure}

Figure~\ref{fig:miscenContours} compares the constraints on the scaling relation parameters from different treatments of miscentring effects. Noticeable shifts in both the scaling relation amplitude $A_{\rm s}$ and scatter $\sigma_{\log M_{\star}^{\rm grp}}$ are observed when we ignore the miscentring effects in the modelling. However, shifts among different statistical models are negligible given the current uncertainties, implying that our current analysis is insensitive to the subtle differences in the assumed miscentring distributions.  

\begin{figure}
  \centering
  \includegraphics[width=\hsize]{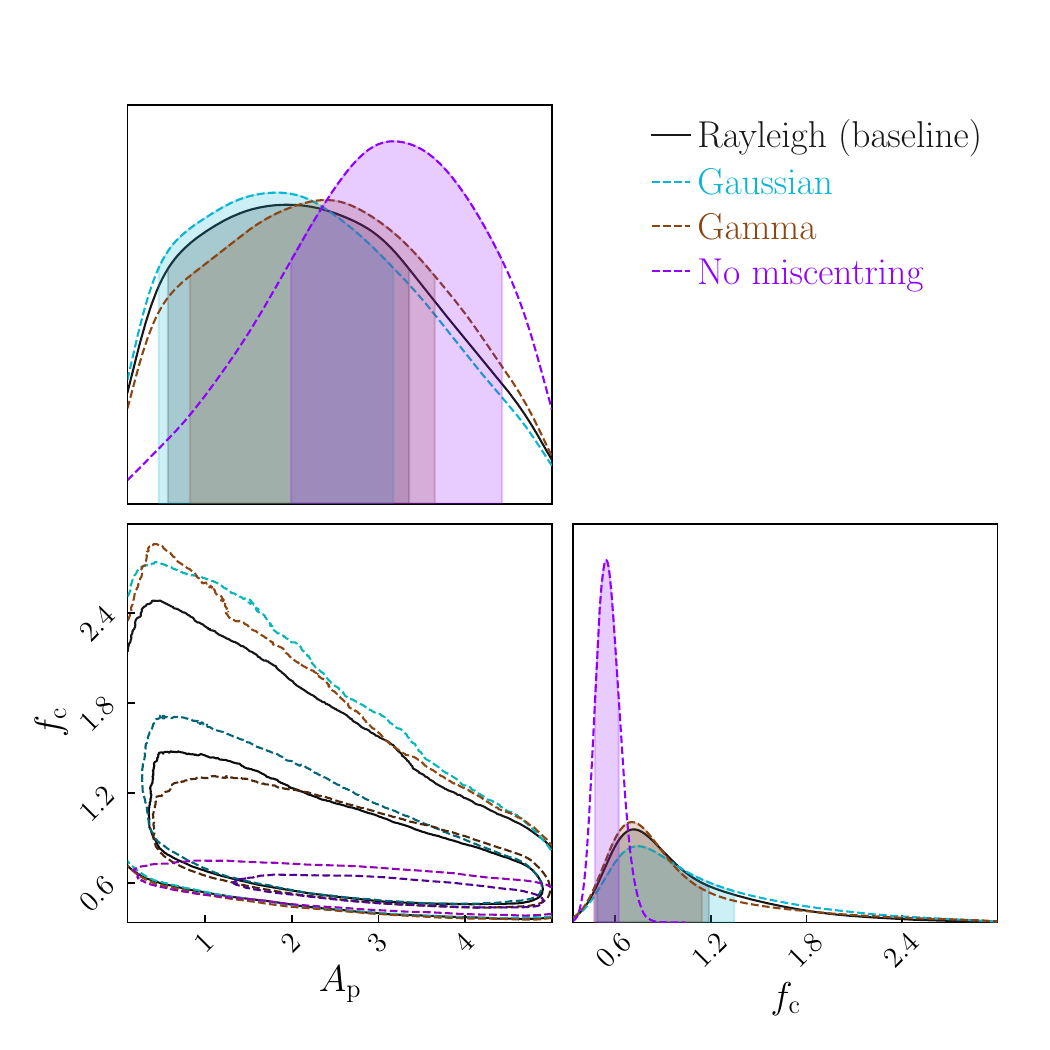}
      \caption{Same comparison as Fig.~\ref{fig:miscenContours} but for the parameters of mass concentration $f_{\rm c}$ and point mass contribution $A_{\rm p}$.} 
         \label{fig:miscenContoursInner}
\end{figure}

These conclusions hold when we check the constraints on the parameters describing the halo inner mass distribution, as is shown in Fig.~\ref{fig:miscenContoursInner}. Without a miscentring model, the scaling parameters for mass-concentration, $f_{\rm c}$, and point mass contribution, $A_{\rm p}$, have much narrower but potentially biased constraints due to the high degeneracy between these parameters and the miscentring parameters. This finding is consistent with the results of \citet{Viola2015MNRAS.452.3529V}, and it cautions against interpreting the halo mass concentration constrained by weak lensing analyses without a realistic miscentring model.

When comparing the reduced $\chi^2$ values of these different models, none stands out as superior to the others. The model with the Gamma distribution has the best reduced $\chi^2$ value of 1.02, while the model without miscentring shows the worst reduced $\chi^2$ value of 1.05. However, it is important to note that the calculated reduced $\chi^2$ value assumes independence among free parameters, even though some degeneracy between parameters is observed in the posterior distributions. Therefore, the reported reduced $\chi^2$ values should be seen as indicative rather than definitive for ruling out models.

In practice, the cause of miscentring in a group sample is more intricate than what the adopted statistical distributions can account for. For example, \citet{Ahad2023MNRAS.518.3685A} found that line-of-sight projections, which result in a discrepancy between the projected and intrinsic luminosity, account for approximately half of the identified miscentred groups in their simulations. Furthermore, the aggregation and fragmentation effects, referring to the phenomena where two small groups are identified as a single larger group, and a single large group is identified as several smaller groups, respectively, are common in real data group-finding algorithms (see Appendix A of \citealt{Jakobs2018MNRAS.480.3338J}), which further complicate the distribution of miscentred BCGs. Developing a sophisticated miscentring model that accounts for these more complicated selection effects is still important but can only be addressed using simulations that include the sample selection from a specific survey, which is beyond the scope of our current study.

\section{Conclusions}
\label{Sec:conclusion}

We conducted a weak lensing analysis using the latest KiDS-1000 shear catalogue (v2) from \citet{Li2023AA...679A.133L} to constrain the scaling relation between baryonic observables and halo masses for galaxy groups identified by GAMA. Using a baseline halo model with seven free parameters, we achieved a good fit to the measured ESD signals, with a reduced $\chi^2$ of 1.03 for stacks based on group stellar masses and 0.84 for stacks based on group $r$-band luminosity.

Compared to previous studies by \citet{Viola2015MNRAS.452.3529V} and \citet{Rana2022MNRAS.510.5408R}, we refined the lens sample selection, updated the shear measurements and calibration, and adopted a new modelling approach based on the conditional stellar mass (or luminosity) function within the halo model framework. Despite these changes, our constraints on the scaling relation between the mean halo mass and group $r$-band luminosity are fully consistent with the ones from previous studies. Specifically, our baseline model yields a normalisation and power-law index combination of $0.96_{-0.15}^{+0.24}$ and $1.08_{-0.09}^{+0.10}$ (Eq.~\ref{eq:scalingRes}), whereas \citet{Viola2015MNRAS.452.3529V} and \citet{Rana2022MNRAS.510.5408R} reported combinations of $0.95\pm 0.14$ and $1.16\pm 0.13$, and $0.81\pm 0.04$ and $1.01\pm 0.07$, respectively.

We further compared the constrained group stellar mass-halo mass relation to predictions from the latest FLAMINGO cosmological simulations, as well as the \textsc{L-Galaxies} SAM implemented in IllustrisTNG gravity-only simulations. We find a general agreement between our measurements and simulation predictions for halos with masses ${\gtrsim}10^{13.5} h_{70}^{-1}{\rm M}_{\odot}$. For halos with masses below this value, the FLAMINGO simulations slightly over-predict group stellar masses, while the \textsc{L-Galaxies} SAM shows better agreement with our measurements. These findings are consistent with those of \citet{Schaye2023MNRAS.526.4978S}, who found that the FLAMINGO stellar-to-halo mass relation for central galaxies is higher at the low-halo-mass end compared to the semi-empirical UniverseMachine model results from \citet{Behroozi2019MNRAS.488.3143B}.

After validating the simulation predictions, we improved our baseline halo model by incorporating the simulation-informed scatter in the group stellar mass distribution as a function of halo mass. Using an exponential equation with a variable amplitude (Eq.~\ref{eq:scatter}), the improved halo model captures the general decreasing trend of scatter with increasing halo masses and accounted for uncertainties among different simulation predictions. Although the current measurement statistics are insufficient to directly constrain the variable scatter, we find that the updated model yields tighter constraints on the scaling relation parameters, highlighting the advantages of simulation-informed halo modelling.

We tested the robustness of our scaling relation results against sensible changes to the miscentring modelling. We verified that including a statistical model to account for the potential miscentring of the selected central galaxies is necessary. Ignoring this miscentring effect would bias not only the estimation of the mass concentration in the inner region of the halo profile but also the scaling relation constraints. When testing different statistical models for miscentring, we observe minor shifts in the scaling relation parameters that are well within the current measurement uncertainties. This suggests that the current data statistics are insufficient to distinguish among different statistical models of miscentring. However, with improved statistics of lens samples from upcoming spectroscopic surveys such as the 4MOST Wide Area VISTA Extragalactic Survey (WAVES; \citealt{Driver2019Msngr.175...46D}) and Hemisphere Survey of the Nearby Universe (4HS; \citealt{Taylor2023Msngr.190...46T}), along with significantly enhanced weak lensing measurements from the ESA \textit{Euclid}~\citep{Euclid2023AA...671A.100E} and \textit{Rubin} LSST~\citep{Ivezic2019ApJ873111I} surveys, we will be able to measure galaxy-galaxy lensing signals down to much smaller scales. To accurately model these small-scale lensing signals, further development of realistic miscentring models that account for observational effects is warranted.
\begin{acknowledgements}
    We thank Andrej Dvornik for his assistance with the KiDS galaxy-galaxy lensing pipeline, Roi Kugel for his help with the FLAMINGO products, and Maciej Bilicki and Giorgio Francesco Lesci for their valuable comments. We acknowledge support from: the Netherlands Research School for Astronomy (SSL); the European Research Council (ERC) under the European Union’s Horizon 2020 research and innovation programme with Grant agreement No. 101053992 (SSL, HHo). The KiDS data used in this paper are based on observations made with ESO Telescopes at the La Silla Paranal Observatory under programme IDs 177.A-3016, 177.A-3017, 177.A-3018 and 179.A-2004, and on data products produced by the KiDS consortium. The KiDS production team acknowledges support from: Deutsche Forschungsgemeinschaft, ERC, NOVA and NWO-M grants; Target; the University of Padova, and the University Federico II (Naples). GAMA is a joint European-Australasian project based around a spectroscopic campaign using the Anglo-Australian Telescope. The GAMA input catalogue is based on data taken from the Sloan Digital Sky Survey and the UKIRT Infrared Deep Sky Survey. Complementary imaging of the GAMA regions is being obtained by a number of independent survey programmes including GALEX MIS, VST KiDS, VISTA VIKING, WISE, Herschel-ATLAS, GMRT and ASKAP providing UV to radio coverage. GAMA is funded by the STFC (UK), the ARC (Australia), the AAO, and the participating institutions. The GAMA website is https://www.gama-survey.org/.

    Author contributions: All authors contributed to the development and writing of this paper. The authorship list is given in two groups: the lead authors (SSL, HHo, KK), followed by an alphabetical group, which includes those who are key contributors to both the scientific analysis and the data products. 
    
\end{acknowledgements}

\bibliographystyle{aa} 
\bibliography{reference}

\begin{thebibliography}{154}
\expandafter\ifx\csname natexlab\endcsname\relax\def\natexlab#1{#1}\fi

\bibitem[{{Abbott} {et~al.}(2022){Abbott}, {Aguena}, {Alarcon}, {Allam}, {Alves}, {Amon}, {Andrade-Oliveira}, {Annis}, {Avila}, {Bacon}, {Baxter}, {Bechtol}, {Becker}, {Bernstein}, {Bhargava}, {Birrer}, {Blazek}, {Brandao-Souza}, {Bridle}, {Brooks}, {Buckley-Geer}, {Burke}, {Camacho}, {Campos}, {Carnero Rosell}, {Carrasco Kind}, {Carretero}, {Castander}, {Cawthon}, {Chang}, {Chen}, {Chen}, {Choi}, {Conselice}, {Cordero}, {Costanzi}, {Crocce}, {da Costa}, {da Silva Pereira}, {Davis}, {Davis}, {De Vicente}, {DeRose}, {Desai}, {Di Valentino}, {Diehl}, {Dietrich}, {Dodelson}, {Doel}, {Doux}, {Drlica-Wagner}, {Eckert}, {Eifler}, {Elsner}, {Elvin-Poole}, {Everett}, {Evrard}, {Fang}, {Farahi}, {Fernandez}, {Ferrero}, {Fert{\'e}}, {Fosalba}, {Friedrich}, {Frieman}, {Garc{\'\i}a-Bellido}, {Gatti}, {Gaztanaga}, {Gerdes}, {Giannantonio}, {Giannini}, {Gruen}, {Gruendl}, {Gschwend}, {Gutierrez}, {Harrison}, {Hartley}, {Herner}, {Hinton}, {Hollowood}, {Honscheid}, {Hoyle}, {Huff}, {Huterer}, {Jain}, {James}, {Jarvis},
  {Jeffrey}, {Jeltema}, {Kovacs}, {Krause}, {Kron}, {Kuehn}, {Kuropatkin}, {Lahav}, {Leget}, {Lemos}, {Liddle}, {Lidman}, {Lima}, {Lin}, {MacCrann}, {Maia}, {Marshall}, {Martini}, {McCullough}, {Melchior}, {Mena-Fern{\'a}ndez}, {Menanteau}, {Miquel}, {Mohr}, {Morgan}, {Muir}, {Myles}, {Nadathur}, {Navarro-Alsina}, {Nichol}, {Ogando}, {Omori}, {Palmese}, {Pandey}, {Park}, {Paz-Chinch{\'o}n}, {Petravick}, {Pieres}, {Plazas Malag{\'o}n}, {Porredon}, {Prat}, {Raveri}, {Rodriguez-Monroy}, {Rollins}, {Romer}, {Roodman}, {Rosenfeld}, {Ross}, {Rykoff}, {Samuroff}, {S{\'a}nchez}, {Sanchez}, {Sanchez}, {Sanchez Cid}, {Scarpine}, {Schubnell}, {Scolnic}, {Secco}, {Serrano}, {Sevilla-Noarbe}, {Sheldon}, {Shin}, {Smith}, {Soares-Santos}, {Suchyta}, {Swanson}, {Tabbutt}, {Tarle}, {Thomas}, {To}, {Troja}, {Troxel}, {Tucker}, {Tutusaus}, {Varga}, {Walker}, {Weaverdyck}, {Wechsler}, {Weller}, {Yanny}, {Yin}, {Zhang}, {Zuntz}, \& {DES Collaboration}}]{Abbott2022PhRvD.105b3520A}
{Abbott}, T.~M.~C., {Aguena}, M., {Alarcon}, A., {et~al.} 2022, \prd, 105, 023520

\bibitem[{{Ahad} {et~al.}(2023){Ahad}, {Bah{\'e}}, \& {Hoekstra}}]{Ahad2023MNRAS.518.3685A}
{Ahad}, S.~L., {Bah{\'e}}, Y.~M., \& {Hoekstra}, H. 2023, \mnras, 518, 3685

\bibitem[{{Allen} {et~al.}(2011){Allen}, {Evrard}, \& {Mantz}}]{Allen2011ARAA..49..409A}
{Allen}, S.~W., {Evrard}, A.~E., \& {Mantz}, A.~B. 2011, \araa, 49, 409

\bibitem[{{Angulo} \& {Hahn}(2022)}]{Angulo2022LRCA....8....1A}
{Angulo}, R.~E. \& {Hahn}, O. 2022, Living Reviews in Computational Astrophysics, 8, 1

\bibitem[{{Ayromlou} {et~al.}(2021){Ayromlou}, {Nelson}, {Yates}, {Kauffmann}, {Renneby}, \& {White}}]{Ayromlou2021MNRAS.502.1051A}
{Ayromlou}, M., {Nelson}, D., {Yates}, R.~M., {et~al.} 2021, \mnras, 502, 1051

\bibitem[{{Bahar} {et~al.}(2022){Bahar}, {Bulbul}, {Clerc}, {Ghirardini}, {Liu}, {Nandra}, {Pacaud}, {Chiu}, {Comparat}, {Ider-Chitham}, {Klein}, {Liu}, {Merloni}, {Migkas}, {Okabe}, {Ramos-Ceja}, {Reiprich}, {Sanders}, \& {Schrabback}}]{Bahar2022AA...661A...7B}
{Bahar}, Y.~E., {Bulbul}, E., {Clerc}, N., {et~al.} 2022, \aap, 661, A7

\bibitem[{{Baltz} {et~al.}(2009){Baltz}, {Marshall}, \& {Oguri}}]{Baltz2009JCAP...01..015B}
{Baltz}, E.~A., {Marshall}, P., \& {Oguri}, M. 2009, \jcap, 2009, 015

\bibitem[{{Barnes} {et~al.}(2021){Barnes}, {Vogelsberger}, {Pearce}, {Pop}, {Kannan}, {Cao}, {Kay}, \& {Hernquist}}]{Barnes2021MNRAS.506.2533B}
{Barnes}, D.~J., {Vogelsberger}, M., {Pearce}, F.~A., {et~al.} 2021, \mnras, 506, 2533

\bibitem[{{Bartelmann} \& {Schneider}(2001)}]{Bartelmann2001PhR...340..291B}
{Bartelmann}, M. \& {Schneider}, P. 2001, \physrep, 340, 291

\bibitem[{{Behroozi} {et~al.}(2019){Behroozi}, {Wechsler}, {Hearin}, \& {Conroy}}]{Behroozi2019MNRAS.488.3143B}
{Behroozi}, P., {Wechsler}, R.~H., {Hearin}, A.~P., \& {Conroy}, C. 2019, \mnras, 488, 3143

\bibitem[{{Behroozi} {et~al.}(2010){Behroozi}, {Conroy}, \& {Wechsler}}]{Behroozi2010ApJ...717..379B}
{Behroozi}, P.~S., {Conroy}, C., \& {Wechsler}, R.~H. 2010, \apj, 717, 379

\bibitem[{{Berlind} \& {Weinberg}(2002)}]{Berlind2002ApJ...575..587B}
{Berlind}, A.~A. \& {Weinberg}, D.~H. 2002, \apj, 575, 587

\bibitem[{{Bertin} \& {Arnouts}(1996)}]{Bertin1996AAS..117..393B}
{Bertin}, E. \& {Arnouts}, S. 1996, \aaps, 117, 393

\bibitem[{{Biffi} {et~al.}(2016){Biffi}, {Borgani}, {Murante}, {Rasia}, {Planelles}, {Granato}, {Ragone-Figueroa}, {Beck}, {Gaspari}, \& {Dolag}}]{Biffi2016ApJ...827..112B}
{Biffi}, V., {Borgani}, S., {Murante}, G., {et~al.} 2016, \apj, 827, 112

\bibitem[{{Blumenthal} {et~al.}(1984){Blumenthal}, {Faber}, {Primack}, \& {Rees}}]{Blumenthal1984Natur.311..517B}
{Blumenthal}, G.~R., {Faber}, S.~M., {Primack}, J.~R., \& {Rees}, M.~J. 1984, \nat, 311, 517

\bibitem[{{Bocquet} {et~al.}(2016){Bocquet}, {Saro}, {Dolag}, \& {Mohr}}]{Bocquet2016MNRAS.456.2361B}
{Bocquet}, S., {Saro}, A., {Dolag}, K., \& {Mohr}, J.~J. 2016, \mnras, 456, 2361

\bibitem[{{Bradshaw} {et~al.}(2020){Bradshaw}, {Leauthaud}, {Hearin}, {Huang}, \& {Behroozi}}]{Bradshaw2020MNRAS.493..337B}
{Bradshaw}, C., {Leauthaud}, A., {Hearin}, A., {Huang}, S., \& {Behroozi}, P. 2020, \mnras, 493, 337

\bibitem[{{Brouwer} {et~al.}(2016){Brouwer}, {Cacciato}, {Dvornik}, {Eardley}, {Heymans}, {Hoekstra}, {Kuijken}, {McNaught-Roberts}, {Sif{\'o}n}, {Viola}, {Alpaslan}, {Bilicki}, {Bland-Hawthorn}, {Brough}, {Choi}, {Driver}, {Erben}, {Grado}, {Hildebrandt}, {Holwerda}, {Hopkins}, {de Jong}, {Liske}, {McFarland}, {Nakajima}, {Napolitano}, {Norberg}, {Peacock}, {Radovich}, {Robotham}, {Schneider}, {Sikkema}, {van Uitert}, {Verdoes Kleijn}, \& {Valentijn}}]{Brouwer2016MNRAS.462.4451B}
{Brouwer}, M.~M., {Cacciato}, M., {Dvornik}, A., {et~al.} 2016, \mnras, 462, 4451

\bibitem[{{Bruzual} \& {Charlot}(2003)}]{Bruzual2003MNRAS.344.1000B}
{Bruzual}, G. \& {Charlot}, S. 2003, \mnras, 344, 1000

\bibitem[{{Cacciato} {et~al.}(2009){Cacciato}, {van den Bosch}, {More}, {Li}, {Mo}, \& {Yang}}]{Cacciato2009MNRAS.394..929C}
{Cacciato}, M., {van den Bosch}, F.~C., {More}, S., {et~al.} 2009, \mnras, 394, 929

\bibitem[{{Cacciato} {et~al.}(2014){Cacciato}, {van Uitert}, \& {Hoekstra}}]{Cacciato2014MNRAS.437..377C}
{Cacciato}, M., {van Uitert}, E., \& {Hoekstra}, H. 2014, \mnras, 437, 377

\bibitem[{{Castro} {et~al.}(2021){Castro}, {Borgani}, {Dolag}, {Marra}, {Quartin}, {Saro}, \& {Sefusatti}}]{Castro2021MNRAS.500.2316C}
{Castro}, T., {Borgani}, S., {Dolag}, K., {et~al.} 2021, \mnras, 500, 2316

\bibitem[{{Cavaliere} \& {Fusco-Femiano}(1976)}]{Cavaliere1976AA....49..137C}
{Cavaliere}, A. \& {Fusco-Femiano}, R. 1976, \aap, 49, 137

\bibitem[{{Chabrier}(2003)}]{Chabrier2003PASP..115..763C}
{Chabrier}, G. 2003, \pasp, 115, 763

\bibitem[{{Chisari} {et~al.}(2018){Chisari}, {Richardson}, {Devriendt}, {Dubois}, {Schneider}, {Le Brun}, {Beckmann}, {Peirani}, {Slyz}, \& {Pichon}}]{Chisari2018MNRAS.480.3962C}
{Chisari}, N.~E., {Richardson}, M.~L.~A., {Devriendt}, J., {et~al.} 2018, \mnras, 480, 3962

\bibitem[{{Cole}(1991)}]{Cole1991ApJ...367...45C}
{Cole}, S. 1991, \apj, 367, 45

\bibitem[{{Cooray}(2006)}]{Cooray2006MNRAS.365..842C}
{Cooray}, A. 2006, \mnras, 365, 842

\bibitem[{{Cooray} \& {Sheth}(2002)}]{Cooray2002PhR...372....1C}
{Cooray}, A. \& {Sheth}, R. 2002, \physrep, 372, 1

\bibitem[{{Coupon} {et~al.}(2015){Coupon}, {Arnouts}, {van Waerbeke}, {Moutard}, {Ilbert}, {van Uitert}, {Erben}, {Garilli}, {Guzzo}, {Heymans}, {Hildebrandt}, {Hoekstra}, {Kilbinger}, {Kitching}, {Mellier}, {Miller}, {Scodeggio}, {Bonnett}, {Branchini}, {Davidzon}, {De Lucia}, {Fritz}, {Fu}, {Hudelot}, {Hudson}, {Kuijken}, {Leauthaud}, {Le F{\`e}vre}, {McCracken}, {Moscardini}, {Rowe}, {Schrabback}, {Semboloni}, \& {Velander}}]{Coupon2015MNRAS.449.1352C}
{Coupon}, J., {Arnouts}, S., {van Waerbeke}, L., {et~al.} 2015, \mnras, 449, 1352

\bibitem[{{Crain} \& {van de Voort}(2023)}]{Crain2023ARAA..61..473C}
{Crain}, R.~A. \& {van de Voort}, F. 2023, \araa, 61, 473

\bibitem[{{Cui} {et~al.}(2016){Cui}, {Power}, {Biffi}, {Borgani}, {Murante}, {Fabjan}, {Knebe}, {Lewis}, \& {Poole}}]{Cui2016MNRAS.456.2566C}
{Cui}, W., {Power}, C., {Biffi}, V., {et~al.} 2016, \mnras, 456, 2566

\bibitem[{{Davis} {et~al.}(1985){Davis}, {Efstathiou}, {Frenk}, \& {White}}]{Davis1985ApJ...292..371D}
{Davis}, M., {Efstathiou}, G., {Frenk}, C.~S., \& {White}, S.~D.~M. 1985, \apj, 292, 371

\bibitem[{{de Graaff} {et~al.}(2022){de Graaff}, {Trayford}, {Franx}, {Schaller}, {Schaye}, \& {van der Wel}}]{Graaff2022MNRAS.511.2544D}
{de Graaff}, A., {Trayford}, J., {Franx}, M., {et~al.} 2022, \mnras, 511, 2544

\bibitem[{{de Jong} {et~al.}(2013){de Jong}, {Verdoes Kleijn}, {Kuijken}, \& {Valentijn}}]{Jong2013ExA....35...25D}
{de Jong}, J. T.~A., {Verdoes Kleijn}, G.~A., {Kuijken}, K.~H., \& {Valentijn}, E.~A. 2013, Experimental Astronomy, 35, 25

\bibitem[{{Debackere} {et~al.}(2020){Debackere}, {Schaye}, \& {Hoekstra}}]{Debackere2020MNRAS.492.2285D}
{Debackere}, S. N.~B., {Schaye}, J., \& {Hoekstra}, H. 2020, \mnras, 492, 2285

\bibitem[{{Debackere} {et~al.}(2021){Debackere}, {Schaye}, \& {Hoekstra}}]{Debackere2021MNRAS.505..593D}
{Debackere}, S. N.~B., {Schaye}, J., \& {Hoekstra}, H. 2021, \mnras, 505, 593

\bibitem[{{Driver} {et~al.}(2022){Driver}, {Bellstedt}, {Robotham}, {Baldry}, {Davies}, {Liske}, {Obreschkow}, {Taylor}, {Wright}, {Alpaslan}, {Bamford}, {Bauer}, {Bland-Hawthorn}, {Bilicki}, {Bravo}, {Brough}, {Casura}, {Cluver}, {Colless}, {Conselice}, {Croom}, {de Jong}, {D'Eugenio}, {De Propris}, {Dogruel}, {Drinkwater}, {Dvornik}, {Farrow}, {Frenk}, {Giblin}, {Graham}, {Grootes}, {Gunawardhana}, {Hashemizadeh}, {H{\"a}u{\ss}ler}, {Heymans}, {Hildebrandt}, {Holwerda}, {Hopkins}, {Jarrett}, {Heath Jones}, {Kelvin}, {Koushan}, {Kuijken}, {Lara-L{\'o}pez}, {Lange}, {L{\'o}pez-S{\'a}nchez}, {Loveday}, {Mahajan}, {Meyer}, {Moffett}, {Napolitano}, {Norberg}, {Owers}, {Radovich}, {Raouf}, {Peacock}, {Phillipps}, {Pimbblet}, {Popescu}, {Said}, {Sansom}, {Seibert}, {Sutherland}, {Thorne}, {Tuffs}, {Turner}, {van der Wel}, {van Kampen}, \& {Wilkins}}]{Driver2022MNRAS.513..439D}
{Driver}, S.~P., {Bellstedt}, S., {Robotham}, A. S.~G., {et~al.} 2022, \mnras, 513, 439

\bibitem[{{Driver} {et~al.}(2011){Driver}, {Hill}, {Kelvin}, {Robotham}, {Liske}, {Norberg}, {Baldry}, {Bamford}, {Hopkins}, {Loveday}, {Peacock}, {Andrae}, {Bland-Hawthorn}, {Brough}, {Brown}, {Cameron}, {Ching}, {Colless}, {Conselice}, {Croom}, {Cross}, {de Propris}, {Dye}, {Drinkwater}, {Ellis}, {Graham}, {Grootes}, {Gunawardhana}, {Jones}, {van Kampen}, {Maraston}, {Nichol}, {Parkinson}, {Phillipps}, {Pimbblet}, {Popescu}, {Prescott}, {Roseboom}, {Sadler}, {Sansom}, {Sharp}, {Smith}, {Taylor}, {Thomas}, {Tuffs}, {Wijesinghe}, {Dunne}, {Frenk}, {Jarvis}, {Madore}, {Meyer}, {Seibert}, {Staveley-Smith}, {Sutherland}, \& {Warren}}]{Driver2011MNRAS.413..971D}
{Driver}, S.~P., {Hill}, D.~T., {Kelvin}, L.~S., {et~al.} 2011, \mnras, 413, 971

\bibitem[{{Driver} {et~al.}(2019){Driver}, {Liske}, {Davies}, {Robotham}, {Baldry}, {Brown}, {Cluver}, {Kuijken}, {Loveday}, {McMahon}, {Meyer}, {Norberg}, {Owers}, {Power}, {Taylor}, \& {WAVES Team}}]{Driver2019Msngr.175...46D}
{Driver}, S.~P., {Liske}, J., {Davies}, L.~J.~M., {et~al.} 2019, The Messenger, 175, 46

\bibitem[{{Duffy} {et~al.}(2008){Duffy}, {Schaye}, {Kay}, \& {Dalla Vecchia}}]{Duffy2008MNRAS.390L..64D}
{Duffy}, A.~R., {Schaye}, J., {Kay}, S.~T., \& {Dalla Vecchia}, C. 2008, \mnras, 390, L64

\bibitem[{{Dvornik} {et~al.}(2017){Dvornik}, {Cacciato}, {Kuijken}, {Viola}, {Hoekstra}, {Nakajima}, {van Uitert}, {Brouwer}, {Choi}, {Erben}, {Fenech Conti}, {Farrow}, {Herbonnet}, {Heymans}, {Hildebrandt}, {Hopkins}, {McFarland}, {Norberg}, {Schneider}, {Sif{\'o}n}, {Valentijn}, \& {Wang}}]{Dvornik2017MNRAS.468.3251D}
{Dvornik}, A., {Cacciato}, M., {Kuijken}, K., {et~al.} 2017, \mnras, 468, 3251

\bibitem[{{Dvornik} {et~al.}(2018){Dvornik}, {Hoekstra}, {Kuijken}, {Schneider}, {Amon}, {Nakajima}, {Viola}, {Choi}, {Erben}, {Farrow}, {Heymans}, {Hildebrandt}, {Sif{\'o}n}, \& {Wang}}]{Dvornik2018MNRAS.479.1240D}
{Dvornik}, A., {Hoekstra}, H., {Kuijken}, K., {et~al.} 2018, \mnras, 479, 1240

\bibitem[{{Eckmiller} {et~al.}(2011){Eckmiller}, {Hudson}, \& {Reiprich}}]{Eckmiller2011AA...535A.105E}
{Eckmiller}, H.~J., {Hudson}, D.~S., \& {Reiprich}, T.~H. 2011, \aap, 535, A105

\bibitem[{{Edge} {et~al.}(2013){Edge}, {Sutherland}, {Kuijken}, {Driver}, {McMahon}, {Eales}, \& {Emerson}}]{Edge2013Msngr.154...32E}
{Edge}, A., {Sutherland}, W., {Kuijken}, K., {et~al.} 2013, The Messenger, 154, 32

\bibitem[{{Elahi} {et~al.}(2019){Elahi}, {Ca{\~n}as}, {Poulton}, {Tobar}, {Willis}, {Lagos}, {Power}, \& {Robotham}}]{Elahi2019PASA...36...21E}
{Elahi}, P.~J., {Ca{\~n}as}, R., {Poulton}, R. J.~J., {et~al.} 2019, \pasa, 36, e021

\bibitem[{{Engler} {et~al.}(2021){Engler}, {Pillepich}, {Joshi}, {Nelson}, {Pasquali}, {Grebel}, {Lisker}, {Zinger}, {Donnari}, {Marinacci}, {Vogelsberger}, \& {Hernquist}}]{Engler2021MNRAS.500.3957E}
{Engler}, C., {Pillepich}, A., {Joshi}, G.~D., {et~al.} 2021, \mnras, 500, 3957

\bibitem[{{Euclid Collaboration} {et~al.}(2023){Euclid Collaboration}, {Castro}, {Fumagalli}, {Angulo}, {Bocquet}, {Borgani}, {Carbone}, {Dakin}, {Dolag}, {Giocoli}, {Monaco}, {Ragagnin}, {Saro}, {Sefusatti}, {Costanzi}, {Le Brun}, {Corasaniti}, {Amara}, {Amendola}, {Baldi}, {Bender}, {Bodendorf}, {Branchini}, {Brescia}, {Camera}, {Capobianco}, {Carretero}, {Castellano}, {Cavuoti}, {Cimatti}, {Cledassou}, {Congedo}, {Conversi}, {Copin}, {Corcione}, {Courbin}, {Da Silva}, {Degaudenzi}, {Douspis}, {Dubath}, {Duncan}, {Dupac}, {Farrens}, {Ferriol}, {Fosalba}, {Frailis}, {Franceschi}, {Galeotta}, {Garilli}, {Gillis}, {Grazian}, {Grupp}, {Haugan}, {Hormuth}, {Hornstrup}, {Hudelot}, {Jahnke}, {Kermiche}, {Kitching}, {Kunz}, {Kurki-Suonio}, {Lilje}, {Lloro}, {Mansutti}, {Marggraf}, {Marulli}, {Meneghetti}, {Merlin}, {Meylan}, {Moresco}, {Moscardini}, {Munari}, {Niemi}, {Padilla}, {Paltani}, {Pasian}, {Pedersen}, {Pettorino}, {Pires}, {Polenta}, {Poncet}, {Popa}, {Pozzetti}, {Raison}, {Rebolo}, {Renzi}, {Rhodes},
  {Riccio}, {Romelli}, {Saglia}, {Sapone}, {Sartoris}, {Schneider}, {Seidel}, {Sirri}, {Stanco}, {Tallada Cresp{\'\i}}, {Taylor}, {Toledo-Moreo}, {Torradeflot}, {Tutusaus}, {Valentijn}, {Valenziano}, {Vassallo}, {Wang}, {Weller}, {Zacchei}, {Zamorani}, {Andreon}, {Bardelli}, {Bozzo}, {Colodro-Conde}, {Di Ferdinando}, {Farina}, {Graci{\'a}-Carpio}, {Lindholm}, {Neissner}, {Scottez}, {Tenti}, {Zucca}, {Baccigalupi}, {Balaguera-Antol{\'\i}nez}, {Ballardini}, {Bernardeau}, {Biviano}, {Blanchard}, {Borlaff}, {Burigana}, {Cabanac}, {Cappi}, {Carvalho}, {Casas}, {Castignani}, {Cooray}, {Coupon}, {Courtois}, {Davini}, {De Lucia}, {Desprez}, {Dole}, {Escartin}, {Escoffier}, {Finelli}, {Ganga}, {Garcia-Bellido}, {George}, {Gozaliasl}, {Hildebrandt}, {Hook}, {Ili{\'c}}, {Kansal}, {Keihanen}, {Kirkpatrick}, {Loureiro}, {Macias-Perez}, {Magliocchetti}, {Maoli}, {Marcin}, {Martinelli}, {Martinet}, {Matthew}, {Maturi}, {Metcalf}, {Morgante}, {Nadathur}, {Nucita}, {Patrizii}, {Peel}, {Popa}, {Porciani}, {Potter},
  {Pourtsidou}, {P{\"o}ntinen}, {S{\'a}nchez}, {Sakr}, {Schirmer}, {Sereno}, {Spurio Mancini}, {Teyssier}, {Valiviita}, {Veropalumbo}, \& {Viel}}]{Euclid2023AA...671A.100E}
{Euclid Collaboration}, {Castro}, T., {Fumagalli}, A., {et~al.} 2023, \aap, 671, A100

\bibitem[{{Euclid Collaboration} {et~al.}(2025){Euclid Collaboration}, {Mellier}, {Abdurro'uf}, {Acevedo Barroso}, {Ach{\'u}carro}, {Adamek}, {Adam}, {Addison}, {Aghanim}, {Aguena}, {Ajani}, {Akrami}, {Al-Bahlawan}, {Alavi}, {Albuquerque}, {Alestas}, {Alguero}, {Allaoui}, {Allen}, {Allevato}, {Alonso-Tetilla}, {Altieri}, {Alvarez-Candal}, {Alvi}, {Amara}, {Amendola}, {Amiaux}, {Andika}, {Andreon}, {Andrews}, {Angora}, {Angulo}, {Annibali}, {Anselmi}, {Anselmi}, {Arcari}, {Archidiacono}, {Aric{\`o}}, {Arnaud}, {Arnouts}, {Asgari}, {Asorey}, {Atayde}, {Atek}, {Atrio-Barandela}, {Aubert}, {Aubourg}, {Auphan}, {Auricchio}, {Aussel}, {Aussel}, {Avelino}, {Avgoustidis}, {Avila}, {Awan}, {Azzollini}, {Baccigalupi}, {Bachelet}, {Bacon}, {Baes}, {Bagley}, {Bahr-Kalus}, {Balaguera-Antolinez}, {Balbinot}, {Balcells}, {Baldi}, {Baldry}, {Balestra}, {Ballardini}, {Ballester}, {Balogh}, {Ba{\~n}ados}, {Barbier}, {Bardelli}, {Baron}, {Barreiro}, {Barrena}, {Barriere}, {Barros}, {Barthelemy}, {Bartolo}, {Basset},
  {Battaglia}, {Battisti}, {Baugh}, {Baumont}, {Bazzanini}, {Beaulieu}, {Beckmann}, {Belikov}, {Bel}, {Bellagamba}, {Bella}, {Bellini}, {Benabed}, {Bender}, {Benevento}, {Bennett}, {Benson}, {Bergamini}, {Bermejo-Climent}, {Bernardeau}, {Bertacca}, {Berthe}, {Berthier}, {Bethermin}, {Beutler}, {Bevillon}, {Bhargava}, {Bhatawdekar}, {Bianchi}, {Bisigello}, {Biviano}, {Blake}, {Blanchard}, {Blazek}, {Blot}, {Bosco}, {Bodendorf}, {Boenke}, {B{\"o}hringer}, {Boldrini}, {Bolzonella}, {Bonchi}, {Bonici}, {Bonino}, {Bonino}, {Bonvin}, {Bon}, {Booth}, {Borgani}, {Borlaff}, {Borsato}, {Bose}, {Botticella}, {Boucaud}, {Bouche}, {Boucher}, {Boutigny}, {Bouvard}, {Bouwens}, {Bouy}, {Bowler}, {Bozza}, {Bozzo}, {Branchini}, {Brando}, {Brau-Nogue}, {Brekke}, {Bremer}, {Brescia}, {Breton}, {Brinchmann}, {Brinckmann}, {Brockley-Blatt}, {Brodwin}, {Brouard}, {Brown}, {Bruton}, {Bucko}, {Buddelmeijer}, {Buenadicha}, {Buitrago}, {Burger}, {Burigana}, {Busillo}, {Busonero}, {Cabanac}, {Cabayol-Garcia}, {Cagliari}, {Caillat},
  {Caillat}, {Calabrese}, {Calabro}, {Calderone}, {Calura}, {Camacho Quevedo}, {Camera}, {Campos}, {Ca{\~n}as-Herrera}, {Candini}, {Cantiello}, {Capobianco}, {Cappellaro}, {Cappelluti}, {Cappi}, {Caputi}, {Cara}, {Carbone}, {Cardone}, {Carella}, {Carlberg}, {Carle}, {Carminati}, {Caro}, {Carrasco}, {Carretero}, {Carrilho}, {Carron Duque}, \& {Carry}}]{Euclid2025AA...697A...1E}
{Euclid Collaboration}, {Mellier}, Y., {Abdurro'uf}, {et~al.} 2025, \aap, 697, A1

\bibitem[{{Evrard} {et~al.}(1996){Evrard}, {Metzler}, \& {Navarro}}]{Evrard1996ApJ...469..494E}
{Evrard}, A.~E., {Metzler}, C.~A., \& {Navarro}, J.~F. 1996, \apj, 469, 494

\bibitem[{{Farrow} {et~al.}(2015){Farrow}, {Cole}, {Norberg}, {Metcalfe}, {Baldry}, {Bland-Hawthorn}, {Brown}, {Hopkins}, {Lacey}, {Liske}, {Loveday}, {Palamara}, {Robotham}, \& {Sridhar}}]{Farrow2015MNRAS.454.2120F}
{Farrow}, D.~J., {Cole}, S., {Norberg}, P., {et~al.} 2015, \mnras, 454, 2120

\bibitem[{{Fenech Conti} {et~al.}(2017){Fenech Conti}, {Herbonnet}, {Hoekstra}, {Merten}, {Miller}, \& {Viola}}]{Conti2017MNRAS.467.1627F}
{Fenech Conti}, I., {Herbonnet}, R., {Hoekstra}, H., {et~al.} 2017, \mnras, 467, 1627

\bibitem[{{Foreman-Mackey} {et~al.}(2013){Foreman-Mackey}, {Hogg}, {Lang}, \& {Goodman}}]{Foreman2013PASP..125..306F}
{Foreman-Mackey}, D., {Hogg}, D.~W., {Lang}, D., \& {Goodman}, J. 2013, \pasp, 125, 306

\bibitem[{{Fortuna} {et~al.}(2021){Fortuna}, {Hoekstra}, {Joachimi}, {Johnston}, {Chisari}, {Georgiou}, \& {Mahony}}]{Fortuna2021MNRAS.501.2983F}
{Fortuna}, M.~C., {Hoekstra}, H., {Joachimi}, B., {et~al.} 2021, \mnras, 501, 2983

\bibitem[{{Genel} {et~al.}(2014){Genel}, {Vogelsberger}, {Springel}, {Sijacki}, {Nelson}, {Snyder}, {Rodriguez-Gomez}, {Torrey}, \& {Hernquist}}]{Genel2014MNRAS.445..175G}
{Genel}, S., {Vogelsberger}, M., {Springel}, V., {et~al.} 2014, \mnras, 445, 175

\bibitem[{{Giblin} {et~al.}(2021){Giblin}, {Heymans}, {Asgari}, {Hildebrandt}, {Hoekstra}, {Joachimi}, {Kannawadi}, {Kuijken}, {Lin}, {Miller}, {Tr{\"o}ster}, {van den Busch}, {Wright}, {Bilicki}, {Blake}, {de Jong}, {Dvornik}, {Erben}, {Getman}, {Napolitano}, {Schneider}, {Shan}, \& {Valentijn}}]{Giblin2021AA...645A.105G}
{Giblin}, B., {Heymans}, C., {Asgari}, M., {et~al.} 2021, \aap, 645, A105

\bibitem[{{Goodman} \& {Weare}(2010)}]{Goodman2010CAMCS...5...65G}
{Goodman}, J. \& {Weare}, J. 2010, Communications in Applied Mathematics and Computational Science, 5, 65

\bibitem[{{Gouin} {et~al.}(2019){Gouin}, {Gavazzi}, {Pichon}, {Dubois}, {Laigle}, {Chisari}, {Codis}, {Devriendt}, \& {Peirani}}]{Gouin2019AA...626A..72G}
{Gouin}, C., {Gavazzi}, R., {Pichon}, C., {et~al.} 2019, \aap, 626, A72

\bibitem[{{Guo} {et~al.}(2011){Guo}, {White}, {Boylan-Kolchin}, {De Lucia}, {Kauffmann}, {Lemson}, {Li}, {Springel}, \& {Weinmann}}]{Guo2011MNRAS.413..101G}
{Guo}, Q., {White}, S., {Boylan-Kolchin}, M., {et~al.} 2011, \mnras, 413, 101

\bibitem[{{Guzik} \& {Seljak}(2002)}]{Guzik2002MNRAS.335..311G}
{Guzik}, J. \& {Seljak}, U. 2002, \mnras, 335, 311

\bibitem[{{Han} {et~al.}(2015){Han}, {Eke}, {Frenk}, {Mandelbaum}, {Norberg}, {Schneider}, {Peacock}, {Jing}, {Baldry}, {Bland-Hawthorn}, {Brough}, {Brown}, {Liske}, {Loveday}, \& {Robotham}}]{Han2015MNRAS.446.1356H}
{Han}, J., {Eke}, V.~R., {Frenk}, C.~S., {et~al.} 2015, \mnras, 446, 1356

\bibitem[{{Hellwing} {et~al.}(2016){Hellwing}, {Schaller}, {Frenk}, {Theuns}, {Schaye}, {Bower}, \& {Crain}}]{Hellwing2016MNRAS.461L..11H}
{Hellwing}, W.~A., {Schaller}, M., {Frenk}, C.~S., {et~al.} 2016, \mnras, 461, L11

\bibitem[{{Henriques} {et~al.}(2015){Henriques}, {White}, {Thomas}, {Angulo}, {Guo}, {Lemson}, {Springel}, \& {Overzier}}]{Henriques2015MNRAS.451.2663H}
{Henriques}, B. M.~B., {White}, S. D.~M., {Thomas}, P.~A., {et~al.} 2015, \mnras, 451, 2663

\bibitem[{{Hern{\'a}ndez-Mart{\'\i}n} {et~al.}(2020){Hern{\'a}ndez-Mart{\'\i}n}, {Schrabback}, {Hoekstra}, {Martinet}, {Hlavacek-Larrondo}, {Bleem}, {Gladders}, {Stalder}, {Stark}, \& {Bayliss}}]{Hern2020AA...640A.117H}
{Hern{\'a}ndez-Mart{\'\i}n}, B., {Schrabback}, T., {Hoekstra}, H., {et~al.} 2020, \aap, 640, A117

\bibitem[{{Heymans} {et~al.}(2006){Heymans}, {Van Waerbeke}, {Bacon}, {Berge}, {Bernstein}, {Bertin}, {Bridle}, {Brown}, {Clowe}, {Dahle}, {Erben}, {Gray}, {Hetterscheidt}, {Hoekstra}, {Hudelot}, {Jarvis}, {Kuijken}, {Margoniner}, {Massey}, {Mellier}, {Nakajima}, {Refregier}, {Rhodes}, {Schrabback}, \& {Wittman}}]{Heymans2006MNRAS.368.1323H}
{Heymans}, C., {Van Waerbeke}, L., {Bacon}, D., {et~al.} 2006, \mnras, 368, 1323

\bibitem[{{Hildebrandt} {et~al.}(2021){Hildebrandt}, {van den Busch}, {Wright}, {Blake}, {Joachimi}, {Kuijken}, {Tr{\"o}ster}, {Asgari}, {Bilicki}, {de Jong}, {Dvornik}, {Erben}, {Getman}, {Giblin}, {Heymans}, {Kannawadi}, {Lin}, \& {Shan}}]{Hildebrandt2021AA...647A.124H}
{Hildebrandt}, H., {van den Busch}, J.~L., {Wright}, A.~H., {et~al.} 2021, \aap, 647, A124

\bibitem[{{Hoekstra} {et~al.}(1998){Hoekstra}, {Franx}, {Kuijken}, \& {Squires}}]{Hoekstra1998ApJ...504..636H}
{Hoekstra}, H., {Franx}, M., {Kuijken}, K., \& {Squires}, G. 1998, \apj, 504, 636

\bibitem[{{Hoekstra} {et~al.}(2015){Hoekstra}, {Herbonnet}, {Muzzin}, {Babul}, {Mahdavi}, {Viola}, \& {Cacciato}}]{Hoekstra2015MNRAS.449..685H}
{Hoekstra}, H., {Herbonnet}, R., {Muzzin}, A., {et~al.} 2015, \mnras, 449, 685

\bibitem[{{Hoekstra} {et~al.}(2021){Hoekstra}, {Kannawadi}, \& {Kitching}}]{Hoekstra2021AA...646A.124H}
{Hoekstra}, H., {Kannawadi}, A., \& {Kitching}, T.~D. 2021, \aap, 646, A124

\bibitem[{{Ivezi{\'c}} {et~al.}(2019){Ivezi{\'c}}, {Kahn}, {Tyson}, {Abel}, {Acosta}, {Allsman}, {Alonso}, {AlSayyad}, {Anderson}, {Andrew}, {Angel}, {Angeli}, {Ansari}, {Antilogus}, {Araujo}, {Armstrong}, {Arndt}, {Astier}, {Aubourg}, {Auza}, {Axelrod}, {Bard}, {Barr}, {Barrau}, {Bartlett}, {Bauer}, {Bauman}, {Baumont}, {Bechtol}, {Bechtol}, {Becker}, {Becla}, {Beldica}, {Bellavia}, {Bianco}, {Biswas}, {Blanc}, {Blazek}, {Blandford}, {Bloom}, {Bogart}, {Bond}, {Booth}, {Borgland}, {Borne}, {Bosch}, {Boutigny}, {Brackett}, {Bradshaw}, {Brandt}, {Brown}, {Bullock}, {Burchat}, {Burke}, {Cagnoli}, {Calabrese}, {Callahan}, {Callen}, {Carlin}, {Carlson}, {Chandrasekharan}, {Charles-Emerson}, {Chesley}, {Cheu}, {Chiang}, {Chiang}, {Chirino}, {Chow}, {Ciardi}, {Claver}, {Cohen-Tanugi}, {Cockrum}, {Coles}, {Connolly}, {Cook}, {Cooray}, {Covey}, {Cribbs}, {Cui}, {Cutri}, {Daly}, {Daniel}, {Daruich}, {Daubard}, {Daues}, {Dawson}, {Delgado}, {Dellapenna}, {de Peyster}, {de Val-Borro}, {Digel}, {Doherty}, {Dubois},
  {Dubois-Felsmann}, {Durech}, {Economou}, {Eifler}, {Eracleous}, {Emmons}, {Fausti Neto}, {Ferguson}, {Figueroa}, {Fisher-Levine}, {Focke}, {Foss}, {Frank}, {Freemon}, {Gangler}, {Gawiser}, {Geary}, {Gee}, {Geha}, {Gessner}, {Gibson}, {Gilmore}, {Glanzman}, {Glick}, {Goldina}, {Goldstein}, {Goodenow}, {Graham}, {Gressler}, {Gris}, {Guy}, {Guyonnet}, {Haller}, {Harris}, {Hascall}, {Haupt}, {Hernandez}, {Herrmann}, {Hileman}, {Hoblitt}, {Hodgson}, {Hogan}, {Howard}, {Huang}, {Huffer}, {Ingraham}, {Innes}, {Jacoby}, {Jain}, {Jammes}, {Jee}, {Jenness}, {Jernigan}, {Jevremovi{\'c}}, {Johns}, {Johnson}, {Johnson}, {Jones}, {Juramy-Gilles}, {Juri{\'c}}, {Kalirai}, {Kallivayalil}, {Kalmbach}, {Kantor}, {Karst}, {Kasliwal}, {Kelly}, {Kessler}, {Kinnison}, {Kirkby}, {Knox}, {Kotov}, {Krabbendam}, {Krughoff}, {Kub{\'a}nek}, {Kuczewski}, {Kulkarni}, {Ku}, {Kurita}, {Lage}, {Lambert}, {Lange}, {Langton}, {Le Guillou}, {Levine}, {Liang}, {Lim}, {Lintott}, {Long}, {Lopez}, {Lotz}, {Lupton}, {Lust}, {MacArthur}, {Mahabal},
  {Mandelbaum}, {Markiewicz}, {Marsh}, {Marshall}, {Marshall}, {May}, {McKercher}, {McQueen}, {Meyers}, {Migliore}, {Miller}, {Mills}, {Miraval}, {Moeyens}, {Moolekamp}, {Monet}, {Moniez}, {Monkewitz}, {Montgomery}, {Morrison}, {Mueller}, {Muller}, {Mu{\~n}oz Arancibia}, {Neill}, {Newbry}, {Nief}, {Nomerotski}, {Nordby}, {O'Connor}, {Oliver}, {Olivier}, {Olsen}, {O'Mullane}, {Ortiz}, {Osier}, {Owen}, {Pain}, {Palecek}, {Parejko}, {Parsons}, {Pease}, {Peterson}, {Peterson}, {Petravick}, {Libby Petrick}, {Petry}, {Pierfederici}, {Pietrowicz}, {Pike}, {Pinto}, {Plante}, {Plate}, {Plutchak}, {Price}, {Prouza}, {Radeka}, {Rajagopal}, {Rasmussen}, {Regnault}, {Reil}, {Reiss}, {Reuter}, {Ridgway}, {Riot}, {Ritz}, {Robinson}, {Roby}, {Roodman}, {Rosing}, {Roucelle}, {Rumore}, {Russo}, {Saha}, {Sassolas}, {Schalk}, {Schellart}, {Schindler}, {Schmidt}, {Schneider}, {Schneider}, {Schoening}, {Schumacher}, {Schwamb}, {Sebag}, {Selvy}, {Sembroski}, {Seppala}, {Serio}, {Serrano}, {Shaw}, {Shipsey}, {Sick}, {Silvestri},
  {Slater}, {Smith}, {Smith}, {Sobhani}, {Soldahl}, {Storrie-Lombardi}, {Stover}, {Strauss}, {Street}, {Stubbs}, {Sullivan}, {Sweeney}, {Swinbank}, {Szalay}, {Takacs}, {Tether}, {Thaler}, {Thayer}, {Thomas}, {Thornton}, {Thukral}, {Tice}, {Trilling}, {Turri}, {Van Berg}, {Vanden Berk}, {Vetter}, {Virieux}, {Vucina}, {Wahl}, {Walkowicz}, {Walsh}, {Walter}, {Wang}, {Wang}, {Warner}, {Wiecha}, {Willman}, {Winters}, {Wittman}, {Wolff}, {Wood-Vasey}, {Wu}, {Xin}, {Yoachim}, \& {Zhan}}]{Ivezic2019ApJ873111I}
{Ivezi{\'c}}, {\v{Z}}., {Kahn}, S.~M., {Tyson}, J.~A., {et~al.} 2019, \apj, 873, 111

\bibitem[{{Jakobs} {et~al.}(2018){Jakobs}, {Viola}, {McCarthy}, {van Waerbeke}, {Hoekstra}, {Robotham}, {Hinshaw}, {Hojjati}, {Tanimura}, {Tr{\"o}ster}, {Baldry}, {Heymans}, {Hildebrandt}, {Kuijken}, {Norberg}, {Schaye}, {Sif{\'o}n}, {van Uitert}, {Valentijn}, {Verdoes Kleijn}, \& {Wang}}]{Jakobs2018MNRAS.480.3338J}
{Jakobs}, A., {Viola}, M., {McCarthy}, I., {et~al.} 2018, \mnras, 480, 3338

\bibitem[{{Jansen} {et~al.}(2024){Jansen}, {Tewes}, {Schrabback}, {Aghanim}, {Amara}, {Andreon}, {Auricchio}, {Baldi}, {Branchini}, {Brescia}, {Brinchmann}, {Camera}, {Capobianco}, {Carbone}, {Cardone}, {Carretero}, {Casas}, {Castellano}, {Cavuoti}, {Cimatti}, {Congedo}, {Conversi}, {Copin}, {Corcione}, {Courbin}, {Courtois}, {Da Silva}, {Degaudenzi}, {Dinis}, {Dubath}, {Dupac}, {Farina}, {Farrens}, {Ferriol}, {Frailis}, {Franceschi}, {Fumana}, {Galeotta}, {Gillis}, {Giocoli}, {Grazian}, {Grupp}, {Haugan}, {Hoekstra}, {Holmes}, {Hormuth}, {Hornstrup}, {Hudelot}, {Jahnke}, {Joachimi}, {Kermiche}, {Kiessling}, {Kilbinger}, {Kitching}, {Kubik}, {Kurki-Suonio}, {Ligori}, {Lilje}, {Lindholm}, {Lloro}, {Maiorano}, {Mansutti}, {Marggraf}, {Markovic}, {Martinet}, {Marulli}, {Massey}, {Medinaceli}, {Mei}, {Melchior}, {Mellier}, {Meneghetti}, {Merlin}, {Meylan}, {Miller}, {Moresco}, {Moscardini}, {Munari}, {Nakajima}, {Niemi}, {Padilla}, {Paltani}, {Pasian}, {Pedersen}, {Pettorino}, {Pires}, {Polenta}, {Poncet},
  {Raison}, {Renzi}, {Rhodes}, {Riccio}, {Romelli}, {Roncarelli}, {Rossetti}, {Saglia}, {Sapone}, {Sartoris}, {Schneider}, {Secroun}, {Seidel}, {Serrano}, {Sirignano}, {Sirri}, {Skottfelt}, {Stanco}, {Tallada-Cresp{\'\i}}, {Tereno}, {Toledo-Moreo}, {Torradeflot}, {Tutusaus}, {Valentijn}, {Valenziano}, {Vassallo}, {Veropalumbo}, {Wang}, {Weller}, {Zamorani}, {Zoubian}, {Colodro-Conde}, \& {Scottez}}]{Jansen2024AA...683A.240J}
{Jansen}, H., {Tewes}, M., {Schrabback}, T., {et~al.} 2024, \aap, 683, A240

\bibitem[{{Johnston} {et~al.}(2007){Johnston}, {Sheldon}, {Wechsler}, {Rozo}, {Koester}, {Frieman}, {McKay}, {Evrard}, {Becker}, \& {Annis}}]{Johnston2007arXiv0709.1159J}
{Johnston}, D.~E., {Sheldon}, E.~S., {Wechsler}, R.~H., {et~al.} 2007, arXiv e-prints, arXiv:0709.1159

\bibitem[{{Kaiser} \& {Squires}(1993)}]{Kaiser1993ApJ...404..441K}
{Kaiser}, N. \& {Squires}, G. 1993, \apj, 404, 441

\bibitem[{{Kaiser} {et~al.}(1995){Kaiser}, {Squires}, \& {Broadhurst}}]{Kaiser1995ApJ...449..460K}
{Kaiser}, N., {Squires}, G., \& {Broadhurst}, T. 1995, \apj, 449, 460

\bibitem[{{Kannawadi} {et~al.}(2019){Kannawadi}, {Hoekstra}, {Miller}, {Viola}, {Fenech Conti}, {Herbonnet}, {Erben}, {Heymans}, {Hildebrandt}, {Kuijken}, {Vakili}, \& {Wright}}]{Kannawadi2019AA...624A..92K}
{Kannawadi}, A., {Hoekstra}, H., {Miller}, L., {et~al.} 2019, \aap, 624, A92

\bibitem[{{Kauffmann} {et~al.}(1999){Kauffmann}, {Colberg}, {Diaferio}, \& {White}}]{Kauffmann1999MNRAS.303..188K}
{Kauffmann}, G., {Colberg}, J.~M., {Diaferio}, A., \& {White}, S. D.~M. 1999, \mnras, 303, 188

\bibitem[{{Kauffmann} {et~al.}(1993){Kauffmann}, {White}, \& {Guiderdoni}}]{Kauffmann1993MNRAS.264..201K}
{Kauffmann}, G., {White}, S.~D.~M., \& {Guiderdoni}, B. 1993, \mnras, 264, 201

\bibitem[{{Kelly} {et~al.}(2024){Kelly}, {Jobel}, {Eiger}, {Abd}, {Jeltema}, {Giles}, {Hollowood}, {Wilkinson}, {Turner}, {Bhargava}, {Everett}, {Farahi}, {Romer}, {Rykoff}, {Wang}, {Bocquet}, {Cross}, {Faridjoo}, {Franco}, {Gardner}, {Kwiecien}, {Laubner}, {McDaniel}, {O'Donnell}, {Sanchez}, {Schmidt}, {Sripada}, {Swart}, {Upsdell}, {Webber}, {Aguena}, {Allam}, {Alves}, {Bacon}, {Brooks}, {Burke}, {Carnero Rosell}, {Carretero}, {Collins}, {Costanzi}, {da Costa}, {Pereira}, {Davis}, {Doel}, {Ferrero}, {Frieman}, {Garc{\'\i}a-Bellido}, {Giannini}, {Gruen}, {Gruendl}, {Hilton}, {Hinton}, {Honscheid}, {James}, {Kuehn}, {Mann}, {Marshall}, {Mena-Fern{\'a}ndez}, {Miller}, {Miquel}, {Myles}, {Palmese}, {Pieres}, {Plazas Malag{\'o}n}, {Rooney}, {Sahlen}, {Sanchez}, {Sanchez Cid}, {Schubnell}, {Sevilla-Noarbe}, {Smith}, {Stott}, {Suchyta}, {Swanson}, {Tarle}, {To}, {Viana}, {Weaverdyck}, {Wiseman}, \& {DES Collaboration}}]{Kelly2024MNRAS.533..572K}
{Kelly}, P.~M., {Jobel}, J., {Eiger}, O., {et~al.} 2024, \mnras, 533, 572

\bibitem[{{Kettula} {et~al.}(2015){Kettula}, {Giodini}, {van Uitert}, {Hoekstra}, {Finoguenov}, {Lerchster}, {Erben}, {Heymans}, {Hildebrandt}, {Kitching}, {Mahdavi}, {Mellier}, {Miller}, {Mirkazemi}, {Van Waerbeke}, {Coupon}, {Egami}, {Fu}, {Hudson}, {Kneib}, {Kuijken}, {McCracken}, {Pereira}, {Rowe}, {Schrabback}, {Tanaka}, \& {Velander}}]{Kettula2015MNRAS.451.1460K}
{Kettula}, K., {Giodini}, S., {van Uitert}, E., {et~al.} 2015, \mnras, 451, 1460

\bibitem[{{Kitching} \& {Deshpande}(2022)}]{Kitching2022OJAp....5E...6K}
{Kitching}, T.~D. \& {Deshpande}, A.~C. 2022, The Open Journal of Astrophysics, 5, 6

\bibitem[{{Kugel} {et~al.}(2023){Kugel}, {Schaye}, {Schaller}, {Helly}, {Braspenning}, {Elbers}, {Frenk}, {McCarthy}, {Kwan}, {Salcido}, {van Daalen}, {Vandenbroucke}, {Bah{\'e}}, {Borrow}, {Chaikin}, {Hu{\v{s}}ko}, {Jenkins}, {Lacey}, {Nobels}, \& {Vernon}}]{Kugel2023MNRAS.526.6103K}
{Kugel}, R., {Schaye}, J., {Schaller}, M., {et~al.} 2023, \mnras, 526, 6103

\bibitem[{{Kuijken} {et~al.}(2019){Kuijken}, {Heymans}, {Dvornik}, {Hildebrandt}, {de Jong}, {Wright}, {Erben}, {Bilicki}, {Giblin}, {Shan}, {Getman}, {Grado}, {Hoekstra}, {Miller}, {Napolitano}, {Paolilo}, {Radovich}, {Schneider}, {Sutherland}, {Tewes}, {Tortora}, {Valentijn}, \& {Verdoes Kleijn}}]{Kuijken2019AA...625A...2K}
{Kuijken}, K., {Heymans}, C., {Dvornik}, A., {et~al.} 2019, \aap, 625, A2

\bibitem[{{Kuijken} {et~al.}(2015){Kuijken}, {Heymans}, {Hildebrandt}, {Nakajima}, {Erben}, {de Jong}, {Viola}, {Choi}, {Hoekstra}, {Miller}, {van Uitert}, {Amon}, {Blake}, {Brouwer}, {Buddendiek}, {Conti}, {Eriksen}, {Grado}, {Harnois-D{\'e}raps}, {Helmich}, {Herbonnet}, {Irisarri}, {Kitching}, {Klaes}, {La Barbera}, {Napolitano}, {Radovich}, {Schneider}, {Sif{\'o}n}, {Sikkema}, {Simon}, {Tudorica}, {Valentijn}, {Verdoes Kleijn}, \& {van Waerbeke}}]{Kuijken2015MNRAS.454.3500K}
{Kuijken}, K., {Heymans}, C., {Hildebrandt}, H., {et~al.} 2015, \mnras, 454, 3500

\bibitem[{{Le Brun} {et~al.}(2014){Le Brun}, {McCarthy}, {Schaye}, \& {Ponman}}]{Brun2014MNRAS.441.1270L}
{Le Brun}, A. M.~C., {McCarthy}, I.~G., {Schaye}, J., \& {Ponman}, T.~J. 2014, \mnras, 441, 1270

\bibitem[{{Leauthaud} {et~al.}(2010){Leauthaud}, {Finoguenov}, {Kneib}, {Taylor}, {Massey}, {Rhodes}, {Ilbert}, {Bundy}, {Tinker}, {George}, {Capak}, {Koekemoer}, {Johnston}, {Zhang}, {Cappelluti}, {Ellis}, {Elvis}, {Giodini}, {Heymans}, {Le F{\`e}vre}, {Lilly}, {McCracken}, {Mellier}, {R{\'e}fr{\'e}gier}, {Salvato}, {Scoville}, {Smoot}, {Tanaka}, {Van Waerbeke}, \& {Wolk}}]{Leauthaud2010ApJ...709...97L}
{Leauthaud}, A., {Finoguenov}, A., {Kneib}, J.-P., {et~al.} 2010, \apj, 709, 97

\bibitem[{{Leauthaud} {et~al.}(2012){Leauthaud}, {Tinker}, {Bundy}, {Behroozi}, {Massey}, {Rhodes}, {George}, {Kneib}, {Benson}, {Wechsler}, {Busha}, {Capak}, {Cort{\^e}s}, {Ilbert}, {Koekemoer}, {Le F{\`e}vre}, {Lilly}, {McCracken}, {Salvato}, {Schrabback}, {Scoville}, {Smith}, \& {Taylor}}]{Leauthaud2012ApJ...744..159L}
{Leauthaud}, A., {Tinker}, J., {Bundy}, K., {et~al.} 2012, \apj, 744, 159

\bibitem[{{Li} {et~al.}(2023{\natexlab{a}}){Li}, {Hoekstra}, {Kuijken}, {Asgari}, {Bilicki}, {Giblin}, {Heymans}, {Hildebrandt}, {Joachimi}, {Miller}, {van den Busch}, {Wright}, {Kannawadi}, {Reischke}, \& {Shan}}]{Li2023AA...679A.133L}
{Li}, S.-S., {Hoekstra}, H., {Kuijken}, K., {et~al.} 2023{\natexlab{a}}, \aap, 679, A133

\bibitem[{{Li} {et~al.}(2023{\natexlab{b}}){Li}, {Kuijken}, {Hoekstra}, {Miller}, {Heymans}, {Hildebrandt}, {van den Busch}, {Wright}, {Yoon}, {Bilicki}, {Bravo}, \& {Lagos}}]{Li2023AA...670A.100L}
{Li}, S.-S., {Kuijken}, K., {Hoekstra}, H., {et~al.} 2023{\natexlab{b}}, \aap, 670, A100

\bibitem[{{Liske} {et~al.}(2015){Liske}, {Baldry}, {Driver}, {Tuffs}, {Alpaslan}, {Andrae}, {Brough}, {Cluver}, {Grootes}, {Gunawardhana}, {Kelvin}, {Loveday}, {Robotham}, {Taylor}, {Bamford}, {Bland-Hawthorn}, {Brown}, {Drinkwater}, {Hopkins}, {Meyer}, {Norberg}, {Peacock}, {Agius}, {Andrews}, {Bauer}, {Ching}, {Colless}, {Conselice}, {Croom}, {Davies}, {De Propris}, {Dunne}, {Eardley}, {Ellis}, {Foster}, {Frenk}, {H{\"a}u{\ss}ler}, {Holwerda}, {Howlett}, {Ibarra}, {Jarvis}, {Jones}, {Kafle}, {Lacey}, {Lange}, {Lara-L{\'o}pez}, {L{\'o}pez-S{\'a}nchez}, {Maddox}, {Madore}, {McNaught-Roberts}, {Moffett}, {Nichol}, {Owers}, {Palamara}, {Penny}, {Phillipps}, {Pimbblet}, {Popescu}, {Prescott}, {Proctor}, {Sadler}, {Sansom}, {Seibert}, {Sharp}, {Sutherland}, {V{\'a}zquez-Mata}, {van Kampen}, {Wilkins}, {Williams}, \& {Wright}}]{Liske2015MNRAS.452.2087L}
{Liske}, J., {Baldry}, I.~K., {Driver}, S.~P., {et~al.} 2015, \mnras, 452, 2087

\bibitem[{{Logan} {et~al.}(2022){Logan}, {Maughan}, {Diaferio}, {Duffy}, {Geller}, {Rines}, \& {Sohn}}]{Logan2022AA...665A.124L}
{Logan}, C. H.~A., {Maughan}, B.~J., {Diaferio}, A., {et~al.} 2022, \aap, 665, A124

\bibitem[{{Loveday} {et~al.}(2012){Loveday}, {Norberg}, {Baldry}, {Driver}, {Hopkins}, {Peacock}, {Bamford}, {Liske}, {Bland-Hawthorn}, {Brough}, {Brown}, {Cameron}, {Conselice}, {Croom}, {Frenk}, {Gunawardhana}, {Hill}, {Jones}, {Kelvin}, {Kuijken}, {Nichol}, {Parkinson}, {Phillipps}, {Pimbblet}, {Popescu}, {Prescott}, {Robotham}, {Sharp}, {Sutherland}, {Taylor}, {Thomas}, {Tuffs}, {van Kampen}, \& {Wijesinghe}}]{Loveday2012MNRAS.420.1239L}
{Loveday}, J., {Norberg}, P., {Baldry}, I.~K., {et~al.} 2012, \mnras, 420, 1239

\bibitem[{{Luppino} \& {Kaiser}(1997)}]{Luppino1997ApJ...475...20L}
{Luppino}, G.~A. \& {Kaiser}, N. 1997, \apj, 475, 20

\bibitem[{{Mandelbaum} {et~al.}(2006){Mandelbaum}, {Seljak}, {Kauffmann}, {Hirata}, \& {Brinkmann}}]{Mandelbaum2006MNRAS.368..715M}
{Mandelbaum}, R., {Seljak}, U., {Kauffmann}, G., {Hirata}, C.~M., \& {Brinkmann}, J. 2006, \mnras, 368, 715

\bibitem[{{McCarthy} {et~al.}(2017){McCarthy}, {Schaye}, {Bird}, \& {Le Brun}}]{McCarthy2017MNRAS.465.2936M}
{McCarthy}, I.~G., {Schaye}, J., {Bird}, S., \& {Le Brun}, A. M.~C. 2017, \mnras, 465, 2936

\bibitem[{{McCarthy} {et~al.}(2010){McCarthy}, {Schaye}, {Ponman}, {Bower}, {Booth}, {Dalla Vecchia}, {Crain}, {Springel}, {Theuns}, \& {Wiersma}}]{McCarthy2010MNRAS.406..822M}
{McCarthy}, I.~G., {Schaye}, J., {Ponman}, T.~J., {et~al.} 2010, \mnras, 406, 822

\bibitem[{{Mead} {et~al.}(2015){Mead}, {Peacock}, {Heymans}, {Joudaki}, \& {Heavens}}]{Mead2015MNRAS.454.1958M}
{Mead}, A.~J., {Peacock}, J.~A., {Heymans}, C., {Joudaki}, S., \& {Heavens}, A.~F. 2015, \mnras, 454, 1958

\bibitem[{{Moster} {et~al.}(2010){Moster}, {Somerville}, {Maulbetsch}, {van den Bosch}, {Macci{\`o}}, {Naab}, \& {Oser}}]{Moster2010ApJ...710..903M}
{Moster}, B.~P., {Somerville}, R.~S., {Maulbetsch}, C., {et~al.} 2010, \apj, 710, 903

\bibitem[{{Mu{\~n}oz-Cuartas} {et~al.}(2011){Mu{\~n}oz-Cuartas}, {Macci{\`o}}, {Gottl{\"o}ber}, \& {Dutton}}]{Munoz2011MNRAS.411..584M}
{Mu{\~n}oz-Cuartas}, J.~C., {Macci{\`o}}, A.~V., {Gottl{\"o}ber}, S., \& {Dutton}, A.~A. 2011, \mnras, 411, 584

\bibitem[{{Murata} {et~al.}(2018){Murata}, {Nishimichi}, {Takada}, {Miyatake}, {Shirasaki}, {More}, {Takahashi}, \& {Osato}}]{Murata2018ApJ...854..120M}
{Murata}, R., {Nishimichi}, T., {Takada}, M., {et~al.} 2018, \apj, 854, 120

\bibitem[{{Navarro} {et~al.}(1997){Navarro}, {Frenk}, \& {White}}]{Navarro1997ApJ...490..493N}
{Navarro}, J.~F., {Frenk}, C.~S., \& {White}, S. D.~M. 1997, \apj, 490, 493

\bibitem[{{Nelson} {et~al.}(2015){Nelson}, {Pillepich}, {Genel}, {Vogelsberger}, {Springel}, {Torrey}, {Rodriguez-Gomez}, {Sijacki}, {Snyder}, {Griffen}, {Marinacci}, {Blecha}, {Sales}, {Xu}, \& {Hernquist}}]{Nelson2015AC....13...12N}
{Nelson}, D., {Pillepich}, A., {Genel}, S., {et~al.} 2015, Astronomy and Computing, 13, 12

\bibitem[{{Nelson} {et~al.}(2019){Nelson}, {Springel}, {Pillepich}, {Rodriguez-Gomez}, {Torrey}, {Genel}, {Vogelsberger}, {Pakmor}, {Marinacci}, {Weinberger}, {Kelley}, {Lovell}, {Diemer}, \& {Hernquist}}]{Nelson2019ComAC...6....2N}
{Nelson}, D., {Springel}, V., {Pillepich}, A., {et~al.} 2019, Computational Astrophysics and Cosmology, 6, 2

\bibitem[{{Oguri} \& {Hamana}(2011)}]{Oguri2011MNRAS.414.1851O}
{Oguri}, M. \& {Hamana}, T. 2011, \mnras, 414, 1851

\bibitem[{{Peacock} \& {Smith}(2000)}]{Peacock2000MNRAS.318.1144P}
{Peacock}, J.~A. \& {Smith}, R.~E. 2000, \mnras, 318, 1144

\bibitem[{{Pei} {et~al.}(2024){Pei}, {Guo}, {Shao}, {He}, \& {Gu}}]{Pei2024MNRAS.531.2262P}
{Pei}, W., {Guo}, Q., {Shao}, S., {He}, Y., \& {Gu}, Q. 2024, \mnras, 531, 2262

\bibitem[{{Pillepich} {et~al.}(2018){Pillepich}, {Springel}, {Nelson}, {Genel}, {Naiman}, {Pakmor}, {Hernquist}, {Torrey}, {Vogelsberger}, {Weinberger}, \& {Marinacci}}]{Pillepich2018MNRAS.473.4077P}
{Pillepich}, A., {Springel}, V., {Nelson}, D., {et~al.} 2018, \mnras, 473, 4077

\bibitem[{{Planck Collaboration} {et~al.}(2016){Planck Collaboration}, {Ade}, {Aghanim}, {Arnaud}, {Ashdown}, {Aumont}, {Baccigalupi}, {Banday}, {Barreiro}, {Bartlett}, {Bartolo}, {Battaner}, {Battye}, {Benabed}, {Beno{\^\i}t}, {Benoit-L{\'e}vy}, {Bernard}, {Bersanelli}, {Bielewicz}, {Bock}, {Bonaldi}, {Bonavera}, {Bond}, {Borrill}, {Bouchet}, {Boulanger}, {Bucher}, {Burigana}, {Butler}, {Calabrese}, {Cardoso}, {Catalano}, {Challinor}, {Chamballu}, {Chary}, {Chiang}, {Chluba}, {Christensen}, {Church}, {Clements}, {Colombi}, {Colombo}, {Combet}, {Coulais}, {Crill}, {Curto}, {Cuttaia}, {Danese}, {Davies}, {Davis}, {de Bernardis}, {de Rosa}, {de Zotti}, {Delabrouille}, {D{\'e}sert}, {Di Valentino}, {Dickinson}, {Diego}, {Dolag}, {Dole}, {Donzelli}, {Dor{\'e}}, {Douspis}, {Ducout}, {Dunkley}, {Dupac}, {Efstathiou}, {Elsner}, {En{\ss}lin}, {Eriksen}, {Farhang}, {Fergusson}, {Finelli}, {Forni}, {Frailis}, {Fraisse}, {Franceschi}, {Frejsel}, {Galeotta}, {Galli}, {Ganga}, {Gauthier}, {Gerbino}, {Ghosh}, {Giard},
  {Giraud-H{\'e}raud}, {Giusarma}, {Gjerl{\o}w}, {Gonz{\'a}lez-Nuevo}, {G{\'o}rski}, {Gratton}, {Gregorio}, {Gruppuso}, {Gudmundsson}, {Hamann}, {Hansen}, {Hanson}, {Harrison}, {Helou}, {Henrot-Versill{\'e}}, {Hern{\'a}ndez-Monteagudo}, {Herranz}, {Hildebrandt}, {Hivon}, {Hobson}, {Holmes}, {Hornstrup}, {Hovest}, {Huang}, {Huffenberger}, {Hurier}, {Jaffe}, {Jaffe}, {Jones}, {Juvela}, {Keih{\"a}nen}, {Keskitalo}, {Kisner}, {Kneissl}, {Knoche}, {Knox}, {Kunz}, {Kurki-Suonio}, {Lagache}, {L{\"a}hteenm{\"a}ki}, {Lamarre}, {Lasenby}, {Lattanzi}, {Lawrence}, {Leahy}, {Leonardi}, {Lesgourgues}, {Levrier}, {Lewis}, {Liguori}, {Lilje}, {Linden-V{\o}rnle}, {L{\'o}pez-Caniego}, {Lubin}, {Mac{\'\i}as-P{\'e}rez}, {Maggio}, {Maino}, {Mandolesi}, {Mangilli}, {Marchini}, {Maris}, {Martin}, {Martinelli}, {Mart{\'\i}nez-Gonz{\'a}lez}, {Masi}, {Matarrese}, {McGehee}, {Meinhold}, {Melchiorri}, {Melin}, {Mendes}, {Mennella}, {Migliaccio}, {Millea}, {Mitra}, {Miville-Desch{\^e}nes}, {Moneti}, {Montier}, {Morgante}, {Mortlock},
  {Moss}, {Munshi}, {Murphy}, {Naselsky}, {Nati}, {Natoli}, {Netterfield}, {N{\o}rgaard-Nielsen}, {Noviello}, {Novikov}, {Novikov}, {Oxborrow}, {Paci}, {Pagano}, {Pajot}, {Paladini}, {Paoletti}, {Partridge}, {Pasian}, {Patanchon}, {Pearson}, {Perdereau}, {Perotto}, {Perrotta}, {Pettorino}, {Piacentini}, {Piat}, {Pierpaoli}, {Pietrobon}, {Plaszczynski}, {Pointecouteau}, {Polenta}, {Popa}, {Pratt}, {Pr{\'e}zeau}, {Prunet}, {Puget}, {Rachen}, {Reach}, {Rebolo}, {Reinecke}, {Remazeilles}, {Renault}, {Renzi}, {Ristorcelli}, {Rocha}, {Rosset}, {Rossetti}, {Roudier}, {Rouill{\'e} d'Orfeuil}, {Rowan-Robinson}, {Rubi{\~n}o-Mart{\'\i}n}, {Rusholme}, {Said}, {Salvatelli}, {Salvati}, {Sandri}, {Santos}, {Savelainen}, {Savini}, {Scott}, {Seiffert}, {Serra}, {Shellard}, {Spencer}, {Spinelli}, {Stolyarov}, {Stompor}, {Sudiwala}, {Sunyaev}, {Sutton}, {Suur-Uski}, {Sygnet}, {Tauber}, {Terenzi}, {Toffolatti}, {Tomasi}, {Tristram}, {Trombetti}, {Tucci}, {Tuovinen}, {T{\"u}rler}, {Umana}, {Valenziano}, {Valiviita}, {Van Tent},
  {Vielva}, {Villa}, {Wade}, {Wandelt}, {Wehus}, {White}, {White}, {Wilkinson}, {Yvon}, {Zacchei}, \& {Zonca}}]{Planck2016AA...594A..13P}
{Planck Collaboration}, {Ade}, P.~A.~R., {Aghanim}, N., {et~al.} 2016, \aap, 594, A13

\bibitem[{{Planck Collaboration} {et~al.}(2020){Planck Collaboration}, {Aghanim}, {Akrami}, {Ashdown}, {Aumont}, {Baccigalupi}, {Ballardini}, {Banday}, {Barreiro}, {Bartolo}, {Basak}, {Battye}, {Benabed}, {Bernard}, {Bersanelli}, {Bielewicz}, {Bock}, {Bond}, {Borrill}, {Bouchet}, {Boulanger}, {Bucher}, {Burigana}, {Butler}, {Calabrese}, {Cardoso}, {Carron}, {Challinor}, {Chiang}, {Chluba}, {Colombo}, {Combet}, {Contreras}, {Crill}, {Cuttaia}, {de Bernardis}, {de Zotti}, {Delabrouille}, {Delouis}, {Di Valentino}, {Diego}, {Dor{\'e}}, {Douspis}, {Ducout}, {Dupac}, {Dusini}, {Efstathiou}, {Elsner}, {En{\ss}lin}, {Eriksen}, {Fantaye}, {Farhang}, {Fergusson}, {Fernandez-Cobos}, {Finelli}, {Forastieri}, {Frailis}, {Fraisse}, {Franceschi}, {Frolov}, {Galeotta}, {Galli}, {Ganga}, {G{\'e}nova-Santos}, {Gerbino}, {Ghosh}, {Gonz{\'a}lez-Nuevo}, {G{\'o}rski}, {Gratton}, {Gruppuso}, {Gudmundsson}, {Hamann}, {Handley}, {Hansen}, {Herranz}, {Hildebrandt}, {Hivon}, {Huang}, {Jaffe}, {Jones}, {Karakci}, {Keih{\"a}nen},
  {Keskitalo}, {Kiiveri}, {Kim}, {Kisner}, {Knox}, {Krachmalnicoff}, {Kunz}, {Kurki-Suonio}, {Lagache}, {Lamarre}, {Lasenby}, {Lattanzi}, {Lawrence}, {Le Jeune}, {Lemos}, {Lesgourgues}, {Levrier}, {Lewis}, {Liguori}, {Lilje}, {Lilley}, {Lindholm}, {L{\'o}pez-Caniego}, {Lubin}, {Ma}, {Mac{\'\i}as-P{\'e}rez}, {Maggio}, {Maino}, {Mandolesi}, {Mangilli}, {Marcos-Caballero}, {Maris}, {Martin}, {Martinelli}, {Mart{\'\i}nez-Gonz{\'a}lez}, {Matarrese}, {Mauri}, {McEwen}, {Meinhold}, {Melchiorri}, {Mennella}, {Migliaccio}, {Millea}, {Mitra}, {Miville-Desch{\^e}nes}, {Molinari}, {Montier}, {Morgante}, {Moss}, {Natoli}, {N{\o}rgaard-Nielsen}, {Pagano}, {Paoletti}, {Partridge}, {Patanchon}, {Peiris}, {Perrotta}, {Pettorino}, {Piacentini}, {Polastri}, {Polenta}, {Puget}, {Rachen}, {Reinecke}, {Remazeilles}, {Renzi}, {Rocha}, {Rosset}, {Roudier}, {Rubi{\~n}o-Mart{\'\i}n}, {Ruiz-Granados}, {Salvati}, {Sandri}, {Savelainen}, {Scott}, {Shellard}, {Sirignano}, {Sirri}, {Spencer}, {Sunyaev}, {Suur-Uski}, {Tauber}, {Tavagnacco},
  {Tenti}, {Toffolatti}, {Tomasi}, {Trombetti}, {Valenziano}, {Valiviita}, {Van Tent}, {Vibert}, {Vielva}, {Villa}, {Vittorio}, {Wandelt}, {Wehus}, {White}, {White}, {Zacchei}, \& {Zonca}}]{Planck2020AA...641A...6P}
{Planck Collaboration}, {Aghanim}, N., {Akrami}, Y., {et~al.} 2020, \aap, 641, A6

\bibitem[{{Pop} {et~al.}(2022){Pop}, {Hernquist}, {Nagai}, {Kannan}, {Weinberger}, {Springel}, {Vogelsberger}, {Nelson}, {Pakmor}, {Pillepich}, \& {Torrey}}]{Pop2022arXiv220511528P}
{Pop}, A.-R., {Hernquist}, L., {Nagai}, D., {et~al.} 2022, arXiv e-prints, arXiv:2205.11528

\bibitem[{{Rana} {et~al.}(2022){Rana}, {More}, {Miyatake}, {Nishimichi}, {Takada}, {Robotham}, {Hopkins}, \& {Holwerda}}]{Rana2022MNRAS.510.5408R}
{Rana}, D., {More}, S., {Miyatake}, H., {et~al.} 2022, \mnras, 510, 5408

\bibitem[{{Rasia} {et~al.}(2006){Rasia}, {Ettori}, {Moscardini}, {Mazzotta}, {Borgani}, {Dolag}, {Tormen}, {Cheng}, \& {Diaferio}}]{Rasia2006MNRAS.369.2013R}
{Rasia}, E., {Ettori}, S., {Moscardini}, L., {et~al.} 2006, \mnras, 369, 2013

\bibitem[{{Robertson} {et~al.}(2024){Robertson}, {Sif{\'o}n}, {Asgari}, {Battaglia}, {Bilicki}, {Richard Bond}, {Devlin}, {Dunkley}, {Giblin}, {Heymans}, {Hildebrandt}, {Hilton}, {Hoekstra}, {Hughes}, {Kuijken}, {Louis}, {Mallaby-Kay}, {Page}, {Partridge}, {Radovich}, {Schneider}, {Shan}, {Spergel}, {Tr{\"o}ster}, {Wollack}, {Vargas}, \& {Wright}}]{Robertson2024AA...681A..87R}
{Robertson}, N.~C., {Sif{\'o}n}, C., {Asgari}, M., {et~al.} 2024, \aap, 681, A87

\bibitem[{{Robotham} {et~al.}(2011){Robotham}, {Norberg}, {Driver}, {Baldry}, {Bamford}, {Hopkins}, {Liske}, {Loveday}, {Merson}, {Peacock}, {Brough}, {Cameron}, {Conselice}, {Croom}, {Frenk}, {Gunawardhana}, {Hill}, {Jones}, {Kelvin}, {Kuijken}, {Nichol}, {Parkinson}, {Pimbblet}, {Phillipps}, {Popescu}, {Prescott}, {Sharp}, {Sutherland}, {Taylor}, {Thomas}, {Tuffs}, {van Kampen}, \& {Wijesinghe}}]{Robotham2011MNRAS.416.2640R}
{Robotham}, A.~S.~G., {Norberg}, P., {Driver}, S.~P., {et~al.} 2011, \mnras, 416, 2640

\bibitem[{{Rowe} {et~al.}(2015){Rowe}, {Jarvis}, {Mandelbaum}, {Bernstein}, {Bosch}, {Simet}, {Meyers}, {Kacprzak}, {Nakajima}, {Zuntz}, {Miyatake}, {Dietrich}, {Armstrong}, {Melchior}, \& {Gill}}]{Rowe2015AC....10..121R}
{Rowe}, B.~T.~P., {Jarvis}, M., {Mandelbaum}, R., {et~al.} 2015, Astronomy and Computing, 10, 121

\bibitem[{{Schaller} {et~al.}(2024){Schaller}, {Borrow}, {Draper}, {Ivkovic}, {McAlpine}, {Vandenbroucke}, {Bah{\'e}}, {Chaikin}, {Chalk}, {Chan}, {Correa}, {van Daalen}, {Elbers}, {Gonnet}, {Hausammann}, {Helly}, {Hu{\v{s}}ko}, {Kegerreis}, {Nobels}, {Ploeckinger}, {Revaz}, {Roper}, {Ruiz-Bonilla}, {Sandnes}, {Uyttenhove}, {Willis}, \& {Xiang}}]{Schaller2024MNRAS.530.2378S}
{Schaller}, M., {Borrow}, J., {Draper}, P.~W., {et~al.} 2024, \mnras, 530, 2378

\bibitem[{{Schaye} {et~al.}(2015){Schaye}, {Crain}, {Bower}, {Furlong}, {Schaller}, {Theuns}, {Dalla Vecchia}, {Frenk}, {McCarthy}, {Helly}, {Jenkins}, {Rosas-Guevara}, {White}, {Baes}, {Booth}, {Camps}, {Navarro}, {Qu}, {Rahmati}, {Sawala}, {Thomas}, \& {Trayford}}]{Schaye2015MNRAS.446..521S}
{Schaye}, J., {Crain}, R.~A., {Bower}, R.~G., {et~al.} 2015, \mnras, 446, 521

\bibitem[{{Schaye} {et~al.}(2010){Schaye}, {Dalla Vecchia}, {Booth}, {Wiersma}, {Theuns}, {Haas}, {Bertone}, {Duffy}, {McCarthy}, \& {van de Voort}}]{Schaye2010MNRAS.402.1536S}
{Schaye}, J., {Dalla Vecchia}, C., {Booth}, C.~M., {et~al.} 2010, \mnras, 402, 1536

\bibitem[{{Schaye} {et~al.}(2023){Schaye}, {Kugel}, {Schaller}, {Helly}, {Braspenning}, {Elbers}, {McCarthy}, {van Daalen}, {Vandenbroucke}, {Frenk}, {Kwan}, {Salcido}, {Bah{\'e}}, {Borrow}, {Chaikin}, {Hahn}, {Hu{\v{s}}ko}, {Jenkins}, {Lacey}, \& {Nobels}}]{Schaye2023MNRAS.526.4978S}
{Schaye}, J., {Kugel}, R., {Schaller}, M., {et~al.} 2023, \mnras, 526, 4978

\bibitem[{{Schneider} {et~al.}(2020){Schneider}, {Stoira}, {Refregier}, {Weiss}, {Knabenhans}, {Stadel}, \& {Teyssier}}]{Schneider2020JCAP...04..019S}
{Schneider}, A., {Stoira}, N., {Refregier}, A., {et~al.} 2020, \jcap, 2020, 019

\bibitem[{{Seitz} \& {Schneider}(1997)}]{Seitz1997AA...318..687S}
{Seitz}, C. \& {Schneider}, P. 1997, \aap, 318, 687

\bibitem[{{Seljak}(2000)}]{Seljak2000MNRAS.318..203S}
{Seljak}, U. 2000, \mnras, 318, 203

\bibitem[{{Semboloni} {et~al.}(2011){Semboloni}, {Hoekstra}, {Schaye}, {van Daalen}, \& {McCarthy}}]{Semboloni2011MNRAS.417.2020S}
{Semboloni}, E., {Hoekstra}, H., {Schaye}, J., {van Daalen}, M.~P., \& {McCarthy}, I.~G. 2011, \mnras, 417, 2020

\bibitem[{{Sif{\'o}n} {et~al.}(2015){Sif{\'o}n}, {Cacciato}, {Hoekstra}, {Brouwer}, {van Uitert}, {Viola}, {Baldry}, {Brough}, {Brown}, {Choi}, {Driver}, {Erben}, {Grado}, {Heymans}, {Hildebrandt}, {Joachimi}, {de Jong}, {Kuijken}, {McFarland}, {Miller}, {Nakajima}, {Napolitano}, {Norberg}, {Robotham}, {Schneider}, \& {Verdoes Kleijn}}]{Sifon2015MNRAS.454.3938S}
{Sif{\'o}n}, C., {Cacciato}, M., {Hoekstra}, H., {et~al.} 2015, \mnras, 454, 3938

\bibitem[{{Smith} {et~al.}(2003){Smith}, {Peacock}, {Jenkins}, {White}, {Frenk}, {Pearce}, {Thomas}, {Efstathiou}, \& {Couchman}}]{Smith2003MNRAS.341.1311S}
{Smith}, R.~E., {Peacock}, J.~A., {Jenkins}, A., {et~al.} 2003, \mnras, 341, 1311

\bibitem[{{Springel}(2010)}]{Springel2010MNRAS.401..791S}
{Springel}, V. 2010, \mnras, 401, 791

\bibitem[{{Springel} {et~al.}(2005){Springel}, {White}, {Jenkins}, {Frenk}, {Yoshida}, {Gao}, {Navarro}, {Thacker}, {Croton}, {Helly}, {Peacock}, {Cole}, {Thomas}, {Couchman}, {Evrard}, {Colberg}, \& {Pearce}}]{Springel2005Natur.435..629S}
{Springel}, V., {White}, S. D.~M., {Jenkins}, A., {et~al.} 2005, \nat, 435, 629

\bibitem[{{Springel} {et~al.}(2001){Springel}, {White}, {Tormen}, \& {Kauffmann}}]{Springel2001MNRAS.328..726S}
{Springel}, V., {White}, S. D.~M., {Tormen}, G., \& {Kauffmann}, G. 2001, \mnras, 328, 726

\bibitem[{{Takada} \& {Jain}(2003)}]{Takada2003MNRAS.340..580T}
{Takada}, M. \& {Jain}, B. 2003, \mnras, 340, 580

\bibitem[{{Taylor} {et~al.}(2023){Taylor}, {Cluver}, {Bell}, {Brinchmann}, {Colless}, {Courtois}, {Hoekstra}, {Kannappan}, {Lagos}, {Liske}, {Tempel}, {Howlett}, {McGee}, {Said}, {Skelton}, {Gunawardhana}, {Bellstedt}, {Hunt}, {Jarrett}, {Lidman}, {Lucey}, {Alam}, {Bilicki}, {de Graaff}, {Hellwing}, {Leslie}, {Loubser}, {Marchetti}, {Maseda}, {Mogotsi}, {Norberg}, {Sonnenfeld}, {Sorce}, \& {4HS Team}}]{Taylor2023Msngr.190...46T}
{Taylor}, E.~N., {Cluver}, M., {Bell}, E., {et~al.} 2023, The Messenger, 190, 46

\bibitem[{{Taylor} {et~al.}(2011){Taylor}, {Hopkins}, {Baldry}, {Brown}, {Driver}, {Kelvin}, {Hill}, {Robotham}, {Bland-Hawthorn}, {Jones}, {Sharp}, {Thomas}, {Liske}, {Loveday}, {Norberg}, {Peacock}, {Bamford}, {Brough}, {Colless}, {Cameron}, {Conselice}, {Croom}, {Frenk}, {Gunawardhana}, {Kuijken}, {Nichol}, {Parkinson}, {Phillipps}, {Pimbblet}, {Popescu}, {Prescott}, {Sutherland}, {Tuffs}, {van Kampen}, \& {Wijesinghe}}]{Taylor2011MNRAS.418.1587T}
{Taylor}, E.~N., {Hopkins}, A.~M., {Baldry}, I.~K., {et~al.} 2011, \mnras, 418, 1587

\bibitem[{{Tinker} {et~al.}(2010){Tinker}, {Robertson}, {Kravtsov}, {Klypin}, {Warren}, {Yepes}, \& {Gottl{\"o}ber}}]{Tinker2010ApJ...724..878T}
{Tinker}, J.~L., {Robertson}, B.~E., {Kravtsov}, A.~V., {et~al.} 2010, \apj, 724, 878

\bibitem[{{Tyson} {et~al.}(1990){Tyson}, {Valdes}, \& {Wenk}}]{Tyson1990ApJ...349L...1T}
{Tyson}, J.~A., {Valdes}, F., \& {Wenk}, R.~A. 1990, \apjl, 349, L1

\bibitem[{{Vale} \& {Ostriker}(2004)}]{Vale2004MNRAS.353..189V}
{Vale}, A. \& {Ostriker}, J.~P. 2004, \mnras, 353, 189

\bibitem[{{van Daalen} {et~al.}(2020){van Daalen}, {McCarthy}, \& {Schaye}}]{Daalen2020MNRAS.491.2424V}
{van Daalen}, M.~P., {McCarthy}, I.~G., \& {Schaye}, J. 2020, \mnras, 491, 2424

\bibitem[{{van Daalen} {et~al.}(2011){van Daalen}, {Schaye}, {Booth}, \& {Dalla Vecchia}}]{Daalen2011MNRAS.415.3649V}
{van Daalen}, M.~P., {Schaye}, J., {Booth}, C.~M., \& {Dalla Vecchia}, C. 2011, \mnras, 415, 3649

\bibitem[{{van den Bosch} {et~al.}(2013){van den Bosch}, {More}, {Cacciato}, {Mo}, \& {Yang}}]{Bosch2013MNRAS.430..725V}
{van den Bosch}, F.~C., {More}, S., {Cacciato}, M., {Mo}, H., \& {Yang}, X. 2013, \mnras, 430, 725

\bibitem[{{van den Busch} {et~al.}(2022){van den Busch}, {Wright}, {Hildebrandt}, {Bilicki}, {Asgari}, {Joudaki}, {Blake}, {Heymans}, {Kannawadi}, {Shan}, \& {Tr{\"o}ster}}]{Busch2022AA...664A.170V}
{van den Busch}, J.~L., {Wright}, A.~H., {Hildebrandt}, H., {et~al.} 2022, \aap, 664, A170

\bibitem[{{van Uitert} {et~al.}(2016){van Uitert}, {Cacciato}, {Hoekstra}, {Brouwer}, {Sif{\'o}n}, {Viola}, {Baldry}, {Bland-Hawthorn}, {Brough}, {Brown}, {Choi}, {Driver}, {Erben}, {Heymans}, {Hildebrandt}, {Joachimi}, {Kuijken}, {Liske}, {Loveday}, {McFarland}, {Miller}, {Nakajima}, {Peacock}, {Radovich}, {Robotham}, {Schneider}, {Sikkema}, {Taylor}, \& {Verdoes Kleijn}}]{Uitert2016MNRAS.459.3251V}
{van Uitert}, E., {Cacciato}, M., {Hoekstra}, H., {et~al.} 2016, \mnras, 459, 3251

\bibitem[{{van Uitert} {et~al.}(2011){van Uitert}, {Hoekstra}, {Velander}, {Gilbank}, {Gladders}, \& {Yee}}]{Uitert2011AA...534A..14V}
{van Uitert}, E., {Hoekstra}, H., {Velander}, M., {et~al.} 2011, \aap, 534, A14

\bibitem[{{Velliscig} {et~al.}(2017){Velliscig}, {Cacciato}, {Hoekstra}, {Schaye}, {Heymans}, {Hildebrandt}, {Loveday}, {Norberg}, {Sif{\'o}n}, {Schneider}, {van Uitert}, {Viola}, {Brough}, {Erben}, {Holwerda}, {Hopkins}, \& {Kuijken}}]{Velliscig2017MNRAS.471.2856V}
{Velliscig}, M., {Cacciato}, M., {Hoekstra}, H., {et~al.} 2017, \mnras, 471, 2856

\bibitem[{{Viola} {et~al.}(2015){Viola}, {Cacciato}, {Brouwer}, {Kuijken}, {Hoekstra}, {Norberg}, {Robotham}, {van Uitert}, {Alpaslan}, {Baldry}, {Choi}, {de Jong}, {Driver}, {Erben}, {Grado}, {Graham}, {Heymans}, {Hildebrandt}, {Hopkins}, {Irisarri}, {Joachimi}, {Loveday}, {Miller}, {Nakajima}, {Schneider}, {Sif{\'o}n}, \& {Verdoes Kleijn}}]{Viola2015MNRAS.452.3529V}
{Viola}, M., {Cacciato}, M., {Brouwer}, M., {et~al.} 2015, \mnras, 452, 3529

\bibitem[{{Vogelsberger} {et~al.}(2013){Vogelsberger}, {Genel}, {Sijacki}, {Torrey}, {Springel}, \& {Hernquist}}]{Vogelsberger2013MNRAS.436.3031V}
{Vogelsberger}, M., {Genel}, S., {Sijacki}, D., {et~al.} 2013, \mnras, 436, 3031

\bibitem[{{Vogelsberger} {et~al.}(2014){Vogelsberger}, {Genel}, {Springel}, {Torrey}, {Sijacki}, {Xu}, {Snyder}, {Nelson}, \& {Hernquist}}]{Vogelsberger2014MNRAS.444.1518V}
{Vogelsberger}, M., {Genel}, S., {Springel}, V., {et~al.} 2014, \mnras, 444, 1518

\bibitem[{{Vogelsberger} {et~al.}(2020){Vogelsberger}, {Marinacci}, {Torrey}, \& {Puchwein}}]{Vogelsberger2020NatRP...2...42V}
{Vogelsberger}, M., {Marinacci}, F., {Torrey}, P., \& {Puchwein}, E. 2020, Nature Reviews Physics, 2, 42

\bibitem[{{von der Linden} {et~al.}(2014){von der Linden}, {Mantz}, {Allen}, {Applegate}, {Kelly}, {Morris}, {Wright}, {Allen}, {Burchat}, {Burke}, {Donovan}, \& {Ebeling}}]{Linden2014MNRAS.443.1973V}
{von der Linden}, A., {Mantz}, A., {Allen}, S.~W., {et~al.} 2014, \mnras, 443, 1973

\bibitem[{{Wang} {et~al.}(2024){Wang}, {Li}, {Zhu}, {Shan}, {Xu}, {Cappellari}, {Gao}, {Li}, {Lu}, {Mao}, {Yao}, \& {Xie}}]{Wang2024MNRAS.527.1580W}
{Wang}, C., {Li}, R., {Zhu}, K., {et~al.} 2024, \mnras, 527, 1580

\bibitem[{{Wechsler} \& {Tinker}(2018)}]{Wechsler2018ARAA..56..435W}
{Wechsler}, R.~H. \& {Tinker}, J.~L. 2018, \araa, 56, 435

\bibitem[{{White} \& {Frenk}(1991)}]{White1991ApJ...379...52W}
{White}, S. D.~M. \& {Frenk}, C.~S. 1991, \apj, 379, 52

\bibitem[{{White} \& {Rees}(1978)}]{White1978MNRAS.183..341W}
{White}, S.~D.~M. \& {Rees}, M.~J. 1978, \mnras, 183, 341

\bibitem[{{Wright} {et~al.}(2016){Wright}, {Robotham}, {Bourne}, {Driver}, {Dunne}, {Maddox}, {Alpaslan}, {Andrews}, {Bauer}, {Bland-Hawthorn}, {Brough}, {Brown}, {Clarke}, {Cluver}, {Davies}, {Grootes}, {Holwerda}, {Hopkins}, {Jarrett}, {Kafle}, {Lange}, {Liske}, {Loveday}, {Moffett}, {Norberg}, {Popescu}, {Smith}, {Taylor}, {Tuffs}, {Wang}, \& {Wilkins}}]{Wright2016MNRAS.460..765W}
{Wright}, A.~H., {Robotham}, A.~S.~G., {Bourne}, N., {et~al.} 2016, \mnras, 460, 765

\bibitem[{{Yang} {et~al.}(2003){Yang}, {Mo}, \& {van den Bosch}}]{Yang2003MNRAS.339.1057Y}
{Yang}, X., {Mo}, H.~J., \& {van den Bosch}, F.~C. 2003, \mnras, 339, 1057

\bibitem[{{Yang} {et~al.}(2008){Yang}, {Mo}, \& {van den Bosch}}]{Yang2008ApJ...676..248Y}
{Yang}, X., {Mo}, H.~J., \& {van den Bosch}, F.~C. 2008, \apj, 676, 248

\bibitem[{{Yang} {et~al.}(2009){Yang}, {Mo}, \& {van den Bosch}}]{Yang2009ApJ...695..900Y}
{Yang}, X., {Mo}, H.~J., \& {van den Bosch}, F.~C. 2009, \apj, 695, 900

\bibitem[{{Zhang} {et~al.}(2019){Zhang}, {Jeltema}, {Hollowood}, {Everett}, {Rozo}, {Farahi}, {Bermeo}, {Bhargava}, {Giles}, {Romer}, {Wilkinson}, {Rykoff}, {Mantz}, {Diehl}, {Evrard}, {Stern}, {Gruen}, {von der Linden}, {Splettstoesser}, {Chen}, {Costanzi}, {Allen}, {Collins}, {Hilton}, {Klein}, {Mann}, {Manolopoulou}, {Morris}, {Mayers}, {Sahlen}, {Stott}, {Vergara Cervantes}, {Viana}, {Wechsler}, {Allam}, {Avila}, {Bechtol}, {Bertin}, {Brooks}, {Burke}, {Carnero Rosell}, {Carrasco Kind}, {Carretero}, {Castander}, {da Costa}, {De Vicente}, {Desai}, {Dietrich}, {Doel}, {Flaugher}, {Fosalba}, {Frieman}, {Garc{\'\i}a-Bellido}, {Gaztanaga}, {Gruendl}, {Gschwend}, {Gutierrez}, {Hartley}, {Honscheid}, {Hoyle}, {Krause}, {Kuehn}, {Kuropatkin}, {Lima}, {Maia}, {Marshall}, {Melchior}, {Menanteau}, {Miller}, {Miquel}, {Ogando}, {Plazas}, {Sanchez}, {Scarpine}, {Schindler}, {Serrano}, {Sevilla-Noarbe}, {Smith}, {Soares-Santos}, {Suchyta}, {Swanson}, {Tarle}, {Thomas}, {Tucker}, {Vikram}, {Wester}, \& {DES
  Collaboration}}]{Zhang2019MNRAS.487.2578Z}
{Zhang}, Y., {Jeltema}, T., {Hollowood}, D.~L., {et~al.} 2019, \mnras, 487, 2578

\end{thebibliography}

\begin{appendix} 

\section{Higher-order shear biases in \textit{lens}fit measurements}
\label{Sec:nonlinear}

In a weak lensing analysis, it is typically assumed that shear bias does not depend on the shear if the signal is small, and there is no cross-talk between the two shear components~(e.g.~\citealt{Heymans2006MNRAS.368.1323H}). However, this assumption warrants reconsideration when studying galaxy clusters and groups, where the shear amplitude near the mass centre can be several times larger than that of cosmic shear. This appendix examines the potential higher-order biases in the KiDS \textit{lens}fit shear measurements in these regions.

\begin{figure}
  \centering
  \includegraphics[width=\hsize]{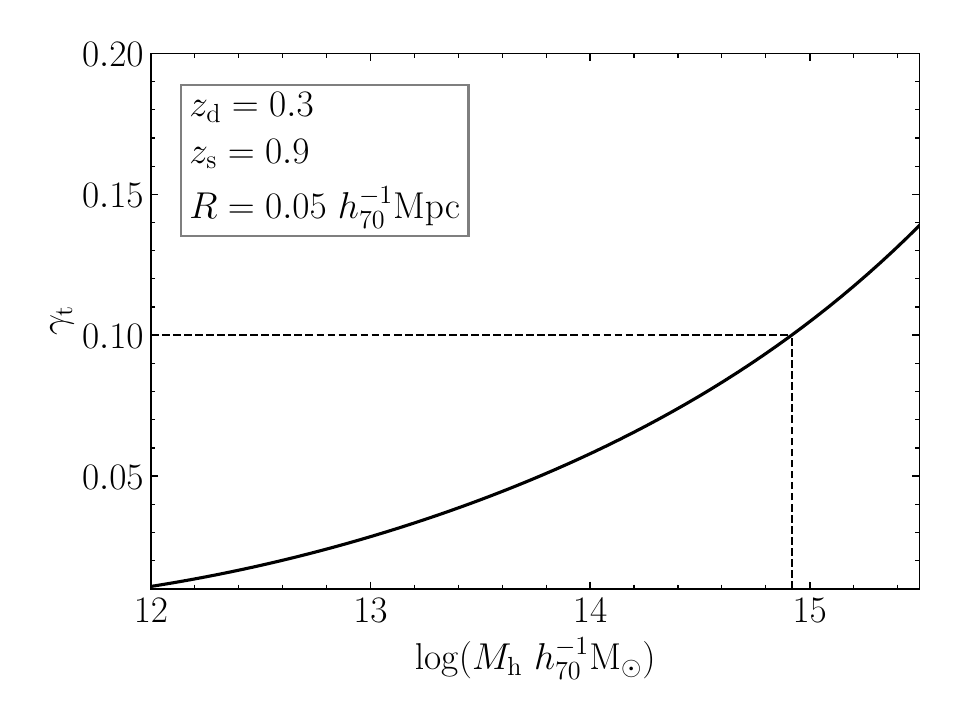}
  \caption{Amplitude of tangential shear as a function of lens halo mass for a toy model based on the NFW profile, with the lens system geometry listed in the figure. The geometry corresponds to the extreme cases from our analysis, where measurements are already heavily affected by blending effects. In typical cases, the majority of shear amplitudes for a given halo mass are much smaller than what is shown here.} 
 \label{fig:shearMass}
\end{figure}

Before investigating the higher-order biases in high-shear regimes, it is instructive to first illustrate the typical shear amplitudes encountered in studies of galaxy clusters and groups. To do this, we constructed a toy model based on the Navarro-Frenk-White (NFW, \citealt{Navarro1997ApJ...490..493N}) profile, with the mass-concentration relation from \citet{Duffy2008MNRAS.390L..64D}. We adopted the lens system geometry corresponding to the extreme cases in our analysis: a lens redshift of $0.3$, a source redshift of $0.9$, and a projected separation of $0.05~h_{70}^{-1}{\rm Mpc}$. Figure~\ref{fig:shearMass} shows the tangential shear amplitude as a function of lens halo mass from this toy model. The shear amplitude reaches 0.1 for a halo mass of ${\sim}10^{14.9}~h_{70}^{-1}{\rm M}_{\odot}$. However, it is important to note that for galaxy group studies using KiDS-like ground-based data, measurements in this regime are already heavily impacted by blending effects, as indicated in Figs.~\ref{fig:ESDMgrp} and \ref{fig:ESDLgrp}, where the innermost point in the highest mass bin shows a drop with large uncertainties.

To study the higher-order shear biases, we generated a set of image simulations covering 37 different shear setups. These include one zero-shear simulation ($0$, $0$) and 36 non-zero shear simulations. For the non-zero shear simulations, the input shear amplitudes range from $0.014$ to $0.14$ per component, with four combinations: ($\pm\gamma$, $\pm\gamma$). Each shear setup contains 18 KiDS $r$-band tile images, paired with counterparts where galaxies are rotated by 90 degrees to reduce shape noise. All simulations were generated using the KiDS \texttt{MultiBand\_ImSim} pipeline~\citep{Li2023AA...670A.100L}\footnote{\url{https://github.com/KiDS-WL/MultiBand_ImSim}}, built on the \texttt{GalSim} package~\citep{Rowe2015AC....10..121R}\footnote{\url{https://github.com/GalSim-developers/GalSim}}.

The zero-shear simulation was used to account for correlations across different shear setups, as they share the same galaxy population, observational conditions, and noise realisations. This approach significantly reduces uncertainties in the shear bias estimates and is valid as we are only concerned with the bias changes relative to the input shear amplitude rather than the absolute shear bias. We denote the zero-shear-subtracted shear estimates as $\tilde{\gamma}_{i}^{\rm obs}$. To capture higher-order biases, we extend the linear shear bias model by including additional terms:
\begin{align}
\label{eq:nonlinear}
\tilde{\gamma}_{i}^{\rm obs} - \gamma_{i}^{\rm input} =&\ \tilde{m}_{i}\ \gamma_{i}^{\rm input} + \tilde{c}_i \notag \\
&+ \tilde{d}_i\ \left(\gamma_{i}^{\rm input}\right)^2 + \tilde{q}_i\ \left(\gamma_{i}^{\rm input}\right)^3 + \tilde{m}_{\perp,i}\ \gamma_{j}^{\rm input} ~,
\end{align}
where subscripts $i$ and $j$ represent different shear components. The equation includes higher-order terms up to the third order, and also has a linear cross-talk term $\tilde{m}_{\perp,i}$. We used tildes on the bias parameters to distinguish our estimates from the true shear bias. 

\begin{figure}
  \centering
  \includegraphics[width=\hsize]{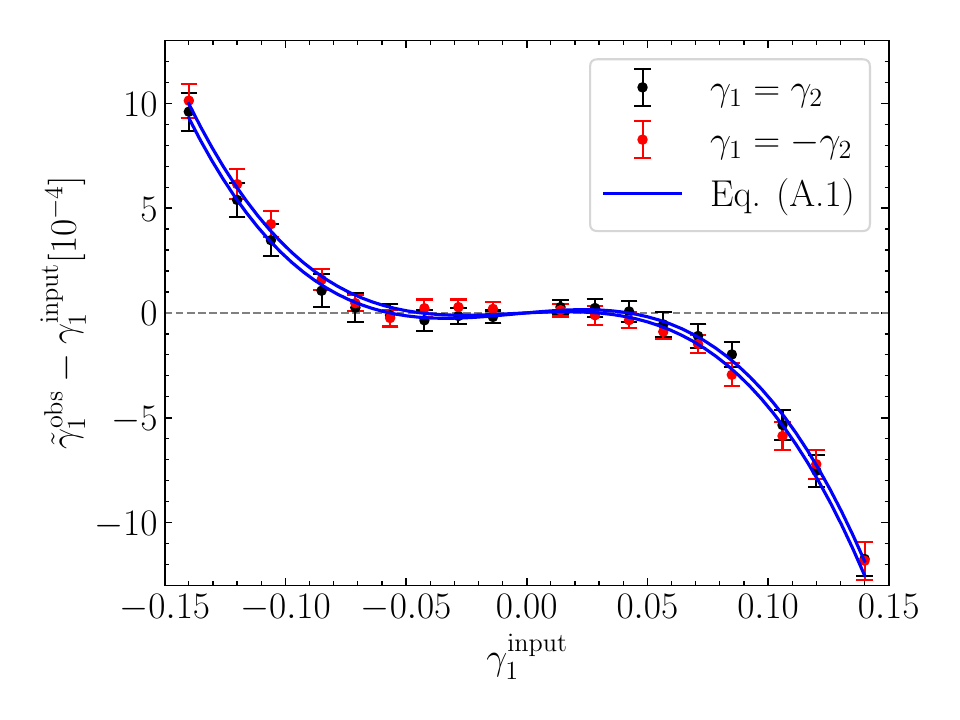}
  \includegraphics[width=\hsize]{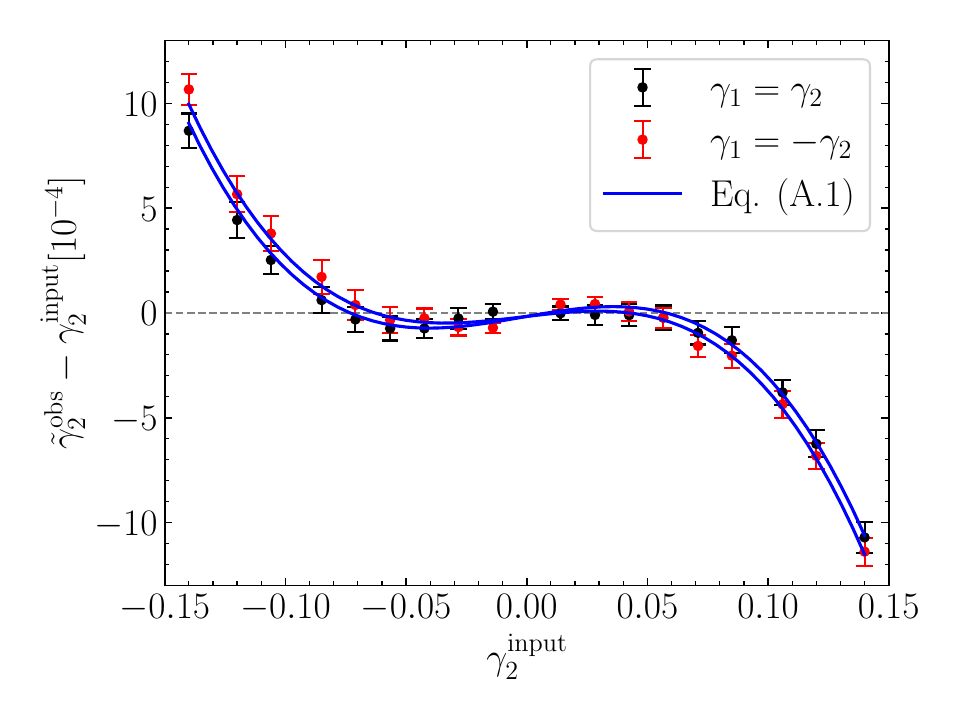}      
  \caption{Difference between the estimated and input shear values as a function of the input shear. The estimated shears are corrected using the zero-shear simulations to account for correlations across different shear setups, as described in the main text. Error bars represent the uncertainties in the weighted mean estimates, calculated by bootstrapping the 18 independent shear estimates for each input shear value. The black and red points distinguish between simulations with the two shear components having the same or opposite signs. The blue lines show the fitting results of Eq.~(\ref{eq:nonlinear}), with the constrained parameter values presented in Table~\ref{table:nonlinear}.} 
 \label{fig:nonlinear}
\end{figure}

Figure~\ref{fig:nonlinear} shows the measured biases along with the fitting results for both components. The corresponding shear bias values are presented in Table~\ref{table:nonlinear}. We observe clear non-linear behaviour for $|\gamma_i| {\gtrsim} 0.1$, which is dominated by the third-order term. However, the fitting results also reveal small but non-zero quadratic and cross-talk terms. 

To further investigate where these higher-order terms arise, we traced back to the sample detection and selection processes, which are also known to introduce biases at the percent level (e.g.~\citealt{Conti2017MNRAS.467.1627F,Kannawadi2019AA...624A..92K,Hern2020AA...640A.117H,Hoekstra2021AA...646A.124H}). Besides, we emphasise that, in higher-shear regimes, some fundamental assumptions based on the small shear signal may no longer hold. Specifically, the transformation between the intrinsic complex ellipticity, $\epsilon^{\rm s}$, and the shear distorted ellipticity,
\begin{equation}
\label{eq:shearRelation}
    \epsilon^{\rm obs} = \frac{\epsilon^{\rm s}+\gamma}{1+\gamma^{*}\epsilon^{\rm s}}~,
\end{equation}
cannot be simplified as $\epsilon^{\rm obs}\approx \epsilon^{\rm s}+\gamma$~\citep{Seitz1997AA...318..687S,Bartelmann2001PhR...340..291B}. The asterisk in the equation denotes the complex conjugate. In other words, the averaged observed ellipticity per component will no longer provide an unbiased estimate of the underlying shear per component, even the full sample has $\langle\epsilon^{\rm s}\rangle{=}0$. 

Given these concerns, we measured the shear biases for three other samples, all based on a perfect galaxy shape measurement algorithm, meaning a direct use of Eq.~(\ref{eq:shearRelation}) to recover the observed galaxy ellipticity. For the first sample, we used all galaxies in our input sample with shape noise cancellation, where $\langle\epsilon^{\rm s}\rangle{=}0$ holds by design. This helps identify biases introduced by the assumption of small shear. For the second sample, we used galaxies detected by SExtractor with observed magnitudes in the range of (20, 24.5). This magnitude range is close to those measurable by the \textit{lens}fit algorithm in the KiDS data. For the third sample, we used galaxies selected by \textit{lens}fit, which removes artefacts, identified stars, poorly resolved objects, blended objects, and so on (see \citealt{Li2023AA...670A.100L} for the detailed selection criteria). 

The shear biases estimated from these samples are also presented in Table~\ref{table:nonlinear}. Consistent with findings from \citet{Kannawadi2019AA...624A..92K} and \citet{Hoekstra2021AA...646A.124H}, we detect percent-level $\tilde{m}_i$ biases from SExtractor detection and \textit{lens}fit selection. Moreover, we find that the quadratic and cross-talk terms are already present at similar levels as in the \textit{lens}fit measurement sample, while the third-order term remains relatively small. This implies that the higher-order biases have different origins, with the former primarily arising from the sample selection while the latter mainly arising from the shape measurement. 

Interestingly, we observe small but non-zero linear and cross-talk biases in the complete sample, confirming the deviation from $\epsilon^{\rm obs}\approx \epsilon^{\rm s}+\gamma$ in high-shear regimes. Notably, about one-quarter of the final cross-talk bias is already present in the complete sample, cautioning against over-interpreting biases at the $10^{-4}$ level when the shear signal is no longer small.

\begin{figure}
  \centering
  \includegraphics[width=\hsize]{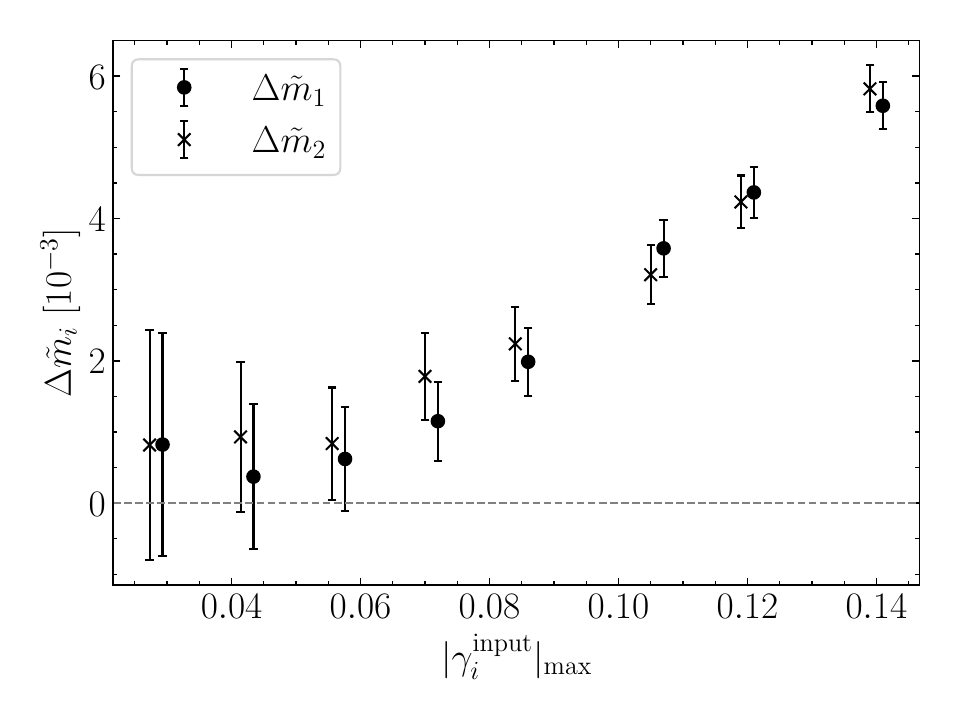}
  \caption{The difference in $\tilde{m}_{i}$ between the non-linear and linear fits as a function of the maximum shear amplitude used in the fit. The difference is defined as $\Delta\tilde{m}_{i} \equiv \tilde{m}_{i}^{\rm non-linear} - \tilde{m}_{i}^{\rm linear}$. Error bars are calculated from bootstrapping the independent shear estimates used in each fit.} 
 \label{fig:nonlinear_dm}
\end{figure}

To quantify the impact of these non-linear effects on the linear shear bias calibration, we compare the multiplicative biases, $\tilde{m}_{i}$, between the non-linear fit and the linear fit, where the higher-order and cross-talk terms in Eq.~(\ref{eq:nonlinear}) are set to zero. Figure~\ref{fig:nonlinear_dm} shows the difference in $\tilde{m}_{i}$ as a function of the maximum shear amplitude used in the fit. Consistent with the results of Fig.~\ref{fig:nonlinear}, we observe that the difference in $\tilde{m}_{i}$ between the two fits increases as more high-shear amplitude simulations are included in the fit. For current KiDS-like analyses, where percent-level accuracy is required, the biases introduced by neglecting these non-linear shear effects are negligible, even in the most extreme cases from our setups. However, these effects become significant for future weak lensing surveys aiming for sub-percent level accuracy, particularly in the presence of high-shear signals ($|\gamma_{i}^{\rm input}| \gtrsim 0.1$).

We note that our results, based on KiDS \textit{lens}fit measurements, show more linear behaviour compared to the recent findings of \citet{Jansen2024AA...683A.240J}, who used the KSB algorithm~\citep{Kaiser1995ApJ...449..460K,Luppino1997ApJ...475...20L,Hoekstra1998ApJ...504..636H} implemented in \texttt{GalSim} and observed significant non-linear effects at $|\gamma_i| \gtrsim 0.05$. However, it is important to recognise that different interpretations of the KSB algorithm can result in subtle variations across various implementations (e.g.~\citealt{Heymans2006MNRAS.368.1323H}). Therefore, the more pronounced non-linear effects reported by \citet{Jansen2024AA...683A.240J} only reflect the performance of the specific KSB implementation in \texttt{GalSim}, rather than the general KSB method. In contrast, Hoekstra et al. (in prep.) find much weaker non-linear effects with a different KSB implementation.

\begin{table*}
    \centering
    \caption{Shear biases constrained from Eq.~(\ref{eq:nonlinear}) for different samples.}
    \renewcommand{\arraystretch}{1.5}
    \label{table:nonlinear}      
    \begin{tabular}{lccccc} 
    \hline\hline      
    Sample & $\tilde{m}_i\ [10^{-3}]$ & $\tilde{c}_i\ [10^{-5}]$ & $\tilde{d}_i\ [10^{-3}]$ & $\tilde{q}_i\ [10^{-1}]$ & $\tilde{m}_{\perp,i}\ [10^{-4}]$ \\ 
    \hline            
     & \multicolumn{5}{c}{$i=1$} \\ 
     \hline
    Complete & $\hphantom{-}0.042\pm0.000$ & $\hphantom{-}0.000\pm0.002$ & $\hphantom{-}0.000\pm0.002$ & $\hphantom{-}0.000\pm0.000$ & $\hphantom{-}0.792\pm0.002$ \\
    SExtractor (20, 24.5) & $-23.22\pm0.069$ & $-1.314\pm0.250$ & $-0.497\pm0.451$ & $\hphantom{-}0.262\pm0.058$ & $\hphantom{-}0.919\pm0.379$ \\
    \textit{lens}fit selection & $-13.63\pm0.142$ & $-0.568\pm0.582$ & $-2.514\pm0.871$ & $-0.041\pm0.116$ & $\hphantom{-}4.289\pm0.693$ \\
    \textit{lens}fit measurement & $\hphantom{-}0.808\pm0.140$ & $\hphantom{-}0.040\pm0.578$ & $-6.505\pm0.936$ & $-4.388\pm0.119$ & $\hphantom{-}2.431\pm0.741$ \\
    \hline            
     & \multicolumn{5}{c}{$i=2$} \\ 
     \hline
    Complete & $-0.042\pm0.000$ & $\hphantom{-}0.000\pm0.002$ & $\hphantom{-}0.000\pm0.002$ & $\hphantom{-}0.002\pm0.000$ & $\hphantom{-}0.795\pm0.002$ \\    
    SExtractor (20, 24.5) & $-23.32\pm0.083$ & $\hphantom{-}1.903\pm0.305$ & $\hphantom{-}1.216\pm0.527$ & $\hphantom{-}0.148\pm0.070$ & $\hphantom{-}2.853\pm0.437$ \\
    \textit{lens}fit selection & $-12.56\pm0.173$ & $-3.208\pm0.644$ & $-3.270\pm1.006$ & $-0.478\pm0.137$ & $\hphantom{-}1.459\pm0.858$ \\
    \textit{lens}fit measurement & $\hphantom{-}1.686\pm0.179$ & $-1.661\pm0.698$ & $-3.147\pm1.044$ & $-4.609\pm0.139$ & $\hphantom{-}3.234\pm0.876$ \\
    \hline                  
    \end{tabular}
    \renewcommand{\arraystretch}{1.0}
    \tablefoot{The first three samples are based on a perfect galaxy shape measurement of Eq.~(\ref{eq:shearRelation}) but with different sample selections. The label `Complete' refers to the full sample with perfect shape noise cancellation, `SExtractor (20, 24.5)' indicates the sample detected by SExtractor with observed magnitudes in the range of (20, 24.5), and `\textit{lens}fit selection' denotes samples further selected by \textit{lens}fit. The `\textit{lens}fit measurement' sample is the usual sample with galaxy shapes measured by \textit{lens}fit.  }
\end{table*}

\end{appendix}

\end{document}